\documentclass{JHEP3}

\usepackage{amsfonts}
\usepackage{amsmath}
\usepackage{array}
\usepackage{feynmp}

\newcommand{\br}[1]{\langle #1 \rangle}
\newcommand{\sq}[1]{[ #1 ]}
\newcommand{\tra}[1]{\hbox{Tr}\left( #1 \right)}

\newcommand{\splitti}[0]{\hbox{Split}}
\newcommand{\ie}[0]{\emph{i.e.}\,}
\newcommand{\eg}[0]{\emph{e.g.}\,}
\newcommand{\etal}[0]{\emph{et al. }}
\newcommand{\cf}[0]{\emph{cf. }}
\newcommand{\soft}[0]{\hbox{Soft}}
\newcommand{\ifi}[0]{\hbox{if }}
\newcommand{\elsi}[0]{\hbox{else }}
\newcommand{\ant}[0]{\hbox{Ant}}

\newcommand{\ha}[0]{\hat a}
\newcommand{\hb}[0]{\hat b}
\newcommand{\hka}[0]{k_{\hat a}}
\newcommand{\hkb}[0]{k_{\hat b}}

\newtheorem{ruler}{Rule}

\author{Claude Duhr, Fabio Maltoni \\ 
Center for Particle Physics and Phenomenology (CP3)\\
Universit\'e catholique de Louvain\\
Chemin du Cyclotron 2\\
B-1348 Louvain-La-Neuve, Belgium\\
E-mail:~\email{claude.duhr@uclouvain.be, fabio.maltoni@uclouvain.be}}

\title{Antenna functions from  MHV rules}

\abstract{QCD amplitudes display a universal behaviour when one or 
more partons are soft and/or collinear. This can expressed 
in terms of antenna functions which are much simpler 
than the full amplitudes and yet correctly embody their 
infrared behaviour. We show how 
antenna functions can be naturally obtained via a 
twistor-inspired MHV approach. As an application, 
we present compact results for MHV and NMHV 
antennas functions valid for any number of gluons.  These are sufficient 
to calculate the complete set of tree-level gluon antenna functions up to
N$^3$LO. As an interesting corollary, we prove that splitting
amplitudes too can be written directly through a MHV diagrammatic approach.
Finally we find that antenna functions, collinear splitting 
amplitudes and eikonal factors satisfy the same kind of 
recursive relation as the full amplitudes.}

\preprint{CP3-08-35}

\keywords{QCD}

\begin{document}

\tableofcontents

\section{Introduction}
\label{sec:intro}

Progress in computations in perturbative QCD has witnessed an
impressive acceleration in recent past: calculations that only a few
years ago were considered out-of-reach have been completed and many
more are now in sight. Particularly impressive are the developments of
new methods for computing one-loop amplitudes with many external legs
and two-loop amplitudes, that bypass the commonly employed Feynman
diagrams techniques and lead to compact expressions and possibly to an
efficient automatization of NLO and NNLO computations. Such glaring
advances have mainly come from a spur of activity initiated in 2003 by
Witten's suggestion of a duality between a string theory in a twistor
space and QCD~\cite{Witten:2003nn}. Starting from the work of Cachazo,
Britto et al.~\cite{Britto:2004nc}, and building up from the previous
seminal work by Bern, Dixon and Kosower~\cite{Bern:1994zx,
Bern:1994cg}, it has been quickly realized that generalised unitarity
methods could provide the philosopher's stone that, loosely speaking,
would turn loop computations into tree-level ones. Since then many new
competing and promising methods to evaluate one-loop amplitudes have
been proposed, some of which have been succesfully implemented in
working codes~\cite{Ossola:2007ax, Berger:2008sj}. Progress in
two-loop calculations has been somewhat slower, but also steady.

In this quest, it is not only important to have efficient tools for
the computations of the necessary ingredients (loop and tree-level
amplitudes) but also a general algorithm that allows their suitable
combination to evaluate infrared-safe observables, such as jet rates.
Several general algorithms for computing observables at NLO are
available since many years, the most popular one being the subtraction
method introduced by Catani and Seymour~\cite{Catani:1996vz}.

Understanding the infrared singular behaviour of tree-level QCD
amplitudes is a prerequisite for computing infrared-finite cross
sections at fixed order in perturbation theory.  In general, when one
or more final state particles are either soft or collinear, the
amplitudes factorise into a product of a scattering amplitude that
depends only on the remaining hard partons in the process (including
any hard partons constructed from an ensemble of unresolved partons)
and a splitting amplitude containing all the singularities due to the
unresolved particles. This factorisation is universal and can be
generalised to more
particles~\cite{Gehrmann-DeRidder:2005cm,Campbell:1997hg,
Catani:1998nv, Catani:1999ss, DelDuca:1999ha} and any number of
loops~\cite{Kosower:1999xi}.

Although several NLO implementations of this technique have already
been around for many years, at NNLO the infrared structure becomes
much more complicated due to an overlapping of the different singular
regions of phase space, and a complete and process-independent
subtraction scheme is still to be worked out.  In this case also,
several proposals have been put forward. A first one by Trocsanyi and
Somogyi resolves the complicated phase space structure by using a
special form of the soft and splitting amplitudes such that there are
no overlapping regions by construction~ \cite{Nagy:2007mn,
Somogyi:2005xz, Somogyi:2006cz, Somogyi:2006da, Somogyi:2006db}. A
second one is based on antenna functions at the amplitude
level~\cite{Kosower:1997zr,Kosower:2002su,Kosower:2003bh} and
amplitude squared level~\cite{Gehrmann-DeRidder:2005cm, GehrmannDeRidder:2004tv}, a
generalisation of the Catani-Seymour dipoles, which interpolate
between the different soft and collinear regions and avoid in this way
the double-counting problem. This method is the only one that has
been successfully applied in a non-trivial calculation so far, \ie, 
the NNLO 3-jet rates in $e^+e^-$
collisions~\cite{GehrmannDeRidder:2007hr,GehrmannDeRidder:2007jk, GehrmannDeRidder:2008ug}. Here subtraction terms
derived from full matrix elements can be viewed as antenna functions,
encapsulating all singular limits due to unresolved partonic emission
between two colour-connected hard
partons~\cite{Gehrmann-DeRidder:2005cm,Campbell:1998nn}.  In
particular, process-independent antenna functions describing arbitrary
QCD multiparticle processes can be directly related to three-parton
matrix elements at NLO (one unresolved parton radiating between two
colour-connected hard partons) and four-parton matrix elements at NNLO
(two unresolved partons radiating between two colour-connected hard
partons)~\cite{Gehrmann-DeRidder:2005hi, Gehrmann-DeRidder:2005aw}.

The main purpose of this work is to show that antenna functions can be
efficiently defined and calculated at the amplitude level by means of
the twistor CSW (or MHV) rules~\cite{Cachazo:2004kj, Risager:2005vk}.  
We apply the MHV rules to study the singular
limits of QCD amplitudes when $n$ gluons become soft and/or collinear,
and present an alternative definition for tree-level antenna
functions. Our work can be considered an extension of the work by
Kosower which is based on the standard recursive Berends-Giele
relations~\cite{Kosower:1997zr,Kosower:2002su,Kosower:2003bh} and that
of Birthwright  \emph{et al.} where MHV rules were used to derive
multi-collinear limits of amplitudes involving quarks and
gluons~\cite{Birthwright:2005ak, Birth:2005vi}.

In Refs.~\cite{Birthwright:2005ak, Birth:2005vi} it has been 
shown that any splitting amplitude can  be extracted from a 
given subset of MHV diagrams ($D^{\rm MHV}_i$), schematically,
\begin{equation}
{\rm Split}(1,\ldots, n) =
\frac{ \lim_{1||\dots||n}  \sum_{i \in {\cal S}}  
D^{\rm MHV}_i |_{(1,\dots,n,a,b,c)}}
{{\rm Born}(1+\dots+n,a,b,c)}\,.
\label{eq:glover}
\end{equation}
where ${\cal S}$ is  a significantly smaller subset of all the possible 
MHV diagrams contributing to a generic amplitude $A_{n+3}(1,\dots,n,a,b,c)$.
In this work we generalise the above approach in various directions. 
First we prove that a similar result holds for the antenna functions, \ie,
\begin{equation}
{\rm Ant}(\hat a,\hat b \leftarrow a, 1, \ldots, n, b) =
\frac{ 
\lim_{1 \sim \dots \sim n}  
\sum_{i \in {\cal A}}  D^{\rm MHV}_i |_{(a, 1,\dots,n, b ,c,d)}
}
{ {\rm Born}(\hat a,\hat b,c,d) }\,.
\label{eq:FACTantenna}
\end{equation}
where we indicated with $\sim$ the possibility of partons being
collinear or soft. Second we are able to prove that the same result can
be obtained in a much more direct way, \ie, without the need of taking any
limit and ratio but by  simply summing over the same subset of MHV
amplitudes but calculated at the shifted momenta $\hat a, \hat b$ 
\begin{equation}
{\rm Ant}(\hat a,\hat b \leftarrow a,1,\ldots, n, b) =
\sum_{i \in {\cal A}}  D^{\rm MHV}_i |_{(\hat a, \hat b, a, 1,\dots,n, b)}\,.
\label{eq:MHVantenna}
\end{equation}
Employing such a diagrammatic approach, very compact general formulas
for MHV and NMHV antenna functions are obtained.  As a corollary we
also prove that splitting amplitudes can be built diagramatically through
a similar formula, which manifestly shows the pole structure of the
splitting amplitudes.

Eq.~(\ref{eq:MHVantenna}) is the main result of this work.  We then
proceed by following the same approach presented in
Ref.~\cite{Duhr:2006iq} to rewrite the usual CSW rules into a
recursive form and we are able to recast Eq.~(\ref{eq:MHVantenna}) in a recursive form. As a result, we find
that the full amplitudes, the antenna functions and the splitting amplitudes
obey formally identical relations, the only differences being in the
definition of vertices and in the initial conditions.  This is our
second main result.

The paper is organised as follows. We start reviewing the basic
concepts and establish the notation for helicity amplitudes in
Section~\ref{sec:twistreview} and for their infrared factorisation
properties in Section~\ref{sec:IRreview}.  In
Section~\ref{sec:antennaCSW} we prove that a power counting argument
similar to that employed in Refs.~\cite{Birthwright:2005ak,
Birth:2005vi} for the splitting amplitudes, can be easily extended to
antenna functions and allows the identification of  
the set of MHV diagrams entering in
Eq.~(\ref{eq:FACTantenna}). In the following sections we first 
prove Eq.~(\ref{eq:MHVantenna}), \ie, that antenna functions 
can be directly calculated by summing over a well 
defined class of MHV diagrams and then provide some closed form results for
MHV and NMHV antenna functions. In Section~\ref{sec:antCSWform} 
we give a simple argument
that proves that also splitting amplitudes can be directly built via a
diagramatic approach. In Section~\ref{sec:antRR} recursive
formulations for both antennas and splitting amplitudes are proposed
and proved.  We summarize our findings in the Conclusion and discuss
some open issues.  The appendices contain some complementary
information, including proofs and explicit results for up to N$^3$LO
antennas.

\section{Tree-level techniques in QCD}
\label{sec:twistreview}

In this section we briefly review the notion of color decomposition of
tree-level QCD amplitudes and spinor helicity formalism. This allows
us to present our conventions and to also underline the aspects that 
will play an important role in the following. 

The basic idea of a color decomposition is to factorise
the information on the gauge structure from the kinematics. As an
example, consider the amplitude for $n$ gluons of colors 
$a_1,a_2,\ldots,a_n$ with $a_i=1,\ldots,N^2-1$. One can easily prove that
at tree level, such an amplitude can be decomposed as~\cite{Mangano:1990by}
\begin{equation}
\label{eq:gcolordecomp}
\begin{cal}A\end{cal}_n\left(\{p_i,h_i,a_i\}\right)=g^{n-2}\sum_{\sigma \in \ S_{n-1}}\tra{T^{a_1}\ldots T^{a_{\sigma (n)}}} A_n\left(1^{h_1},\ldots ,\sigma (n^{h_n})\right),
\end{equation}
where $T^a$ are the fundamental-representation matrices of $SU(N)$, and the sum is over all
$(n-1)!$ permutations of $(2, . . . , n)$.~\footnote{For alternative 
color decompositions see Refs.~\cite{Maltoni:2002mq,DelDuca:1999rs}.} 
Each trace corresponds to a particular color structure. The factor
associated with each color structure, $A$, is gauge invariant and is called a color-ordered
amplitude.\footnote{Also referred to as a dual amplitude or partial
amplitude.} It depends on the four-momenta $p_i$ and polarization
vectors $\epsilon_i$ of the $n$ gluons, represented simply by $i$ in
its argument. The color-ordered amplitudes are far simpler to
calculate than the full amplitude $\begin{cal}A\end{cal}$ due to the
smaller number of Feynman diagrams contributing to them.

As the partial amplitudes are functions only of the helicities and the
momenta of the particles, they can be most easily calculated using
a spinor-helicity formalism. All partial amplitudes can then be expressed
as rational functions of spinor products, defined by
\begin{equation}
\br{ij}\equiv \overline{u_-\left(k_i\right)}\,u_+\left(k_j\right) \qquad \sq{ij}\equiv \overline{u_+\left(k_i\right)}\,u_-\left(k_j\right),
\end{equation}
where $u_\pm(k)$ are the helicity-projected solutions of the massless
Dirac equation in momentum space.  The spinor products fulfill several
useful properties among which we recall the Schouten-identity
\begin{equation}
\label{eq:schoutenid}
\br{ij}\br{kl}=\br{ik}\br{jl}+\br{il}\br{kj}.
\end{equation}
Although the calculation of the partial amplitudes may in general be a
very hard task, there are special classes of partial amplitudes for
which the result can be written in a very compact form:
\begin{enumerate}
\item
Using the effective supersymmetry of QCD at high energy, it is
possible to derive supersymmetric Ward identities for tree-level QCD
amplitudes. These identities imply that the partial amplitudes where
all the particles have the same helicity or only one particle has a
different helicity vanish,
\begin{equation}
\label{eq:SUSYWard}
\begin{split}
A_n\left(1^\pm,2^+,\ldots,n^+\right) & =  0,\\
A_n\left(1^\mp,2^-,\ldots,n^-\right) & =  0.
\end{split}
\end{equation}
Note that the above relations hold at any order in perturbation theory in supersymmetric gauge theories.
\item
The first class of non vanishing partial amplitudes are those with
exactly two negative helicity gluons (independently of their position). 
These amplitudes are the so-called maximally helicity violating 
(MHV) amplitudes, and were first conjectured by Parke and Taylor in 
Ref.~\cite{Parke:1986gb}, and later proven by Berends and Giele 
using their recursive relations~\cite{Berends:1987me}. 
MHV-amplitudes at tree-level have a very simple analytic structure given by
\begin{equation}\label{eq:MHVamp}
A_n\left(1^+,\ldots,i^-,\ldots,j^-,\ldots,n^+\right)=\frac{\br{ij}^4}{\br{12}\br{23}\ldots\br{(n-1)n}\br{n1}},
\end{equation}
and using parity
\begin{equation}\label{eq:googlyMHVamp}
A_n\left(1^-,\ldots,i^+,\ldots,j^+,\ldots,n^-\right)=(-1)^n\frac{\sq{ij}^4}{\sq{12}\sq{23}\ldots\sq{(n-1)n}\sq{n1}}.
\end{equation}
It can be easily seen that all amplitudes for $n=4$ and $n=5$ can be obtained from the above formulas. However, starting from $n=6$ on, no general formula valid for any helicity configuration is known.
\end{enumerate}
Until a few years ago the two main techniques used to calculate partial 
amplitudes were the color-ordered Feynman rules~\cite{Dixon:1996wi} and the Berends-Giele recursive
relations~\cite{Berends:1987me}. In 2003, E.~Witten conjectured a
duality between QCD and a string theory in twistor
space~\cite{Witten:2003nn}. The two main outcomes of the twistor
approach to QCD at tree-level were the so-called BCF recursive
relations~\cite{Britto:2004ap, Britto:2005fq} and the CSW
formalism\footnote{Also known as
MHV rules.}~\cite{Cachazo:2004kj, Risager:2005vk}.  In the
following we will recall briefly the CSW formalism.

The CSW formalism was originally introduced as a conjecture in
Ref.~\cite{Cachazo:2004kj} and later proven recursively in
Ref.~\cite{Risager:2005vk}. It states that all
tree-level color-ordered amplitudes can be built up by connecting
MHV-amplitudes by scalar propagators. However, one then needs to
consider MHV-amplitudes where some of the external legs are
off-shell. It was shown in Ref.~\cite{Cachazo:2004kj} that for an
off-shell line carrying momentum $P_{a\dot a}$, one can choose
\begin{equation}
P_{a\dot a}\eta^{\dot a},
\end{equation}
where $\eta^{\dot a}$ is an arbitrary antiholomorphic spinor. 
The MHV rules can be used to calculate tree-level color-ordered amplitudes 
and are very simple:~\cite{Cachazo:2004kj}
\begin{enumerate}
\item Write down all possible ways to decompose the amplitude into
      MHV-amplitudes connected by scalar propagators $1/P^2$.
\item Introduce a term $P_{a\dot a}\eta^{\dot a}$ for each off-shell leg in 
an MHV-amplitude.
\end{enumerate}
Note that as the vertices in the CSW formalism correspond to
MHV-amplitudes, Eq.~(\ref{eq:MHVamp}), they are completely holomorphic
objects\footnote{This is equivalent to state that as long as the
$\eta$-dependence is kept general, the antiholomorphic spinor products
coming from the off-shell continuation must cancel out in the
end.}. Thus, in the CSW formalism the only possible source of
antiholomorphic spinor products in an amplitude are the scalar
propagators. This important property will be extensively used in
Section~\ref{sec:antennaCSW}.


\section{Infrared factorisation of tree-level QCD amplitudes}
\label{sec:IRreview}

It is well known that amplitudes with massless particles in the final
state exhibit infrared singularities in the limit where a particle
becomes soft or where two or more particles become collinear.  A
precise knowledge of the infrared behavior of the QCD amplitudes is
thus needed to handle the infrared divergencies that appear in QCD
calculations beyond LO.  The infrared factorisation of QCD amplitudes
is universal and the divergencies can be described by universal
quantities, known as splitting amplitudes and soft
factors. The knowledge of these universal functions is at the basis
of the subtraction methods, which handle the infrared
divergencies and allow a numerical implementation~\cite{Catani:1996vz, Gehrmann-DeRidder:2005cm,  Nagy:2007mn,
 Somogyi:2006cz, Somogyi:2006da,
Somogyi:2006db, Frixione:1995ms, Catani:2007vq}.  In this section we review the different
factorisation properties of a QCD gluon amplitude. We start by
recalling the soft and collinear limits, and finally we also review
the antenna factorisation, which provides a tool to describe in a
unified way all the infrared singularities contained in an amplitude.

Collinear singularities arise if two or more adjacent particles, say
$1,\ldots,n$, in a color-ordered amplitude become collinear. In this
limit, the amplitude factorises as
\begin{equation}
\label{eq:colfac}
A_m(1,\ldots,n,\ldots,m)\sim \sum_h\splitti_{-h}(1,\ldots,n)A_{m-n+1}(P^h,n+1,\ldots,m),
\end{equation}
where $P$ denotes the combined momentum $p_1+\ldots+p_n$, \ie,
denotes the collinear direction.  The factorisation
property~(\ref{eq:colfac}) implies that in the region of phase space
where the particles $1,\ldots,n$ become collinear, the amplitude
$A_m(1,\ldots,n,\ldots,m)$ can be approximated by the right-hand side
of Eq.~(\ref{eq:colfac}).

In the calculation of the splitting amplitudes, one is generally interested
to have the simplest of all cases $m-n+1=4$, \ie, the factorisation over 
the four-point amplitude, 
\begin{equation}
\label{eq:colfac2}
A_{n+3}(1,\ldots,n,a^+,b^{-h},c^-) \sim  \splitti_{-h}(1,\ldots,n)A_{4}(P^h,a^+,b^{-h},c^-).
\end{equation}
The first important observation is that, in this case, 
due to the supersymmetric Ward identities~(\ref{eq:SUSYWard}), no sum over 
the helicities of the intermediate particle has to be carried out. 
Furthermore, from Eq.~(\ref{eq:MHVamp}) the hard four-point amplitude 
always has  the simple analytic structure of an MHV-amplitude. The
splitting amplitudes can therefore be extracted from an $(n+3)$-point
amplitude using a simple algorithm, based on a simple power counting:
\begin{enumerate}
\item
If $i$ and $j$ are particles from the collinear set, 
rescale the corresponding spinor products as
\begin{equation}
\begin{split}
\br{ij} & \rightarrow  t\,\br{ij},\\
\sq{ij} & \rightarrow  t\,\sq{ij}.
\end{split}
\end{equation}
\item
Expand the amplitude in powers of  $t$,
\begin{equation}
A_{n+3}(1,\ldots,n,a^+,b^{-h},c^-)=\begin{cal}O\end{cal}\left(1/t^{n-1}\right)+\ldots,
\end{equation}
where the dots indicate terms that are less divergent than $1/t^{n-1}$. 
The splitting amplitude corresponds to the term of maximal divergence, 
\ie, if $i$ is in the collinear set and $a$ is not in the collinear set, 
then the splitting amplitude corresponds to the term in $1/t^{n-1}$, with 
\begin{equation}
\label{eq:colrule}
\begin{split}
\br{ai} & \rightarrow  \sqrt{z_i}\br{aP},\\
\sq{ai} & \rightarrow  \sqrt{z_i}\sq{aP},
\end{split}
\end{equation}
where $z_i$ denote the longitudinal momentum fractions.
 \end{enumerate}
 
The factorisation of the amplitude in the soft limit is very different from
the collinear one. Consider a pure gluon color-ordered amplitude at tree-level. This amplitude displays an infrared divergence as some of the gluons become
soft, \ie, some of the gluon energies vanish. Two different cases might occur,
the color-connected and the color-disconnected,
that need to be considered separately~\cite{Campbell:1997hg}.

Consider first the color-connected case  where the soft gluons are 
all adjacent in the color-ordered amplitude, say $s_1,\ldots,s_m$. 
The amplitude then factorises as
\begin{equation}
\label{eq:csoft}
A_n(1,\ldots,a,s_1,\ldots,s_m,b,\ldots,n)\sim \soft(a,s_1,\ldots,s_m,b)A_{n-m}(1,\ldots,a,b,\ldots,n).
\end{equation}
This factorisation property implies that in the region of phase space
where the particles $s_1,\ldots,s_m$ become soft, the amplitude
$A_n(1,\ldots,a,s_1,\ldots,s_m,b,\ldots,n)$ can be approximated by the
right-hand side of Eq.~(\ref{eq:csoft}).  The (color-ordered) soft
factor $\soft(a,s_1,\ldots,s_m,b)$ defined in this way depends on the
momenta and the helicities of the soft particles, and also on the
particles $a$ and $b$ which are next to it
in the color-ordered amplitude. The soft factors
are, however, independent of the helicities of the particles $a$ and
$b$~\cite{Dixon:1996wi}.

The color-disconnected case corresponds to the situation where the
soft particles are not all adjacent, but separated by at least one
hard gluon. In this case the factorisation is, \eg\,for two sets of
soft particles,
\begin{eqnarray}
 & &
 A_n(1,\ldots,a,s_1,\ldots,s_m,b,\ldots,c,t_1,\ldots,t_l,d\ldots,n)\sim
 \\
& & \qquad
\soft(a,s_1,\ldots,s_m,b)\,\soft(c,t_1,\ldots,t_l,d)\,A_{n-m-l}(1,\ldots,a,b,\ldots,c,d,\dots,n),\nonumber
\end{eqnarray}
\ie, the color-disconnected soft factors are products of the corresponding color-connected soft factors. For the rest of this work we will thus only deal with the color-connected case.

Similar to the case of splitting amplitudes, the calculation of soft
factors can be simplified by choosing the hard amplitude to be an
MHV-amplitude,
\begin{equation}
\label{eq:csoftmhv}
A_{n+4}\big(a^+,1,\ldots,n,b^+,c^-,d^-\big)\sim \soft(a,1,\ldots,m,b)A_{4}\big(a^+,b^+,c^-,d^-\big).
\end{equation}
Note that as the soft factor is independent of the helicities of the hard particles $a$ and $b$, it is unaffected by the helicity assignment in the hard part. The soft factors can be obtained by simple power counting, following a set of simple rules:
\begin{enumerate}
\item
Rescale all spinor products according to
\begin{eqnarray}
\br{ij} & \rightarrow & t^2\, \br{ij},\nonumber\\
\br{aj} & \rightarrow & t\, \br{aj},\nonumber\\
\sq{ij} & \rightarrow & t^2\, \sq{ij},\label{eq:brij}\\
\sq{aj} & \rightarrow & t\, \sq{aj}, \nonumber\\
s_{ij} & \rightarrow & t^4\, s_{ij},\nonumber\\
s_{aj} & \rightarrow & t^2\, s_{aj}\nonumber.
\end{eqnarray}
\item
Expand the amplitude in  power of $t$, 
\begin{equation}
A_n(1,\ldots,a,s_1,\ldots,s_m,b,\ldots,n)=\begin{cal}O\end{cal}\left(1/t^{2m}\right)+\ldots,
\end{equation}
where the dots indicate terms which are less divergent than $1/t^{2m}$. The soft factor then corresponds to the ``maximal divergence'', \ie, to the term in $1/t^{2m}$.
\end{enumerate}

The collinear and soft factors presented above are
sufficient to describe all of the infrared divergencies of a QCD
amplitude. However, it is sometimes more convenient to define a
general class of objects that describe all the infrared singularities
in one single function, the so-called antenna
function~\cite{Kosower:1997zr,Kosower:2002su,Kosower:2003bh}. Let us
consider an $N$-point pure gluon amplitude
$A_m(a,1,\ldots,n,b,\ldots,m)$. The singular limit for the particles
$1,\ldots,n$ is defined as the limit where the particles $1,\ldots,n$
are either collinear to $a$, collinear to $b$ or soft.

An antenna function is defined as any function
$\ant(\ha^{h_{\ha}},\hb^{h_{\hb}}\leftarrow a,1,\ldots,n,b)$ that
reproduces the correct singular limits. Note that, at variance with the
splitting amplitudes and the soft factors, the definition above
does not lead to unique functions. In the singular regions of phase space the
amplitude can be approximated by
\begin{equation}
A_m(a,1,\ldots,n,b,\ldots,m) \sim \sum_{h_{\ha},h_{\hb}}\ant(\ha^{h_{\ha}},\hb^{h_{\hb}}\leftarrow a,1,\ldots,n,b)\,A_{m-n}({\hat a}^{-h_{\ha}},{\hat b}^{-h_{\hb}},\ldots,m),
\label{eq:antfac}
\end{equation}
where $k_{\hat a}$ and $k_{\hat b}$ are the so-called reconstruction functions, 
\begin{equation}
\begin{split}
\hka & =  f_{\hat a}(a,1,\ldots,n,b),\\
\hkb & =  f_{\hat b}(a,1,\ldots,n,b).
\end{split}
\end{equation}
The reconstruction functions satisfy the following properties:
\begin{enumerate}
\item On-shellness, $k_{\hat a}^2=k_{\hat b}^2=0$.
\item Momentum conservation, $\hka+\hkb= k_a+k_1+\ldots+k_n+k_b$.
\item They reduce to the right expressions in the various singular
      limits, \eg\,if $1,\ldots,n$ become collinear to $a$, then
      $\hka\rightarrow k_a+k_1+\ldots+k_n$ and $\hkb\rightarrow k_b$.
\item They leave the leading pole unchanged.
\end{enumerate}
Explicit expressions for the reconstruction functions 
are collected in Appendix~\ref{app:recfun}. As we already pointed out, 
the definition of an antenna function is not strict enough to 
uniquely fix its functional form and several
definitions can be found in the literature. In
Ref.~\cite{Kosower:2002su}, Kosower presented a way to build an
antenna function, based on the well known Berends-Giele recursive
relations.  In Refs.~\cite{Gehrmann-DeRidder:2005cm, Gehrmann-DeRidder:2005hi,
Gehrmann-DeRidder:2005aw}, Gehrmann,
Gehrmann-De Ridder and Glover defined antenna functions as ratios
between the full and the hard squared matrix elements.  In the following, 
we will present an alternative definition based on the CSW formalism and
the properties special to this approach.


\section{Power counting for antenna functions in the CSW formalism}
\label{sec:antennaCSW}

The CSW formalism provides a very powerful method to efficiently calculate
partial amplitudes. It is then natural to wonder whether
 it can also be used to calculate splitting amplitudes and/or  soft factors.
In Refs.~\cite{Birthwright:2005ak,Birth:2005vi}, Birthwright \etal 
showed that it is indeed possible to derive very simple and compact 
formulas for several classes of splitting amplitudes, using the fact that 
in the CSW formalism the pole structure of an amplitude is manifest. 
They derived rules to identify the CSW diagrams that contribute 
to a given collinear limit just by looking at the CSW pole structure.

In the following we show that it is possible to generalize the rules
of Refs.~\cite{Birthwright:2005ak,Birth:2005vi} such that they do not
only describe the collinear behavior of an amplitude, but the full
infrared behavior, \ie, the antenna functions.

We start by briefly reviewing the calculation of 
splitting amplitudes using the CSW formalism presented in 
Refs.~\cite{Birthwright:2005ak,Birth:2005vi}. 
The CSW pole structure of splitting amplitudes is given by 
\begin{eqnarray}\label{eq:CSWPoleSplit}
\begin{split}
\splitti_- & \sim & \frac{1}{[\quad ]^{n_-}}\,f(\br{\quad}),\\
\splitti_+ & \sim & \frac{1}{[\quad ]^{n_--1}}\,f(\br{\quad}),
\end{split}
\end{eqnarray}
where $n_-$ is the number of negative-helicity gluons in the collinear
set. In this case, if the hard amplitude is chosen to be a four-point
amplitude, Eq.~(\ref{eq:colfac2}), the CSW diagrams that contribute in
the collinear limit are exactly those where all CSW propagators go
on-shell. Furthermore, the rules introduced in the previous section,
Eq.~(\ref{eq:colrule}), in the CSW formalism become
\begin{equation}\label{eq:collim}
\begin{split}
\br{a\,P_{i,j}} & \rightarrow  \br{a\, P} \sq{P\,\eta} \sum_{\alpha=i}^{j}z_\alpha,\\
\br{k\, P_{i,j}} & \rightarrow  \sq{P\, \eta} \sum_{\alpha=i}^{j}\sqrt{z_\alpha}\br{k\,\alpha},\\
\br{P_{i,j}\, P_{k,l}} & \rightarrow  \sq{P\, \eta}^2 \sum_{\alpha=i}^j\sum_{\beta=k}^l\sqrt{z_\alpha z_\beta}\br{\alpha\,\beta},
\end{split}
\end{equation}
where $a $ is a particle which is not in the collinear set, $k$ is in the collinear set, and $P_{i,j}$ and $P_{k,l}$ only contain particles from the collinear set.
Introducing the following notation\footnote{These notations differ slightly from those used in Refs.~\cite{Birthwright:2005ak,Birth:2005vi}.}
\begin{equation}
\label{eq:delta}
\begin{split}
\Delta(i,j;k) & =  \sum_{\alpha=i}^{j}\sqrt{z_\alpha}\br{k\,\alpha},\\
\Delta(i,j;k,l) & =  \sum_{\alpha=i}^j\sum_{\beta=k}^l\sqrt{z_\alpha z_\beta}\br{\alpha\,\beta}\,,
\end{split}
\end{equation}
Eqs.~(\ref{eq:collim}) become
\begin{equation}
\label{eq:collimdelta}
\begin{split}
\br{k\, P_{i,j}} & \rightarrow  \sq{P\, \eta} \Delta(i,j;k),\\
\br{P_{i,j}\, P_{k,l}} & \rightarrow  \sq{P\, \eta}^2 \Delta(i,j;k,l).
\end{split}
\end{equation}
As already stated in Section~\ref{sec:twistreview}, 
$\eta$ is an arbitrary spinor that can be thought as parametrising 
the gauge-dependence of a quantity. As in the collinear limit the only $\eta$ dependence is in $\sq{P\eta}$, we can neglect these spinor products, \ie, we set $\sq{P\eta}=1$.
The procedure presented in Refs.~\cite{Birthwright:2005ak, Birth:2005vi} to extract splitting amplitudes goes as follows:
\begin{ruler}[Collinear limits in the CSW formalism]
\label{rule:glover}
\begin{enumerate}
\item[]$\phantom{abba}$
\item Consider all CSW diagrams contributing to the collinear limit of the $n+3$-point amplitude \mbox{$A_{n+3}\big(1,\ldots,n,(n+1)^+,(n+2)^{-h},(n+3)^-\big)$}. This set is obtained by including 
the diagrams where all the scalar propagators go on-shell in the collinear limit, or equivalently, the diagrams where $n+1$, $n+2$ and $n+3$ are attached to the same CSW vertex.
\item Go to the collinear limit by applying the rules in Eq.~(\ref{eq:collim}).
\item Divide by the hard four-point amplitude $A_4\big(P^{h},(n+1)^+,(n+2)^{-h},(n+3)^-\big)$.
\end{enumerate}
\end{ruler}
In the rest of this section we show how it is possible to extend this
procedure to the extraction of antenna functions. To do so, first one
has to find  the CSW diagrams that contribute to a given singular limit, \ie,
the CSW pole structures of the antenna function has to be identified. 

Let us start with the soft limits in the CSW construction. To be
concrete, let us consider an $(n+4)$-point gluon amplitude
$A_{n+4}(a,1,\ldots,n,b,c,d)$, where the gluons $1, \ldots,n$ become
soft. The amplitude then factorises according to Eq.~(\ref{eq:csoftmhv}). 
The soft factor can be easily calculated using the following result~\cite{Catani:1999ss},
\begin{ruler}[Gluon insertion rule]
\label{rule:gluoninsertionrule}
~\\[10pt]
In the soft limit, only Feynman diagrams where 
the soft gluons are radiated from the external legs of the 
hard amplitude contribute.
\end{ruler}
The proof of the gluon insertion rule, based on simple 
power counting arguments, is presented in Appendix~\ref{app:appendixa}.

\FIGURE[!t]{
            \begin{fmffile}{softemission}
           \parbox{30mm}{\begin{fmfgraph*}(50,50)
            \fmfleft{i1,i2,i3}
            \fmfright{o1,o2,o3}
            \fmflabel{$c^-$}{i1}
            \fmflabel{$d^-$}{i3}
            \fmflabel{$b^+$}{o1}
            \fmflabel{$a^+$}{o3}
            \fmflabel{$-$}{o2}
            \fmf{plain}{i1,va1,v1,v2,o1}
            \fmf{plain}{i3,va2,v1,va3,o3}
            \fmffreeze
            \fmf{dashes}{v2,o2}
            \end{fmfgraph*}}
             \parbox{30mm}{\begin{fmfgraph*}(50,50)
            \fmfleft{i1,i2,i3}
            \fmfright{o1,o2,o3}
            \fmflabel{$c^-$}{i1}
            \fmflabel{$d^-$}{i3}
            \fmflabel{$b^+$}{o1}
            \fmflabel{$a^+$}{o3}
            \fmflabel{$-$}{o2}
            \fmf{plain}{i1,va1,v1,v2,o1}
            \fmf{plain}{i3,va2,v1,va3,o3}
            \fmffreeze
            \fmf{dashes}{va3,o2}
            \end{fmfgraph*}}
             \parbox{30mm}{\begin{fmfgraph*}(50,50)
            \fmfleft{i1,i2,i3}
            \fmfright{o1,o2,o3}
            \fmflabel{$c^-$}{i1}
            \fmflabel{$d^-$}{i3}
            \fmflabel{$b^+$}{o1}
            \fmflabel{$a^+$}{o3}
            \fmflabel{$+$}{o2}
            \fmf{plain}{i1,va1,v1,v2,o1}
            \fmf{plain}{i3,va2,v1,va3,o3}
            \fmffreeze
            \fmf{dashes}{v1,o2}
            \end{fmfgraph*}}
                     \end{fmffile}
                     \vspace{3mm}
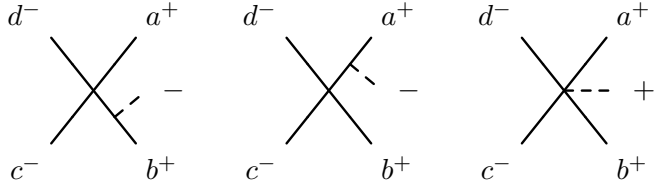
\caption{\label{fig:attachgluon}Possible CSW diagrams where a soft gluon is radiated between the external legs $a$ and $b$. The dashed line corresponds to a soft gluon.}
}

Rule~\ref{rule:gluoninsertionrule} allows an easy identification of 
the Feynman diagrams corresponding to the emission of additional soft gluons. 
In order to establish the CSW pole structure of the soft emission, we need to
generalize this result to CSW diagrams.  
Let us consider an $(n+4)$-point gluon amplitude
$A_{n+4}(a,1,\ldots,n,b,c,d)$, where the gluons $1, \ldots,n$ become
soft.  Rule~\ref{rule:gluoninsertionrule} states that in the soft
limit only those Feynman diagrams contribute where the soft particles
are radiated from the external legs $a$ and $b$. However, it is easy
to see that, given the helicity configuration, only
negative-helicity gluons can be radiated from the external legs in the
CSW formalism (a positive-helicity gluon would lead to a three-point
CSW vertex with two positive-helicity gluons). The positive-helicity
gluons must thus be radiated  only from the CSW
vertex which builds up the hard amplitude. Note that this is 
consistent with the gluon insertion rule (\cf the soft
factorisation of a pure MHV-amplitude). Similar considerations hold
true for different helicity assignments of the particles $a$, $b$,
$c$, $d$ in the hard amplitude.
 
This leads us to the following new formulation of the gluon insertion rule, 
valid for CSW diagrams  (See Fig.~\ref{fig:attachgluon}),
\begin{ruler}[Gluon insertion rule in the CSW formalism]
\label{ruler:gluoninsertionrule}
~\\[10pt]
In the soft limit, only CSW diagrams  where
negative-helicity soft gluons are radiated from the external legs $a$
and $b$ or positive-helicity soft gluons from the CSW vertex that
forms the hard amplitude contribute.
\end{ruler}

Note that the gluon insertion rule is a very restrictive on the possible
diagrams (Fig.~\ref{fig:attachgluon}), because it forces the hard 
gluons $c$ and $d$ to be attached to the same CSW vertex. 
Thus only a small number of CSW diagrams contributes in
the soft limit. Furthermore it is easy to see that, similar to
Rule~\ref{rule:glover}, this class of diagrams is exactly that
where all scalar propagators go on-shell.

We now turn to the CSW pole structure of soft limits. Consider the
situation where the set $\{1,\ldots,n\}$ contains $n_s$
negative-helicity gluons. The set $\{a,b,c,d\}$ contains two
additional negative-helicity gluons, so that the hard amplitude
$A_4(a,b,c,d)$ is an MHV-amplitude. The number of CSW propagators in
the $(n+4)$-point amplitude is then $p=(n_s+2)-2=n_s$. As the hard
four-point amplitude does not contain any CSW propagator, we come the
conclusion that a soft factor containing $n_s$ negative-helicity
gluons has the following CSW pole structure
\begin{equation}
\soft \sim \frac{1}{[\quad ]^{n_s}}\,f(\br{\quad}).
\end{equation}

We can extend this result to the case of color-connected soft/collinear limits\footnote{The color-disconnected case is trivial.}. Let us consider to this effect the amplitude $A_{n+3}(a,1,\ldots,k,\ldots,n,b,c)$ in the limit where $1,\ldots,k$ are soft and $k+1,\ldots,n$ are collinear. In this limit the amplitude can be approximated by~\cite{Campbell:1997hg}
\begin{equation}
\begin{split}
A_{n+3}(&a,1,\ldots,k,\ldots,n,b,c)\\
&\sim \sum_h\,\begin{cal}S\end{cal}(a;1,\ldots,k;k+1\dots,n)\splitti_{-h}(k+1,\ldots,n)A_{4}\left(a,P^h,b,c\right),
\end{split}
\end{equation}
where $\begin{cal}S\end{cal}$ denotes a universal soft factor. In Ref.~\cite{Campbell:1997hg} it was shown that $\begin{cal}S\end{cal}$ can be obtained by taking the leading soft behavior in the splitting amplitude $\splitti_{-h}(1,\ldots,n)$, \ie
\begin{equation}
\splitti_{-h}(1,\ldots,n)  \sim\begin{cal}S\end{cal}(a;1,\ldots,k;k+1\dots,n)\splitti_{-h}(k+1,\ldots,n).
\end{equation}
We will now determine the CSW pole structure by applying the gluon insertion rule~\ref{ruler:gluoninsertionrule} to the splitting amplitude $\splitti_{-h}(k+1,\ldots,n)$. From Rule~\ref{rule:glover} we know that in the limit where $k+1,\ldots,n$ are collinear, only those CSW diagrams of $A_{n-k+3}(a,k+1,\ldots,n,b,c)$ contribute where $a,b,c$ are attached to the same CSW vertex. We now apply the Rule~\ref{ruler:gluoninsertionrule} to attach soft gluons to this set of diagrams. It is easy to see that
\begin{itemize}
\item As soft negative-helicity gluons can only be emitted from the external lines, we add a new divergent propagator to the diagram each time we add a soft negative-helicity gluon. 
\item For the emission of soft positive-helicity gluons, the number of CSW propagators is left unchanged.
\end{itemize}
Hence, if $n_s$ denotes the number of soft negtive-helicity gluons in the set $\{1,\ldots,k\}$, we add $n_s$ new divergent propagators do the diagrams, and so
\begin{itemize}
\item if $h=+1$,
\begin{equation}
\begin{cal}S\end{cal}(a;1,\ldots,k;k+1\dots,n)\splitti_{-}(k+1,\ldots,n)\sim\frac{1}{\sq{\quad}^{n_s}}\,\frac{1}{\sq{\quad}^{n_-}}\sim\frac{1}{\sq{\quad}^{n_s+n_-}},
\end{equation}
\item if $h=-1$,
\begin{equation}
\begin{cal}S\end{cal}(a;1,\ldots,k;k+1\dots,n)\splitti_{+}(k+1,\ldots,n)\sim\frac{1}{\sq{\quad}^{n_s}}\,\frac{1}{\sq{\quad}^{n_--1}}\sim\frac{1}{\sq{\quad}^{(n_s+n_-)-1}},
\end{equation}
where $n_-$ is the number of negative-helicity gluons in the collinear set $\{k+1,\ldots,n\}$.
\end{itemize}

We can now turn to the case of antenna functions.  By definition, an
antenna function contains all possible infrared singularities, both
soft and collinear, that an amplitude can have if $n$ particles become soft
or collinear. We analyze the CSW pole structures of
the antenna functions, using the fact that they reproduce the known
CSW pole structures in the various soft and collinear limits.

Let us start with $\ant(\ha^-,\hb^- \leftarrow a,1,\ldots,n,b)$,
and consider the CSW pole structure of the different soft and
collinear singularities contained in this antenna function. Let $n_-$
denote the number of negative helicities in the set $\{1,\ldots,n\}$
and $N_-$ the number of negative helicities in the set
$\{a,1,\ldots,n,b\}$. Then we have the following possibilities:
\begin{itemize}
\item
$k_1,\ldots,k_n \parallel k_a$\\
In this limit we have $\hka \rightarrow P \equiv k_a+k_1+\ldots+k_n$ and $\hkb\rightarrow k_b$, and
\begin{equation}
\ant(\ha^-,\hb^- \leftarrow a,1,\ldots,n,b) \rightarrow \splitti_-(a,1,\ldots,n) \sim \frac{1}{\sq{\quad}^{n_a}} \sim \frac{1}{\sq{\quad}^{N_-}},
\end{equation}
where $n_a$ is the number of negative helicities in the set $\{a,1,\ldots,n\}$, and $n_a = N_-$.
Note that if $\hkb\rightarrow k_b$, then  $h_b=-h_{\hat b} = +1$. 
\item
$k_1,\ldots,k_n \parallel k_b$\\
In this limit we have $\hka \rightarrow k_a$ and $\hkb\rightarrow P \equiv k_1+\ldots+k_n+k_b$, and
\begin{equation}
\ant(\ha^-,\hb^- \leftarrow a,1,\ldots,n,b) \rightarrow \splitti_-(1,\ldots,n,b) \sim \frac{1}{\sq{\quad}^{n_b}} \sim \frac{1}{\sq{\quad}^{N_-}},
\end{equation}
where $n_b$ is the number of negative helicities in the set $\{1,\ldots,n,b\}$, and $n_b = N_-$.
Since  $\hka\rightarrow k_a$, then  $h_a=-h_{\hat a} = +1$.
\item
$k_1,\ldots,k_n \rightarrow 0$\\
In this limit we have $\hka\rightarrow k_a$ and $\hkb\rightarrow k_b$, and
\begin{equation}
\ant(\ha^-,\hb^- \leftarrow a,1,\ldots,n,b) \rightarrow \soft(a,1,\ldots,n,b) \sim \frac{1}{\sq{\quad}^{n_-}} \sim \frac{1}{\sq{\quad}^{N_-}},
\end{equation}
and $n_- = N_-$ because if $\hka\rightarrow k_a$, then $h_a=-h_{\hat a} = +1$ and 
if $\hkb\rightarrow k_b$, then  $h_b=-h_{\hat b} = +1$.
\item
$k_1,\ldots,k_{i-1} \parallel k_a$ and $k_i,\ldots,k_j \rightarrow 0$ and $k_{j+1},\ldots,k_n \parallel k_b$.\\
In this limit we have $\hka \rightarrow k_a+k_1+\ldots+k_{i-1}$ and $\hkb\rightarrow k_{j+1}+\ldots+k_n+k_b$, and
\begin{equation}
\begin{split}
\ant(\ha^-,\hb^-\leftarrow a,1,\ldots,n,b)\rightarrow &\,\splitti_-(a,1,\ldots,i-1)\,\splitti_-(j+1,\ldots,n,b)\\
&\times\begin{cal}S\end{cal}(a,1,\ldots,i-1;i,\ldots,j;j+1,\ldots,n,b) \\
  \sim& \frac{1}{\sq{\quad}^{n_a}}\,\frac{1}{\sq{\quad}^{n_s}}\, \frac{1}{\sq{\quad}^{n_b}} \sim \frac{1}{\sq{\quad}^{N_-}},
\end{split}
\end{equation}
where $n_a$, $n_s$ and $n_b$ are the numbers of negative-helicity gluons in the sets $\{a,1,\ldots,i-1\}$, $\{i,\ldots,j\}$ and $\{j+1,\ldots,n,b\}$ respectively, and $n_a+n_s+n_b = N_-$.
\end{itemize} 
We are therefore able to conclude that
\begin{equation}\label{eq:CSWPoleAnt1}
\ant(\ha^-,\hb^- \leftarrow a,1,\ldots,n,b)\sim \frac{1}{\sq{\quad}^{N_-}}\,f(\br{\quad}).
\end{equation}
Similar arguments for the other helicity assignments lead to
\begin{equation}\label{eq:CSWPoleAnt2}
\begin{split}
\ant(\ha^-,\hb^+ \leftarrow a,1,\ldots,n,b) & \sim  \frac{1}{\sq{\quad}^{N_--1}}\,f(\br{\quad}),\\
\ant(\ha^+,\hb^- \leftarrow a,1,\ldots,n,b) & \sim  \frac{1}{\sq{\quad}^{N_--1}}\,f(\br{\quad}),\\
\ant(\ha^+,\hb^+ \leftarrow a,1,\ldots,n,b) & \sim \frac{1}{\sq{\quad}^{N_--2}}\,f(\br{\quad}).
\end{split}
\end{equation}
Eqs.~(\ref{eq:CSWPoleAnt1} - \ref{eq:CSWPoleAnt2}) are the analogues for antenna functions of Eq.~(\ref{eq:CSWPoleSplit}) for splitting amplitudes. We show now that, as it was already the case for splitting amplitudes, the CSW pole structure allows us to identify a priori the set of CSW diagrams that contributes to the singular limit.

Let us start with $\ant(\ha^-,\hb^- \leftarrow a,1,\ldots,n,b)$,
and consider an $(n+4)$-point amplitude $A_{n+4}(a,1,\ldots,n,b,c^-,d^-)$. 
In the limit where the particles $1,\ldots,n$ become singular,
the amplitude exhibits the factorisation property, Eq.~(\ref{eq:antfac}),
\begin{equation}
\label{eq:mmant}
A_{n+4}(a,1,\ldots,n,b,c^-,d^-) \sim \ant(\ha^-,\hb^-  \leftarrow a,1,\ldots,n,b)\,A_{4}(\ha^+,\hb^+,c^-, d^-).
\end{equation}
If $N_-$ is the number of negative-helicity gluons in the set
$\{a,1,\ldots,n,b\}$, then the number of CSW propagators in the
$(n+4)$-point amplitude is $p=(N_-+2)-2=N_-$. From Eq.~(\ref{eq:CSWPoleAnt1}) we know that the CSW
pole structure of this antenna function is
\begin{equation}
\ant(\hat a^-,\hat b^- \leftarrow a,1,\ldots,n,b) \sim \frac{1}{\sq{\quad}^{N_-}}\,f(\br{\quad}),
\end{equation}
and thus the CSW diagrams contributing to the antenna function are exactly those where all scalar propagators go on-shell. Similar conclusions can be drawn for the other helicity configurations. This brings us to the first important conclusion that Rule~\ref{rule:glover} can be generalized to antenna functions:
\begin{quote}
\emph{The CSW diagrams that contribute to the singular limit of  the $(n+4)$-point ampltitude $A_{n+4}(a,1,\ldots,n, b,c^-,d^-)$ are exactly those were all scalar propagators go on-shell.}
\end{quote}


\section{Diagrammatic evaluation of antenna functions in the CSW formalism}
\label{sec:antCSWform}

In the previous section we showed how we can identify the CSW diagrams that contribute to the singular limit just by examining the pole structure of the CSW diagrams that build up the amplitude. The simple idea is that
not all the CSW diagrams that contribute to an amplitude are relevant in the computation
of a given singular limit. This result turns out to be very similar to the corresponding result for splitting amplitudes obtained in Refs.~\cite{Birthwright:2005ak, Birth:2005vi}.

For the antenna functions, however, it is possible to go one step further. We 
note that keeping only those CSW diagrams where all
propagators go on-shell is equivalent to requiring that
$c$ and $d$ must be attached to the same $m$-point CSW vertex with
$m\ge 4$. Let us consider now a specific CSW diagram satisfying this
condition. Then the CSW vertex with $c$ and $d$  attached  gives
a contribution
\begin{equation}
\label{eq:cd1}
\frac{\br{cd}^3}{\br{P_{j,b}\,c}\br{d\,P_{a,i}}},
\end{equation}
with $(i,j)\neq (b,a)$ because otherwise $c$ and $d$ would be attached to a 3-point CSW vertex, and $i=a$ and / or $j=b$  if $a$ and $b$ are attached to the same CSW vertex as $c$ and $d$. Consider the following Schouten identity, Eq.~(\ref{eq:schoutenid}),
\begin{equation}
\label{eq:Schouten}
\br{\hat b\hat a}\br{ca}=\br{\hat ba}\br{c\hat a}+\br{c\hat b}\br{a\hat a}
\Rightarrow \frac{\br{ca}}{\br{c\hat a}}=\frac{\br{\hat ba}}{\br{\hat b\hat a}}+\frac{\br{c\hat b}\br{a\hat a}}{\br{c\hat a}\br{\hat b\hat a}},
\end{equation}
where $\hat a$ and $\hat b$ denote the reconstruction functions defined in the Appendix~\ref{app:recfun}. It is manifest that, if   $\br{a\hat a}$ goes to zero in the singular limit, then Eq.~(\ref{eq:Schouten}) drastically simplifies. To show this, we go into the frame where $k_a$ and $k_b$ are aligned along the same direction.
As recalled in Appendix~\ref{app:recfun}, the reconstruction functions can be chosen to be of the form
\begin{equation}
\hka = A\,k_a+ B\, k_b+ \sum_{j=1}^nk_jr_j,
\end{equation}
where $A$, $B$ and $r_j$ are functions of invariants in the particle momenta, so they are unaffected by this specific choice of reference frame. As $\hka$ is by definition lightlike, we can switch to lightcone coordinates. In the specific reference frame we chose, we get
\begin{eqnarray}
\hka^+ & = & A\,k_a^++\sum_{j=1}^nk_j^+r_j,\nonumber\\
\hka^- & = & B\,k_b^-+\sum_{j=1}^nk_j^-r_j,\\
k_{\ha\bot} & = & \sum_{j=1}^nk_{j\bot}r_j.\nonumber
\end{eqnarray}
Furthermore it is easy to see that in this frame we can write
\begin{equation}
\label{eq:singularframe}
k_j \hbox{ singular} \,\Leftrightarrow\, k_{j\bot}\rightarrow 0,\qquad\forall j = 1,\ldots,n.
\end{equation}
The spinor product $\br{a\ha}$ can now be evaluated in this frame using the lightcone coordinates~\cite{DelDuca:1999ha},
\begin{equation}
\br{a\ha} = -i\sqrt{\frac{-k_a^+}{\hka^+}}\,k_{\ha\bot}=-i\sqrt{\frac{-k_a^+}{A\,k_a^++\sum_{j=1}^nk_j^+r_j}}\,\sum_{j=1}^nk_{j\bot}r_j.
\end{equation}
By definition of the reconstruction functions, $\hka$ and $k_a$ are
collinear in every singular limit, and so the square root gives just a
phase in the singular limit. Thus, due to
Eq.~(\ref{eq:singularframe}), $\br{a\ha}$ goes to zero in the singular
limit unless there are poles in the coefficients $r_j$ that could
prevent the product $r_j\,k_{j\bot}$ from going to zero as
$k_{j\bot}\rightarrow 0$. Recalling the analytic expression for the
coefficients, Eq.~(\ref{eq:coeff}), we see that $r_j$ may contain a
pole if $k_j\cdot K\rightarrow 0$.  As $K$ always contains the momenta
of the two hard particles $a$ and $b$, $k_j\cdot K \rightarrow 0$ if
and only if $k_j \rightarrow 0$. However, a quick look at
Eq.~(\ref{eq:coeff}) shows that $r_j$ does not contain a pole in this
limit, so $r_j$ does not contain a pole in any limit. This concludes
the proof that $\br{a\ha}$ goes to zero in the singular limit, and
allows us to conclude that in the singular limit
\begin{equation}
\frac{\br{ca}}{\br{c\ha}}\rightarrow \frac{\br{\hb a}}{\br{\hb \ha}}.
\end{equation} 
Similar conclusions can of course be drawn for $\br{b\hb}$.
We can also analyze what happens to the spinor product $\br{P_{j,b}\,c}$ in the singular limit. 
\begin{itemize}
\item[-]
If $j=b$, then $\br{P_{j,b}\,c} = \br{bc}$, and in the singular limit
\begin{equation}
\label{eq:singbc}
\br{bc}\rightarrow \br{\hb c}\frac{\br{\ha b}}{\br{\ha\hb}}.
\end{equation}
\item[-]
If $j\neq b$, then $\br{P_{j,b}\,c}=\br{bc}\sq{b\eta}+\sum_{k=j}^n\br{kc}\sq{k\eta}$. 
The first term has already been dealt with in Eq.~(\ref{eq:singbc}). The remaining terms can be rewritten using Schouten-identity
\begin{equation}
\br{kc}=\br{\hb c}\frac{\br{\ha k}}{\br{\ha\hb}}+\br{\ha c}\frac{\br{k\hb}}{\br{\ha\hb}}.
\end{equation}
The second term on the right-hand side, proportional to $\br{k\hb}$, does not contribute in the singular limit. 
To see this let us first see what happens in the limit  where $k_j,\ldots,k_\ell \parallel k_a$ and $k_{\ell+1},\ldots,k_n \parallel k_b$, with $j<\ell\le b$. Then this CSW diagram would have a non-divergent scalar propagator. As we are only looking for those diagrams where all scalar propagators go on-shell, we can neglect this case. So we only need to analyze the situation where $k_j,\ldots,k_n \parallel k_b$. Using lightcone coordinates, we find, $\forall\, j\le\ell\le n$,
\begin{equation}
\label{eq:brellhb}
\br{\ell\hb}=k_{\ell\bot}\sqrt{\frac{\hkb^+}{k_\ell^+}}-k_{\hb\bot}\sqrt{\frac{k_\ell^+}{\hkb^+}}.
\end{equation}
We know already that in the singular limit
\begin{equation}
\label{eq:ktolim}
k_{\hb}^+\rightarrow 0, \qquad k_{\hb\bot}\rightarrow 0,\qquad k_{\ell\bot}\rightarrow 0,\quad\forall 1\le\ell\le n\,,
\end{equation}
and $\hkb$ is collinear to $k_b$ in every singular limit.
Let us have a closer look at the first term. From Eq.~(\ref{eq:ktolim}) it follows that in the singular limit $k_{\ell\bot},\, k_\ell^+,\,k_{\hb}^+\rightarrow 0$, where we used the fact that in our specific choice of reference frame $k_b^+=0$. Furthermore, all external particles must fulfill the on-shell condition $k^+\,k^- = |k_\bot|^2$, and so 
\begin{equation}
\frac{k_{\ell\bot}}{\sqrt{k_\ell^+}}\sim \sqrt{k_\ell^-},
\end{equation}
and $k_\ell^-\neq 0$ if $k_j \parallel k_b$. So the first term goes to zero as $\sqrt{k_{\hb}^+}$. Similar arguments show that also the second term goes to zero in the singular limit, and so we can conclude that $\br{k\hb}$ vanishes.
Putting everything together, we obtain
\begin{equation}
\br{P_{j,b}\,c} \rightarrow \br{\hb c}\frac{\br{\ha b}}{\br{\ha\hb}}\sq{b\eta}+\sum_{k=j}^{n}\br{\hb c}\frac{\br{\ha k}}{\br{\ha\hb}}\sq{k\eta}=\br{\hb c}\frac{\br{\ha P_{j,b}}}{\br{\ha\hb}}.
\end{equation}
\end{itemize}
Both cases can be summarized as
\begin{equation}
\br{P_{j,b}\,c} \rightarrow \br{\hb c}\frac{\br{\ha P_{j,b}}}{\br{\ha\hb}}.
\end{equation}
Similarly one finds 
\begin{equation}
\br{d\,P_{a,i}} \rightarrow \br{d\ha}\frac{\br{P_{a,i}\,\hb}}{\br{\ha\hb}}.
\end{equation}
The contribution from Eq.~(\ref{eq:cd1}) to the singular limit therefore becomes
\begin{equation}
\label{eq:hardcut1}
\frac{\br{cd}^3}{\br{P_{j,b}\,c}\br{d\,P_{a,i}}}\rightarrow \frac{\br{cd}^3}{\br{\ha\hb}\br{\hb c}\br{d\ha}}\,\frac{\br{\ha\hb}^3}{\br{P_{j,b}\,\ha}\br{\hb\,P_{a,i}}}.
\end{equation}
The first factor on the right-hand side of Eq.~(\ref{eq:hardcut1}) corresponds to the hard four-point amplitude in Eq.~(\ref{eq:mmant}). The second factor has the same functional form as the left-hand side, with $c$ and $d$ replaced by $\ha$ and $\hb$. This proves the following result:
The antenna function $\ant(\ha^-,\hb^- \leftarrow a,1,\ldots,n,b)$ can be calculated by evaluating all CSW diagrams that contribute to the $(n+4)$-point amplitude $A_{n+4}(a,1,\ldots,n,b,\ha^-,\hb^-)$ and where $\ha$ and $\hb$ are attached to the same $m$-point CSW vertex, with $m\ge 4$.

In the rest of this section we show how this result can be generalized to the remaining antenna functions. The proof for $\ant(\ha^+,\hb^+ \leftarrow a,1,\ldots,n,b)$ is similar to the previous case, and so we do not give it explicitly here.

Let us turn to $\ant(\hat a^+,\hat b^- \leftarrow a,1,\ldots,n,b)$ and let us consider the $(n+4)$-point amplitude $A_{n+4}(a,1,\ldots,n,b,c^-,d^+)$. From Eq.~(\ref{eq:antfac}) we know that in the singular limit for the particles $1,\ldots,n$, the amplitude exhibits the factorisation property
\begin{equation}
\label{eq:antfac2}
A_{n+4}(a,1,\ldots,n,b,c^-,d^+) \longrightarrow \sum_h\ant(\ha^h,\hb^{-h} \leftarrow a,1,\ldots,n,b)\,A_{4}({\hat a}^{-h},{\hat b}^h,c^-, d^+).
\end{equation}
If $N_-$ is the number of negative-helicity gluons in the set $\{a,1,\ldots,n,b\}$, then the number of CSW propagators in the $(n+4)$-point amplitude is $p=(N_-+1)-2=N_--1$. Furthermore, from the considerations in Section~\ref{sec:antennaCSW} we know that the CSW pole structure of this antenna function is
\begin{equation}
\ant(\ha^+,\hb^- \leftarrow a,1,\ldots,n,b) \sim \frac{1}{\sq{\quad}^{N_--1}},
\end{equation}
and thus the CSW diagrams contributing to the antenna function are exactly those where $c$ and $d$ are attached to the same $n$-point CSW vertex, with $n\ge 4$. The contribution from the CSW vertex that contains $c$ and $d$ is of the form
\begin{equation}
\label{eq:pmcontribution}
\frac{\br{c\,P_{\alpha\beta}}^4}{\br{P_{j,b}\,c}\br{cd}\br{d\,P_{a,i}}},
\end{equation} 
where $P_{\alpha\beta}$ denotes the momentum of the second negative helicity leg attached to this vertex, and $\alpha=\beta$ for an external leg.
We know already that in the singular limit we have
\begin{eqnarray}
\br{P_{j,b}\,c} & \rightarrow & \br{\hb c}\frac{\br{\ha P_{j,b}}}{\br{\ha\hb}},\\
\br{d\,P_{a,i}} & \rightarrow & \br{d\ha}\frac{\br{P_{a,i}\,\hb}}{\br{\ha\hb}}.
\end{eqnarray}
Applying the Schouten-identity~(\ref{eq:Schouten}), we get for the spinor product in the numerator
\begin{eqnarray}
\label{eq:ant3}
\br{c\,P_{\alpha\beta}}& = & \sum_{k=\alpha}^{\beta}\br{cx}\sq{k\eta}\nonumber\\
 & = & \sum_{k=\alpha}^{\beta}\left(\br{c\ha}\frac{\br{k\hb}}{\br{\ha \hb}}+\br{c\hb}\frac{\br{k\ha}}{\br{\hb \ha}}\right)\sq{k\eta}\\
 & = &\br{c\ha}\frac{\br{P_{\alpha\beta}\,\hb}}{\br{\ha \hb}}+\br{c\hb}\frac{\br{\ha\,P_{\alpha\beta}}}{\br{\ha \hb}}\nonumber
\end{eqnarray}
Inserting Eq.~(\ref{eq:ant3}) into Eq.~(\ref{eq:pmcontribution}), we get five terms:
\begin{enumerate}
\item
a term proportional to $\br{P_{\alpha\beta}\,\hb}^4$.
\item
a term proportional to $\br{P_{\alpha\beta}\,\ha}^4$.
\item
three ``mixed" terms of the form $\br{P_{\alpha\beta}\,\ha}^q\br{P_{\alpha\beta}\,\hb}^{4-q}$, $q=1,2,3$.
\end{enumerate}
We separately analyze the different limits:
\begin{itemize}
\item[-]
If $k_{\alpha},\ldots,k_{\beta}\parallel k_a$, then $\hka\rightarrow P$ and $k_{j}\rightarrow z_{j}P$, $\forall \alpha\le j\le \beta $ and so $\br{P_{\alpha\beta}\, \ha}\rightarrow 0$.
\item[-]
If $k_{\alpha},\ldots,k_{\beta}\parallel k_b$, then $\hkb\rightarrow P$ and $k_{j}\rightarrow z_{j}P$, $\forall \alpha \le j \le \beta$ and so $\br{P_{\alpha\beta}\, \hb}\rightarrow 0$.
\item[-]
If $k_{\alpha},\ldots,k_{m}\parallel k_a$ and $k_{m'},\ldots,k_{\beta}\parallel k_b$, $\alpha\le m<m'\le \beta$, then the propagator $1/P_{\alpha\beta}\,^2$ is not divergent, and thus diagrams with a propagator $1/P_{\alpha\beta}\,^2$ are not divergent enough to contribute to this limit.
\item[-]
If $k_{\alpha},\ldots,k_{\beta}\rightarrow 0$, the situation is more subtle. We show in Appendix~\ref{app:appendixc} that in this limit only those diagrams contribute where $\alpha=a$, and $P_{a\beta}\,\rightarrow k_a$. 
Thus in this limit $\br{P_{\alpha\beta}\, \ha}\sim \br{a\ha}\rightarrow 0$.
\end{itemize}
Finally, we see that in any situation either $\br{P_{\alpha\beta}\, \ha}$ or $\br{P_{\alpha\beta}\, \hb}$ go the zero, and thus we can drop the ``mixed" terms. So in the singular limit Eq.~(\ref{eq:pmcontribution}) becomes
\begin{eqnarray}
\frac{\br{P_{\alpha\beta}\,c}^4}{\br{P_{j,b}\,c}\br{cd}\br{d\,P_{a,i}}}&\rightarrow&
\frac{\br{c\ha}^4}{\br{\ha\hb}\br{\hb c}\br{cd}\br{d\ha}}\,\frac{\br{P_{\alpha\beta}\,\hb}^4}{\br{P_{j,b}\,\ha}\br{\ha\hb}\br{\hb\,P_{a,i}}}\nonumber\\
 & & +\frac{\br{c\hb}^4}{\br{\ha\hb}\br{\hb c}\br{cd}\br{d\ha}}\,\frac{\br{P_{\alpha\beta}\,\ha}^4}{\br{P_{j,b}\,\ha}\br{\ha\hb}\br{\hb\,P_{a,i}}},\\
 &\rightarrow&
A_4(\ha^-,\hb^+,c^-,d^+)\,\frac{\br{P_{\alpha\beta}\,\hb}^4}{\br{P_{j,b}\,\ha}\br{\ha\hb}\br{\hb\,P_{a,i}}}\nonumber\\
 & & +A_4(\ha^+,\hb^-,c^-,d^+)\,\frac{\br{P_{\alpha\beta}\,\ha}^4}{\br{P_{j,b}\,\ha}\br{\ha\hb}\br{\hb\,P_{a,i}}}\nonumber.
\end{eqnarray}
We see that the first term contributes to $\ant(\ha^+,\hb^- \leftarrow a,1,\ldots,n,b)$ and the second term contributes to $\ant(\ha^-,\hb^+ \leftarrow a,1,\ldots,n,b)$. Note that again the second factor in each term has the same functional form as the left-hand side, so we proved our claim that \mbox{$\ant(\ha^{h_{\ha}},\hb^{h_{\hb}} \leftarrow a,1,\ldots,n,b)$} can be calculated by evaluating all CSW diagrams that contribute to the $(n+4)$-point amplitude $A_{n+4}(a,1,\ldots,$
$n,b,\ha^{h_{\ha}},\hb^{h_{\hb}})$ and where $\ha$ and $\hb$ are attached to the same $m$-point CSW vertex, with $m\ge 4$.

\section{Discussion and results}
\label{sec:results}

In the previous section we proved the main result of this paper which states that in the CSW formalism it is possible to define in a natural way antenna functions as a sum of MHV diagrams. This can be summarized as follows:
\begin{ruler}[Antenna functions in the CSW formalism]
\label{rule:ant}
~\\[10pt]
The antenna function \mbox{$\ant(\ha^{h_{\ha}},\hb^{h_{\hb}} \leftarrow a,1,\ldots,n,b)$} can be calculated by evaluating all CSW diagrams that contribute to the $(n+4)$-point amplitude $A_{n+4}(a,1,\ldots,n,b,\ha^{h_{\ha}},\hb^{h_{\hb}})$ and where $\ha$ and $\hb$ are attached to the same $m$-point CSW vertex, with $m\ge 4$.
\end{ruler}

This result allows to directly build all antenna functions using the
CSW formalism and the related CSW diagrams. It is possible to identify a
priori the CSW diagrams that contribute to the singular limit, and
thus it is not necessary to evaluate all the CSW diagrams that
contribute to the full amplitude $A_{n+4}(a,1,\ldots,n,b,c,d)$.
This class of CSW diagrams is uniquely defined by the
requirement that all the CSW propagators must go on-shell in the
singular limit, which turns out to be very similar to
Rule~\ref{rule:glover} for splitting amplitudes. The advantage of our
result is, however, that the functional form of the antenna functions
is exactly given by the CSW diagram, whereas in Rule~\ref{rule:glover}
one still has to go to the collinear limit. Antenna functions 
can be represented as a special class of CSW diagrams, the
class of diagrams being the one defined by Rule~\ref{rule:ant}. Let us
make a few comments about the antenna functions obtained from
Rule~\ref{rule:ant}:
\begin{enumerate}
\item
The antenna functions may still contain non-divergent
pieces, coming from CSW diagrams that are not divergent in any
limit. For example, let us consider the antenna function
$\ant(\ha^{-},\hb^{+} \leftarrow a^-,1^-,2^+,b^+)$. In
Ref.~\cite{Kosower:2002su} it was shown that this antenna function is
zero. Using Rule~\ref{rule:ant} we generate a set of four non zero CSW
diagrams. It is easy to check however that all four diagrams are
finite in all the singular limits we are interested in.
\item 
As the antenna functions are built from CSW diagrams, it is easy to see that the antenna functions obtained from Rule~\ref{rule:ant} will always fulfill the reflection identity
\begin{equation}
\label{eq:antreflec}
\ant(\ha^{h_{\ha}},\hb^{h_{\hb}} \leftarrow a,1,\ldots, n,b) = (-1)^n\ant(\hb^{h_{\hb}},\ha^{h_{\ha}} \leftarrow b,n,\ldots, 1,a).
\end{equation}
\item
The antenna functions are in general dependent of an arbitrary spinor
$\eta$. In practise, however, $\eta$ cannot be chosen among the
momenta appearing inside the antenna functions, because this would be
inconsistent with the assumption that there are no infrared poles in
the amplitude coming from antiholomorphic spinor products of the form
$\sq{.\eta}$, \eg\,if we chose $\eta = a$, then $\sq{1\eta}=\sq{1a}$
could lead to a pole in the limit where these two particles become
collinear, which is not included in the class of diagrams defined by
Rule~\ref{rule:ant}.
\item
The proof of our result crucially relies on the fact that $\br{a\ha}$ goes to zero in the singular limit. We were able to show this explicitely for the reconstruction functions given in Eqs~(\ref{eq:reconstruct1}-\ref{eq:reconstruct2}). However, if different reconstruction functions are chosen, it should be checked that this statement still holds true for the new analytic expressions.
\end{enumerate}

In Ref.~\cite{Duhr:2006iq} a recursive formulation of the CSW
formalism in terms of single and double-line currents was
introduced. A review of this recursive algorithm is given in
Appendix~\ref{appendix:B}. As Rule~\ref{rule:ant} allows us to
identify a priori the CSW diagrams that contribute to a given singular
limit, we can use this result to write down a recursive algorithm for
antenna functions in terms of the single and double-line currents. As
the antenna functions contain all kinds of collinear and soft
singularities, it is straightforward to derive the corresponding
recursions for splitting amplitudes and soft factors. This algorithm is
discussed in Section~\ref{sec:antRR}.

\subsection*{General formulas for NMHV antenna functions}
\label{sec:genresults}
In this section we apply Rule~\ref{rule:ant} to derive generic formulas for all MHV and next to MHV (NMHV) type antenna functions. We give explicitly the results for antenna functions of the form $\ant(\ha^{-},\hb^{-} \leftarrow a,1,\ldots,n,b)$ and $\ant(\ha^{-},\hb^{+} \leftarrow a,1,\ldots,n,b)$. The remaining antenna functions are related to the previous ones by parity and reflection identity~(\ref{eq:antreflec}).

\paragraph{MHV-type antenna functions}
\label{subsec:MHVant}
The simplest antenna functions are those obtained from a single CSW vertex. Applying Rule~\ref{rule:ant} we immediately find
\begin{eqnarray}
\ant(\ha^{-},\hb^{-} \leftarrow a^+,1^+,\ldots,n^+,b^+) & = & A_{n+4} (a^+,1^+,\ldots,n^+,b^+,\ha^{-},\hb^{-})\nonumber\\ 
 & =  &\frac{\br{\ha\hb}^3}{\br{a1}\br{12}\ldots\br{nb}\br{b\ha}\br{\hb a}},\\
\ant(\ha^{-},\hb^{+} \leftarrow a^+,1^+,\ldots,j^-,\ldots,n^+,b^+) & = & A_{n+4} (a^+,1^+,\ldots,j^-,\ldots,n^+,b^+,\ha^{-},\hb^{+}) \nonumber\\
 &  = & \frac{\br{\ha j}^4}{\br{a1}\br{12}\ldots\br{nb}\br{b\ha}\br{\ha\hb}\br{\hb a}}.
\end{eqnarray}

\paragraph{NMHV-type antenna functions}
\label{subsec:NMHVant}
The NMHV-type antenna functions are derived from CSW diagrams containing exactly one propagator. The CSW diagrams obtained from Rule~\ref{rule:ant} are shown in Fig.~\ref{fig:NMHV1} and Fig.~\ref{fig:NMHV2}. The generic formulas for NMHV-type antenna functions are
\begin{align}
\label{eq:NMHV1}
&\ant(\ha^{-},\hb^{-} \leftarrow  a^+,1^+,\ldots,j^-,\ldots,n^+,b^+) = \\
&\phantom{+}\sum_{l=j}^{n}\frac{\br{\ha\hb}^4\br{jP_{a,l}}^4}{s_{a,l}\br{P_{a,l}(l+1)}\br{\br{(l+1)\hb}}\br{\hb P_{a,l}}\br{P_{a,l}a}\br{\br{al}}\br{lP_{a,l}}}\nonumber\\
&+\sum_{k=a}^{j-1}\frac{\br{\ha\hb}^4\br{jP_{k+1,b}}^4}{s_{k+1,b}\br{P_{k+1,b}\ha}\br{\br{\ha k}}\br{kP_{k+1,b}}\br{P_{k+1,b}(k+1)}\br{\br{(k+1)b}}\br{bP_{k+1,b}}}\nonumber\\
&+\sum_{k=a}^{j-1}\sum_{l=j}^n\frac{\br{\ha\hb}^4\br{jP_{k+1,l}}^4}{s_{k+1,l}\br{P_{k+1,l}(l+1)}\br{\br{(l+1) k}}\br{kP_{k+1,l}}\br{P_{k+1,l}(k+1)}\br{\br{(k+1)l}}\br{lP_{k+1,l}}}\nonumber,
\end{align}
\FIGURE[!t]{
   \begin{fmffile}{NMHV}
               \parbox{45mm}{
                  \begin{fmfgraph*}(70,70)
                     \fmfcurved
                     \fmfsurround{o2,o3,i1,i2,i3,i4,o1}
                     \fmf{plain}{i1,a1,v1}
                     \fmf{plain}{i2,a2,v1}
                     \fmf{plain}{i3,a3,v1}
                     \fmf{plain}{i4,a4,v1}
                     \fmf{plain}{v1,v2}
                     \fmf{plain}{v2,a5,o1}
                     \fmf{plain}{v2,a6,o2}
                     \fmf{plain}{v2,a7,o3}
                     \fmffreeze
                     \fmf{phantom}{a1,d1,a2}
                     \fmfdot{d1}
                     \fmf{phantom}{a5,d2,a6}
                     \fmfdot{d2}
                     \fmf{phantom}{a6,d3,a7}
                     \fmfdot{d3}
                     \fmflabel{$k$}{i1}
                     \fmflabel{$a$}{i2}
                     \fmflabel{$\hat{b}^-$}{i3}
                     \fmflabel{$\hat{a}^-$}{i4}
                     \fmflabel{$b$}{o1}
                     \fmflabel{$j^-$}{o2}
                     \fmflabel{$k+1$}{o3}
                 \end{fmfgraph*}}
                 \vspace{10mm}
              \parbox{45mm}{
              \begin{fmfgraph*}(70,70)
                 \fmfcurved
                 \fmfsurround{o2,o3,i1,i2,i3,i4,o1}
                 \fmf{plain}{i1,a1,v1}
                 \fmf{plain}{i2,a2,v1}
                 \fmf{plain}{i3,a3,v1}
                 \fmf{plain}{i4,a4,v1}
                 \fmf{plain}{v1,v2}
                 \fmf{plain}{v2,a5,o1}
                 \fmf{plain}{v2,a6,o2}
                 \fmf{plain}{v2,a7,o3}
                 \fmffreeze
                 \fmf{phantom}{a3,d1,a4}
                 \fmfdot{d1}
                 \fmf{phantom}{a5,d2,a6}
                 \fmfdot{d2}
                 \fmf{phantom}{a6,d3,a7}
                 \fmfdot{d3}
                 \fmflabel{$\hat{b}^-$}{i1}
                 \fmflabel{$\hat{a}^-$}{i2}
                 \fmflabel{$b$}{i3}
                 \fmflabel{$\ell+1$}{i4}
                 \fmflabel{$\ell$}{o1}
                 \fmflabel{$j^-$}{o2}
                 \fmflabel{$a$}{o3}
             \end{fmfgraph*}}
          \parbox{45mm}{\begin{fmfgraph*}(70,70)
            \fmfcurved
            \fmfsurround{o2,o3,i1,i2,i3,i4,i45,i5,o1}
          \fmf{plain}{i1,a1,v1}
          \fmf{plain}{i2,a2,v1}
          \fmf{plain}{i3,a3,v1}
          \fmf{plain}{i4,a4,v1}
                    \fmf{plain}{i45,a45,v1}
                    \fmf{plain}{i5,a8,v1}
          \fmf{plain}{v1,v2}
          \fmf{plain}{v2,a5,o1}
          \fmf{plain}{v2,a6,o2}
          \fmf{plain}{v2,a7,o3}
          \fmffreeze
                    \fmf{phantom}{a1,d1,a2}
                    \fmfdot{d1}
                                        \fmf{phantom}{a45,d4,a8}
                    \fmfdot{d4}
                    \fmf{phantom}{a5,d2,a6}
                    \fmfdot{d2}
                    \fmf{phantom}{a6,d3,a7}
                    \fmfdot{d3}
          \fmflabel{$k$}{i1}
          \fmflabel{$a$}{i2}
          \fmflabel{$\hat{b}^-$}{i3}
          \fmflabel{$\hat{a}^-$}{i4}
                    \fmflabel{$b$}{i45}
                                        \fmflabel{$\ell+1$}{i5}
          \fmflabel{$\ell$}{o1}
          \fmflabel{$j^-$}{o2}
          \fmflabel{$k+1$}{o3}
            \end{fmfgraph*}}
                             \end{fmffile}
                     \vspace{3mm}
                     \caption{CSW diagrams contributing to $\ant(\ha^{-},\hb^{-} \leftarrow  a^+,1^+,\ldots,j^-,\ldots,n^+,b^+)$.}
\label{fig:NMHV1}                    }
\begin{align}
&\ant(\ha^{-},\hb^{+} \leftarrow  a^+,1^+,\ldots,i^-,\ldots,j^-,\ldots,n^+,b^+) = \\
&\phantom{+}\sum_{k=a}^{i-1}\frac{\br{\ha P_{k+1,b}}^4\br{ij}^4}{s_{k+1,b}\br{P_{k+1,b}\ha}\br{\br{\ha k}}\br{k P_{k+1,b}}\br{P_{k+1,b}(k+1)}\br{\br{(k+1)b}}\br{bP_{k+1,b}}}\nonumber\\
&+\sum_{l=j}^{n}\frac{\br{\ha P_{a,l}}^4\br{ij}^4}{s_{a,l}\br{P_{a,l}(l+1)}\br{\br{(l+1)\hb}}\br{\hb P_{a,l}}\br{P_{a,l}a}\br{\br{al}}\br{lP_{a,l}}}\nonumber\\
&+\sum_{k=a}^{i-1}\sum_{l=j}^n\frac{\br{\ha P_{k+1,l}}^4\br{ij}^4}{s_{k+1,l}\br{P_{k+1,l}(l+1)}\br{\br{(l+1)k}}\br{k P_{k+1,l}}\br{P_{k+1,l}(k+1)}\br{\br{(k+1)l}}\br{lP_{k+1,l}}}\nonumber\\
&+\sum_{k=i}^{j-1}\sum_{l=j}^b\frac{\br{\ha i}^4\br{j P_{k+1,l}}^4}{s_{k+1,l}\br{P_{k+1,l}(l+1)}\br{\br{(l+1)k}}\br{k P_{k+1,l}}\br{P_{k+1,l}(k+1)}\br{\br{(k+1)l}}\br{lP_{k+1,l}}}\nonumber\\
&+\sum_{k=\ha}^{i-1}\sum_{l=i}^{j-1}\frac{\br{\ha j}^4\br{i P_{k+1,l}}^4}{s_{k+1,l}\br{P_{k+1,l}(l+1)}\br{\br{(l+1)k}}\br{k P_{k+1,l}}\br{P_{k+1,l}(k+1)}\br{\br{(k+1)l}}\br{lP_{k+1,l}}}\nonumber,
\end{align}
\FIGURE[!t]{
    \begin{fmffile}{NMHV2}
            \parbox{45mm}{
            \begin{fmfgraph*}(70,70)
            \fmfcurved
            \fmfsurround{o22,o3,o4,i1,i2,i3,i4,o1,o2}
          \fmf{plain}{i1,a1,v1}
          \fmf{plain}{i2,a2,v1}
          \fmf{plain}{i3,a3,v1}
          \fmf{plain}{i4,a4,v1}
          \fmf{plain}{v1,v2}
          \fmf{plain}{v2,a5,o1}
          \fmf{plain}{v2,a6,o2}
          \fmf{plain}{v2,a7,o3}
                    \fmf{plain}{v2,a8,o4}
          \fmffreeze
                    \fmf{phantom}{a1,d1,a2}
                    \fmfdot{d1}
                    \fmf{phantom}{a5,d2,a6}
                    \fmfdot{d2}
                    \fmf{phantom}{a6,d3,a7}
                    \fmfdot{d3}
                                        \fmf{phantom}{a7,d4,a8}
                    \fmfdot{d4}
          \fmflabel{$k$}{i1}
          \fmflabel{$a$}{i2}
          \fmflabel{$\hat{b}^-$}{i3}
          \fmflabel{$\hat{a}^-$}{i4}
          \fmflabel{$b$}{o1}
          \fmflabel{$j^-$}{o2}
          \fmflabel{$i^-$}{o3}
          \fmflabel{$k+1$}{o4}
            \end{fmfgraph*}}
            \vspace{10mm}
             \parbox{45mm}{
            \begin{fmfgraph*}(70,70)
            \fmfcurved
            \fmfsurround{o22,o3,o4,i1,i2,i3,i4,o1,o2}
          \fmf{plain}{i1,a1,v1}
          \fmf{plain}{i2,a2,v1}
          \fmf{plain}{i3,a3,v1}
          \fmf{plain}{i4,a4,v1}
          \fmf{plain}{v1,v2}
          \fmf{plain}{v2,a5,o1}
          \fmf{plain}{v2,a6,o2}
          \fmf{plain}{v2,a7,o3}
                    \fmf{plain}{v2,a8,o4}
          \fmffreeze
                    \fmf{phantom}{a3,d1,a4}
                    \fmfdot{d1}
                    \fmf{phantom}{a5,d2,a6}
                    \fmfdot{d2}
                    \fmf{phantom}{a6,d3,a7}
                    \fmfdot{d3}
                                        \fmf{phantom}{a7,d4,a8}
                    \fmfdot{d4}
          \fmflabel{$\hat{b}^-$}{i1}
          \fmflabel{$\hat{a}^-$}{i2}
          \fmflabel{$b$}{i3}
                    \fmflabel{$\ell+1$}{i4}
          \fmflabel{$\ell$}{o1}
          \fmflabel{$j^-$}{o2}
          \fmflabel{$i^-$}{o3}
          \fmflabel{$a$}{o4}
            \end{fmfgraph*}}
            \vspace{10mm}
             \parbox{45mm}{
            \begin{fmfgraph*}(70,70)
            \fmfcurved
            \fmfsurround{o22,o3,o4,i1,i2,i3,i4,i5,i6,o1,o2}
          \fmf{plain}{i1,a1,v1}
          \fmf{plain}{i2,a2,v1}
          \fmf{plain}{i3,a3,v1}
          \fmf{plain}{i4,a4,v1}
                    \fmf{plain}{i5,a35,v1}
          \fmf{plain}{i6,a45,v1}
          \fmf{plain}{v1,v2}
          \fmf{plain}{v2,a5,o1}
          \fmf{plain}{v2,a6,o2}
          \fmf{plain}{v2,a7,o3}
                    \fmf{plain}{v2,a8,o4}
          \fmffreeze
                    \fmf{phantom}{a1,d1,a2}
                    \fmfdot{d1}
                                        \fmf{phantom}{a35,d5,a45}
                    \fmfdot{d5}
                    \fmf{phantom}{a5,d2,a6}
                    \fmfdot{d2}
                    \fmf{phantom}{a6,d3,a7}
                    \fmfdot{d3}
                                        \fmf{phantom}{a7,d4,a8}
                    \fmfdot{d4}
                    \fmflabel{$k$}{i1}
                    \fmflabel{$a$}{i2}
          \fmflabel{$\hat{b}^-$}{i3}
          \fmflabel{$\hat{a}^-$}{i4}
          \fmflabel{$b$}{i5}
                    \fmflabel{$\ell+1$}{i6}
          \fmflabel{$\ell$}{o1}
          \fmflabel{$j^-$}{o2}
          \fmflabel{$i^-$}{o3}
          \fmflabel{$k+1$}{o4}
            \end{fmfgraph*}}
            \vspace{10mm}
             \parbox{45mm}{
            \begin{fmfgraph*}(70,70)
            \fmfcurved
            \fmfsurround{o2,o3,i1,i2,i3,i4,i5,i6,o1}
          \fmf{plain}{i1,a1,v1}
          \fmf{plain}{i2,a2,v1}
          \fmf{plain}{i3,a3,v1}
          \fmf{plain}{i4,a4,v1}
                    \fmf{plain}{i5,a35,v1}
          \fmf{plain}{i6,a45,v1}
          \fmf{plain}{v1,v2}
          \fmf{plain}{v2,a5,o1}
          \fmf{plain}{v2,a6,o2}
          \fmf{plain}{v2,a7,o3}
          \fmffreeze
                    \fmf{phantom}{a1,d1,a2}
                    \fmfdot{d1}
                                        \fmf{phantom}{a2,d5,a3}
                    \fmfdot{d5}
                                                          \fmf{phantom}{a35,d4,a45}
                    \fmfdot{d4}
                    \fmf{phantom}{a5,d2,a6}
                    \fmfdot{d2}
                    \fmf{phantom}{a6,d3,a7}
                    \fmfdot{d3}
                    \fmflabel{$k$}{i1}
                    \fmflabel{$i^-$}{i2}
          \fmflabel{$\hat{b}^-$}{i3}
          \fmflabel{$\hat{a}^-$}{i4}
          \fmflabel{$b$}{i5}
                    \fmflabel{$\ell+1$}{i6}
          \fmflabel{$\ell$}{o1}
          \fmflabel{$j^-$}{o2}
          \fmflabel{$k+1$}{o3}
            \end{fmfgraph*}}
            \vspace{10mm}
             \parbox{45mm}{
            \begin{fmfgraph*}(70,70)
            \fmfcurved
            \fmfsurround{o2,o3,i1,i2,i3,i4,i5,i6,o1}
          \fmf{plain}{i1,a1,v1}
          \fmf{plain}{i2,a2,v1}
          \fmf{plain}{i3,a3,v1}
          \fmf{plain}{i4,a4,v1}
                    \fmf{plain}{i5,a35,v1}
          \fmf{plain}{i6,a45,v1}
          \fmf{plain}{v1,v2}
          \fmf{plain}{v2,a5,o1}
          \fmf{plain}{v2,a6,o2}
          \fmf{plain}{v2,a7,o3}
          \fmffreeze
                    \fmf{phantom}{a1,d1,a2}
                    \fmfdot{d1}
                                        \fmf{phantom}{a4,d5,a35}
                    \fmfdot{d5}
                                                          \fmf{phantom}{a35,d4,a45}
                    \fmfdot{d4}
                    \fmf{phantom}{a5,d2,a6}
                    \fmfdot{d2}
                    \fmf{phantom}{a6,d3,a7}
                    \fmfdot{d3}
                    \fmflabel{$k$}{i1}
                    \fmflabel{$a$}{i2}
          \fmflabel{$\hat{b}^-$}{i3}
          \fmflabel{$\hat{a}^-$}{i4}
          \fmflabel{$j^-$}{i5}
                    \fmflabel{$\ell+1$}{i6}
          \fmflabel{$\ell$}{o1}
          \fmflabel{$i^-$}{o2}
          \fmflabel{$k+1$}{o3}
            \end{fmfgraph*}}        
            \end{fmffile}
\caption{CSW diagrams contributing to $\ant(\ha^{-},\hb^{+} \leftarrow  a^+,1^+,\ldots,i^-,\ldots,j^-,\ldots,n^+,b^+)$.}
\label{fig:NMHV2}
}
where we introduced the notations
\begin{equation}
\begin{split}
s_{i,j}&=(p_i+p_{i+1}+\ldots+p_j)^2,\\
\br{\br{ij}}&=\prod_{k=i}^{j-1}\br{k(k+1)}.
\end{split}
\end{equation}
Note that the above generic formulas for NMHV-type antenna functions
(together with the results for the MHV-type antenna functions) are
sufficient to derive the full set of NNLO and N$^3$LO gluon antenna
functions, listed in Appendix~\ref{appendix:results}. 
We checked explicitly that these
antenna functions reproduce the correct infrared limits and that the
limits are numerically independent of $\eta$ using the
\verb+Mathematica+ package \verb+S@M+~\cite{Maitre:2007jq}. Note
that our antenna functions have a slightly simpler and more compact
analytic form than those presented in Ref.~\cite{Kosower:2002su}. This
fact might simplify the phase space integration of the counterterms
when antenna functions are used in a subtraction method.

\section{From antenna functions to splitting amplitudes}

In this section we show how the antenna functions obtained from Rule~\ref{rule:ant} can be used to derive Rule~\ref{rule:glover} for splitting amplitudes presented in the previous section. 
Let us start with  $\splitti_-(a,1,\ldots,n)$. We know that for the antenna function $\ant(\ha^-,\hb^- \leftarrow a,1,\ldots,n,b)$, we have in the limit where $1,\ldots,n$ become collinear to $a$,
\begin{equation}
\ant(\ha^-,\hb^- \leftarrow a,1,\ldots,n,b) \longrightarrow \splitti_-(a,1,\ldots,n).
\end{equation}
Furthermore, we know that the CSW pole structure of the antenna function built from Rule~\ref{rule:ant} is
\begin{equation}
\ant(\ha^-,\hb^- \leftarrow a,1,\ldots,n,b)\sim \frac{1}{\sq{\quad}^{N_-}}\,f(\br{\quad}).
\end{equation}
In the limit under consideration, we have $N_-=n_a$ because in this limit $\hkb \rightarrow k_b$, and so $h_b=+1$. Thus only those CSW diagrams in $\ant(\ha^-,\hb^- \leftarrow a,1,\ldots,n,b)$ contribute where all $N_-=n_a$ CSW propagators go on-shell in the collinear limit. These diagrams correspond exactly to those where $b$ is attached to 
the same CSW vertex as $\ha$ and $\hb$, which is equivalent to Rule~\ref{rule:glover}. The derivation of the corresponding result for $\splitti_+$ is
similar to the $\splitti_-$ case, so we do not report its
derivation. 

Note that unlike antenna functions, the splitting amplitudes are uniquely defined. The arbitrariness in the definition of the antenna function is lost when a specific collinear limit is taken, because the class of CSW diagrams contributing to the antenna function can be divided into two different subclasses:
\begin{enumerate}
\item
The CSW diagrams where $b$ is attached to the same CSW vertex as $\ha$ and $\hb$: These diagrams are divergent in the collinear limit.\footnote{Note that these diagrams may still contain subleading pieces.}
\item
The CSW diagrams where $b$ is not attached to the same CSW vertex as $\ha$ and $\hb$: These diagrams are not divergent enough and are omitted. They contribute to the arbitrary piece from the antenna function.
\end{enumerate}

\subsection*{Diagrammatic approach to splitting amplitudes}
\label{sec:splitdiag}
In the previous section we showed that in the limit where $1,\ldots,n$ become collinear to $a$ only those CSW diagrams of the antenna contribute where $\hat a$, $\hat b$ and $b$ are attached to the same CSW vertex, which is equivalent to the result derived in Ref.~\cite{Birthwright:2005ak, Birth:2005vi}. In this section we show how this result can be interpreted in terms of CSW diagrams, where the vertices are modified.

\FIGURE[!t]{
\begin{fmffile}{splitexp}
\quad
\parbox{20mm}{\begin{fmfgraph}(50,40)
\fmfcurved
\fmfleft{i1,i2,i3}
\fmfright{o1}
\fmf{plain}{i1,v1}
\fmf{plain}{i2,v1}
\fmf{plain}{i3,v1}
\fmf{plain}{v1,o1}
\fmfblob{4mm}{o1}
\end{fmfgraph}}
        +
\parbox{20mm}{\begin{fmfgraph}(50,40)
\fmfcurved
\fmfleft{i1,i2,i3}
\fmfright{u1,o1,u2,o2,u3}
\fmf{plain}{i1,v1}
\fmf{plain}{i2,v1}
\fmf{plain}{i3,v1}
\fmf{plain}{v1,o1}
\fmf{plain}{v1,o2}
\fmfblob{4mm}{o1}
\fmfblob{4mm}{o2}
\end{fmfgraph}}
+            
\parbox{20mm}{\begin{fmfgraph}(50,40)
\fmfcurved
\fmfleft{i1,i2,i3}
\fmfright{o1,o2,o3}
\fmf{plain}{i1,v1}
\fmf{plain}{i2,v1}
\fmf{plain}{i3,v1}
\fmf{plain}{v1,o1}
\fmf{plain}{v1,o2}
\fmf{plain}{v1,o3}
\fmfblob{4mm}{o1}
\fmfblob{4mm}{o2}
\fmfblob{4mm}{o3}
\end{fmfgraph}}
+
\parbox{20mm}{\begin{fmfgraph}(50,40)
\fmfcurved
\fmfleft{i1,i2,i3}
\fmfright{o1,o2,o3,o4}
\fmf{plain}{i1,v1}
\fmf{plain}{i2,v1}
\fmf{plain}{i3,v1}
\fmf{plain}{v1,o1}
\fmf{plain}{v1,o2}
\fmf{plain}{v1,o3}
\fmf{plain}{v1,o4}
\fmfblob{4mm}{o1}
\fmfblob{4mm}{o2}
\fmfblob{4mm}{o3}
\fmfblob{4mm}{o4}
\end{fmfgraph}}
+\quad\ldots      
\end{fmffile}
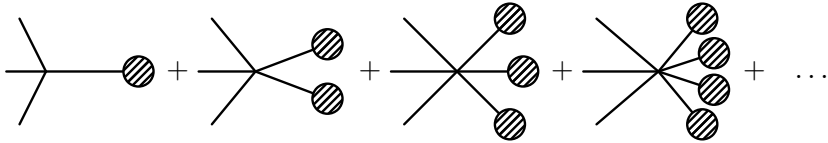
\caption{\label{fig:splitexpand} Diagrammatic expansion of the splitting amplitude. The blobs indicate CSW diagrams with a smaller number of legs.}
}

The set of diagrams defined  by this condition can be easily expanded in terms of the structure of the collinear poles appearing in the diagram \mbox{(~\ie, $1/s_{a,n}$, $1/s_{a,k}s_{k+1,n}$, \emph{etc.})}. This expansion is shown in Fig.~\ref{fig:splitexpand}. We find
\begin{equation}
\label{eq:splitexp}
\splitti_{-h}(a,1,\ldots,n)A_4(P^h,b,c,d)\sim\sum_{\textrm{partitions}}\sum_jV_{\pi,j}\prod_{\pi_i}\frac{1}{s_{\pi_i}}D^{\rm MHV}_{j,(\pi_i)},
\end{equation}
where the first sum goes over all partitions (including 
the set $\{a,1,\ldots, n\}$ itself) 
$\pi = (\pi_i)$ of the set $\{a,1,\ldots,n\}$, and
the second sum runs over all diagrams corresponding to this
partition. $V_{\pi,j}$ is the MHV vertex which $\hat a$, $\hat b$ and
$b$ are attached to and $s_{\pi_i}$ denotes the invariant formed out of
the momenta which are in the subset $\pi_i$. Note that in this way we
reproduce the pole structure of the splitting amplitude: For the
partition into one subset, $\pi = \{a,1,\ldots,n\}$, the pole is
$1/s_{a,n}$, for the partition into two subsets, $\pi =
\{(a,1,\ldots,k),(k+1,\ldots,n)\}$, the pole is $1/s_{a,k}s_{k+1,n}$,
and so on.

The diagrams $D^{\rm MHV}_{j, (\pi_i)}$ are CSW diagrams. They however still contain pieces that are subleading in the collinear limit. We show how it is possible to modify the definition of the CSW vertices such that the diagrams in Eq.~(\ref{eq:splitexp}) only contain the leading collinear pole.
Let us consider a specific diagram $D^{\rm MHV}_{j,(\pi_i)}$. 
All the vertices in $D^{\rm MHV}_{j,(\pi_i)}$ only depend on 
\begin{itemize}
\item[-]
particles from the collinear set $a,1,2,\ldots,n$.
\item[-] 
off-shell legs of the form $P_{k,\ell}$, where $k$ and $\ell$ are in the collinear set.
\end{itemize}
Let us first consider the case of a vertex which only depends on off-shell continued legs. We will give as an example the four-point vertex. The generalization is straightforward. An example of a four-point MHV-vertex with all legs continued off-shell is
\begin{equation}
A_4\big({I}^-,\,{J}^-,\,{K}^+,\,{L}^+\big)=\frac{\br{IJ}^4}{\br{IJ}\br{JK}\br{KL}\br{LI}},
\end{equation}
where we used the multiindex notation introduced in Appendix~\ref{sec:CSWRR}, \ie,$A_4\big({I}^-,\,{J}^-,\,{K}^+,\,{L}^+\big)$ has to be understood as $A_4\big(P_{I}^-,\,P_{J}^-,\,P_{K}^+,\,P_{L}^+\big)$.
Applying Eq.~(\ref{eq:collimdelta}), this vertex gives the following contribution in the collinear limit
\begin{equation}
A_4\big({I}^-,\,{J}^-,\,{K}^+,\,{L}^+\big)\rightarrow \frac{\Delta(I,J)^4}{\Delta(I,J)\Delta(J,K)\Delta(K,L)\Delta(L,I)},
\label{eq:A4collim}
\end{equation}
The $\Delta$-function has been defined in Eq.~(\ref{eq:delta}), with
\begin{equation}
\Delta(I,J) \equiv \Delta(i_1, i_2; j_1,j_2).
\end{equation}
We would now like to define the object on the right-hand side of Eq.~(\ref{eq:A4collim}) as an effective MHV vertex in the collinear limit, from which splitting amplitudes could be built. The idea rests on the following observations:
\begin{enumerate}
\item An on-shell particle with momentum $p_i$ can be seen as labelled by the multiindex $\bar{\imath}=(i,i)$, defined in Eq.~(\ref{eq:onshellmultind}).
\item The $\Delta$-functions are not independent, but it is easy to see that
\begin{eqnarray}
\Delta(\bar{\imath},k) & = & \sqrt{z_i}\br{ki},\nonumber \\
\label{eq:dlink}
\Delta(\bar{\imath},J) & = & \sqrt{z_i}\Delta(J,i) = -\Delta(J,\bar{\imath}),\\
\Delta(\bar{\imath},\bar{\jmath}) & = & \sqrt{z_i}\Delta(\bar{\jmath},i) = \sqrt{z_iz_j}\br{ij}.\nonumber
\end{eqnarray}
\end{enumerate}

Thus we would like to extend the definition Eq.~(\ref{eq:A4collim}) to the case  where on-shell particles are present in the vertex. For example, if the first particle (with negative helicity) in $A_4$ is on-shell, $I=\bar{\imath}=(i,i)$, then, applying Eq.~(\ref{eq:dlink}),
\begin{eqnarray}
\label{eq:A4onshell1}
A_4\big(i^-,\,J^-,\,K^+,\,L^+\big) & = & A_4\big(\bar{\imath}^-,\,J^-,\,K^+,\,L^+\big)\\
& \rightarrow &\frac{\Delta(\bar{\imath},J)^4}{\Delta(\bar{\imath},J)\Delta(J,K)\Delta(K,L)\Delta(L,\bar{\imath})}\nonumber\\
   & \rightarrow & z_i\, \frac{-\Delta(J,i)^4}{\Delta(J,i)\Delta(J,K)\Delta(K,L)\Delta(L,i)}.\nonumber
\end{eqnarray}
On the other hand, a direct application of the collinear rules~(\ref{eq:collim}) leads to
\begin{equation}
\label{eq:A4onshell2}
A_4\big(i^-,\,J^-,\,K^+,\,L^+\big)\rightarrow \frac{-\Delta(J,i)^4}{\Delta(J,i)\Delta(J,K)\Delta(K,L)\Delta(L,i)}.
\end{equation}
The  difference between Eqs.~(\ref{eq:A4onshell1}) and (\ref{eq:A4onshell2})
amounts to a factor $z_i$. This can reabsorbed into the  one-point current attached to the vertex, by defining the ``wave function''
\begin{equation}
J_{-}(i) = \frac{1}{z_i},
\label{eq:temp1}
\end{equation}
which cancels the factor $z_i$ in Eq.~(\ref{eq:A4onshell1}). 
It is easy to see that for an on-shell particle with positive helicity the wave function must then be defined as
\begin{equation}
J_{+}(i) = z_i.
\label{eq:temp2}
\end{equation}
and finally that Eq.~(\ref{eq:temp1}) and (\ref{eq:temp2}) can be 
summarised as 
\begin{equation}
J_{h}(i) = z_i^h.
\end{equation}
The diagram $D^{\rm MHV}_{j,(\pi_i)}$ has however an additional external leg, corresponding to the incoming momentum $P_{\pi_i}$. We define the  wave function of this external leg as
\begin{equation}
J_h(P_{\pi_i})=1.
\end{equation} 

We now turn to the vertex $V_{\pi,j}$ in Eq.~(\ref{eq:splitexp}). 
Two cases are to be identified, corresponding to the value of $h$ in Eq.~(\ref{eq:splitexp}).
\begin{enumerate}
\item If $h=-1$, then 
\begin{equation}
V_{\pi,j}=\frac{\br{\ha\hb}^3}{\br{\hb P_{\pi_1}}\br{\br{P_{\pi_1}P_{\pi_k}}}\br{P_{\pi_k}b}\br{b\ha}},
\end{equation}
where $k$ denotes the length of the partition.
Using the properties of the reconstruction functions and the definition of the $\Delta$-function, we find
\begin{equation}
V_{\pi,j}\rightarrow \frac{1}{z_{\pi_1}z_{\pi_k}}\prod_{\ell=1}^{k-1}\frac{1}{\Delta(P_{\pi_\ell}, P_{\pi_{\ell+1}})},
\end{equation}
where we defined
\begin{equation}
z_{\pi_j}\equiv\sum_{\ell\in\pi_j}z_\ell.
\end{equation}
\item If $h=+1$, then 
\begin{equation}
V_{\pi,j}=\frac{\br{M\hb}^3}{\br{\hb P_{\pi_1}}\br{\br{P_{\pi_1}P_{\pi_k}}}\br{P_{\pi_k}b}\br{b\ha}},
\end{equation}
where now $M$ denotes the propagator with negative helicity attached to $V_{\pi,j}$. This yields
\begin{equation}
V_{\pi,j}\rightarrow \frac{z_M}{z_{\pi_1}z_{\pi_k}}\prod_{\ell=1}^{k-1}\frac{1}{\Delta(P_{\pi_\ell}, P_{\pi_{\ell+1}})},
\end{equation}
\end{enumerate}

Putting everything together, we can write down the following diagrammatic formula for splitting amplitudes:
\begin{equation}
\splitti_{-h}(a,1,\ldots,n)=\sum_{\textrm{partitions}}\sum_j\begin{cal}V\end{cal}_{\pi,j}^{(h)}\prod_{\pi_i}\frac{1}{s_{\pi_i}}D^{(\pi_i)}_j,
\end{equation}
where 
\begin{equation}
\begin{cal}V\end{cal}_{\pi,j}^{(h)}= \frac{z_M^{(1+h)/2}}{z_{\pi_1}z_{\pi_k}}\prod_{\ell=1}^{k-1}\frac{1}{\Delta(P_{\pi_\ell}, P_{\pi_{\ell+1}})}.
\end{equation}


\section{Recursive relations }
\label{sec:antRR}

\subsection*{Recursive relations for antenna functions}

In this section we apply the recursive formulation of the CSW
formalism introduced in Ref.~\cite{Duhr:2006iq} and reviewed in
Appendix~\ref{appendix:B} to the calculation of the antenna
function. From Rule~\ref{rule:ant}, we build the antenna
function $\ant(\ha^{h_{\ha}},\hb^{h_{\hb}} \leftarrow
a,1,\ldots,n,b)$ by calculating all CSW diagrams that
contribute to $A_{n+4}\big(a,1,\ldots,n,b,
\ha^{h_{\ha}}, \hb^{h_{\hb}}\big)$ and where $\ha$ and $\hb$ are attached to the
same $n$-point CSW vertex, $n\ge 4$. In the language of the recursive
formulation of the CSW formalism this amplitude can be built
recursively using Eq.~(\ref{eq:rec1} - \ref{eq:rec2}), and by putting the
off-shell leg on-shell.
This recursion relies on the fact that an arbitrary $n$-point MHV amplitude can be built recursively out of a small set of building blocks, and we introduce
a diagrammatic reprsentation for this construction in terms of double-lines (See Fig.~\ref{fig:MHVdecomp}). Hence, the diagrammatic representation of the recursion contains two types of lines, single-lines representing ordinary CSW propagators as well as double-lines arising from the recursive construction of the MHV vertices.

The part of this single-line current where $\ha$ and $\hb$ are attached to the same $n$-point CSW vertex, $n\ge 4$, corresponds exactly to those diagrams where $\ha$ and $\hb$ are attached to the same double-line current (See Fig.~\ref{fig:ant1}). We can then write immediately
\begin{equation}
\label{eq:antform}
\begin{split}
  \ant(\ha^{h_{\ha}},&\hb^{h_{\hb}} \leftarrow a,1,\ldots,n,b) = \\
 & \sum_{\substack{U<V \\ v_2=b}}\sum_M V_4(U,V,\ha,\hb;M_1,M_2)\,\epsilon^{h_{\ha}\,h_{\hb}}J_{UV}^{(2+h_{\ha}+h_{\hb})/2}(a,1,\ldots,n,b)
\end{split}
\end{equation}
where 
\begin{equation}
\epsilon^{h_{\ha}\,h_{\hb}}=\left\{\begin{array}{ll}
\delta_{M_1}^{\ha}\,\delta_{M_2}^{\hb} & , \ifi h_{\ha}=h_{\hb}=-1,\\
\delta_{M_2}^{\ha} & , \ifi h_{\ha}=-h_{\hb}=-1,\\
\delta_{M_2}^{\hb} & , \ifi h_{\ha}=-h_{\hb}=+1,\\
1 & , \ifi h_{\ha}=h_{\hb}=+1.
\end{array}\right.
\end{equation}
Note that this recursive relation follows the same spirit as that proposed by Kosower in Ref.~\cite{Kosower:2002su},
\begin{enumerate}
\item Build all the single and double-line currents using the recursive relations Eqs.~(\ref{eq:rec1} - \ref{eq:rec2}).
\item Calculate the antenna function by using Eq.~(\ref{eq:antform}).
\end{enumerate}
Note also that we could have used the recursive relations in a different way to calculate the antenna function:
\begin{enumerate}
\item
First build the full amplitude $A_{n+4}(a, 1, \ldots, n, b, c, d)$.
\item
Second extract the infrared divergent piece in the limit were $1, \ldots, n$ are unresolved, \ie, the antenna function $\ant(\ha, \hb \leftarrow a, 1, \dots, n, b)$.
\end{enumerate}
The calculation of the full amplitude $A_{n+4}(a, 1, \ldots, n, b, c, d)$ needs the evaluation of the $(n+3)$-point single-line current $J(a,1,\ldots,n,b,c)$, which contains $(n+2)$-point double-line currents as subcurrents. Eq.~(\ref{eq:antform}) proves that we only need to evaluate the $(n+2)$-point double-line current $J_{UV}(a,1,\ldots,n,b)$. This current contains all the information that is needed to build the antenna functions, \ie, we do not need to solve the full recursion, but we stop after we have built the infrared divergent piece.

Having a recursive formula for antenna functions we can go on and derive recursive relations for splitting amplitudes, by taking the corresponding limits on Eq.~(\ref{eq:antform}). The rest of this section will be devoted to this task.
\FIGURE[!t]{
            \begin{fmffile}{antennacurrent}
\parbox{40mm}{\begin{fmfgraph}(90,90)
\fmfleft{i1}
\fmfright{o1}
\fmf{phantom}{i1,o1}
\end{fmfgraph}}
          \parbox{90mm}{  \begin{fmfgraph*}(150,150)
        \fmfcurved
        \fmfleft{i1,i2,i3,i4}
        \fmfright{o1,o2,o3,o4,o5}
        \fmf{plain}{i3,v1} 
        \fmf{plain}{i2,v1}    
        \fmf{dbl_plain}{v1,v2} 
        \fmf{plain}{v2,o2}
        \fmf{plain}{v2,o4}
        \fmffreeze
        \fmfblob{15mm}{v2} 
        \fmfv{decor.shape=cross,decor.angle=45,decor.size=3mm}{i3}
        \fmfdot{o3}
        \fmflabel{$\hb^-$}{i3}
        \fmflabel{$\ha^-$}{i2}
        \fmflabel{$b$}{o2}
        \fmflabel{$a$}{o4}
        \end{fmfgraph*}}
\end{fmffile}
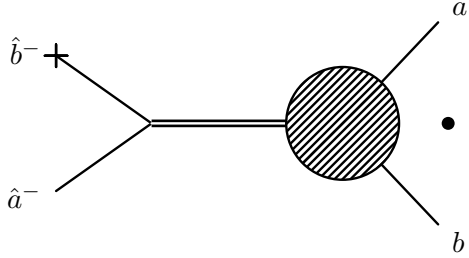
\caption{\label{fig:ant1}Contribution of  $J_+(a,1,\ldots,n,b,c^-)$ in the singular limit. The off-shell leg is denoted by a cross, and the double-line represents the CSW vertex to which $\ha$ and $\hb$ are attached.}
}

\subsection*{Recursive relations for splitting amplitudes} 
\label{splitRR}

In this section we derive the recursive relations for splitting
amplitudes by taking the collinear limit of the corresponding recursive
relations for the antenna functions, Eq.~(\ref{eq:antform}). We
already know that in the limit where $a,1,\ldots, n$ are collinear,
only those CSW diagrams contribute where $\ha$, $\hb$ and $b$ are
attached to the same CSW vertex. In terms of the single and
double-line currents, this means that $\ha$, $\hb$ and $b$ must be
attached to the same double-line current (See
Fig.~\ref{fig:antsplitm1}). In addition we can recast the splitting
function as a sum over products of CSW diagrams, where the CSW
vertices are replaced by $\Delta$ functions and the external particles
have ``wave functions'' corresponding to the momentum fraction $z_i$.

Combining these two observations with the recursive relations for the antenna functions introduced in the previous section, we are naturally led to a set of recursive relations for the splitting amplitudes themselves. We find
\begin{equation}
\label{eq:splitRR}
\splitti_{-h}(a,1,\ldots,n)=J_h(a,1,\ldots,n) + \sum_{\substack{U<V,M \\ v_2=n}}V_c^{-h}(U,V)\,J_{UV}^{(1-h)/2}(a,1,\ldots,n).
\end{equation}
Let us conclude this section with a  few comments.
\begin{enumerate}
\item The recursive relations state that a splitting amplitude can be written as a sum of a single and a double-line current. These currents fulfill formally the same recursive relations as gluon amplitudes and antenna functions, the vertices however correspond to the CSW vertices in the collinear limit introduced in Eq.~(\ref{eq:A4collim}) and the one-point currents correspond to the ``wave functions'' introduced in Section~\ref{sec:splitdiag},
\begin{equation}
J_h(i) = z_i^h.
\end{equation}
\item  
Once again we could use the recursive relations, Eqs.~(\ref{eq:rec1} - \ref{eq:rec2}), in a different way to calculate an $n$-point splitting amplitude:
\begin{enumerate}
\item
First calculate the full $n+3$ point amplitude $A_{n+3}(1,\ldots,n,a,b,c)$.
\item
Then extract the splitting amplitudes according to 
\begin{equation}
A_{n+3}(1,\ldots,n,a,b,c)\sim \splitti_{-h}(1,\ldots,n)\,A_4(P^h,a,b,c).
\end{equation}
\end{enumerate}
The calculation of the full amplitude needs the calculation of an $(n+2)$-point single-line current $J(1,\ldots,n,a,b)$, which contains the $(n+1)$-point double-line current as a subcurrent. On the other hand, Eq.~(\ref{eq:splitRR}) tells us that it is sufficient to evaluate $n$-point single and double-line currents to obtain an $n$-point splitting amplitude.
\item We checked the splitting amplitudes calculated using these recursive relations against the pure gluon splitting amplitudes obtained in Refs.~\cite{DelDuca:1999ha, Birthwright:2005ak}. For all of them we found complete agreement.
\end{enumerate}

\FIGURE[!t]{
\begin{fmffile}{splitcurrent}
   \parbox{50mm}{\begin{fmfgraph*}(100,100)
        \fmfcurved
        \fmfleft{i1,i2,i3,i4}
        \fmfright{o1,o2,o3,o4,o5}
        \fmfbottom{u1,u2,u3,u4}
        \fmf{plain}{i3,v1} 
        \fmf{plain}{i2,v1}    
        \fmf{phantom}{v1,a1,a2,v2} 
        \fmf{dbl_plain}{v1,a1}
        \fmf{plain}{a1,v2}
        \fmf{plain}{v2,o2}
        \fmf{plain}{v2,o4}
        \fmffreeze
                \fmf{phantom}{u2,ua1,a1}
                \fmf{plain}{ua1,a1}
        \fmfblob{10mm}{v2} 
        \fmfv{decor.shape=cross,decor.angle=45,decor.size=3mm}{i3}
        \fmfdot{o3}
        \fmflabel{$\hat{b}^-$}{i3}
        \fmflabel{$\hat{a}^-$}{i2}
        \fmflabel{$b$}{ua1}
        \fmflabel{$n$}{o2}
        \fmflabel{$a$}{o4}
        \end{fmfgraph*}}
        \parbox{50mm}{\begin{fmfgraph*}(100,100)
        \fmfcurved
        \fmfleft{i1,i2,i3,i4}
        \fmfright{o1,o2,o3,o4,o5}
        \fmfbottom{u1,u2,u3,u4}
        \fmf{plain}{i3,v1} 
        \fmf{plain}{i2,v1}    
        \fmf{phantom}{v1,a1,a2,v2} 
        \fmf{dbl_plain}{v1,a1}
        \fmf{dbl_plain}{a1,v2}
        \fmf{plain}{v2,o2}
        \fmf{plain}{v2,o4}
        \fmffreeze
                \fmf{phantom}{u2,ua1,a1}
                \fmf{plain}{ua1,a1}
        \fmfblob{10mm}{v2} 
        \fmfv{decor.shape=cross,decor.angle=45,decor.size=3mm}{i3}
        \fmfdot{o3}
        \fmflabel{$\hat{b}^-$}{i3}
        \fmflabel{$\hat{a}^-$}{i2}
        \fmflabel{$b$}{ua1}
        \fmflabel{$n$}{o2}
        \fmflabel{$a$}{o4}
        \end{fmfgraph*}}
        \end{fmffile}
        \caption{\label{fig:antsplitm1}The contribution from $\ant(\hat a^-,\hat b^- \leftarrow a,1,\ldots,n,b)$ in the limit $ k_1,\ldots,k_n\parallel k_a$. The off-shell leg is denoted by a cross, and the double-line represents the CSW vertex to which $\ha$ and $\hb$ are attached.}
}

\section{Conclusion}
\label{sec:conclusion}

In this paper we have applied the techniques inspired by the twistor
approach to the computation of antenna functions that describe the
complete infrared behaviour of multi-gluon amplitudes. Our findings
can be summarised as follows.

Our first important result is that antenna functions can be directly
written as a sum of a well-defined and limited subset of MHV diagrams
among those contributing to a full QCD amplitude.  This method
leads to very compact expressions for the antennas, which might be
more suitable for analytic integration over the singular phase space
regions, needed in higher order computations. 
It also gives a very natural way of parametrizing, by means
of an antiholomorphic spinor $\eta$, the intrinsic arbitrariness of
the antenna function outside the singular regions.  As an application,
we have shown that it is straightforward to provide explicit expressions for
MHV and NMHV antenna functions at all orders, which are sufficient to
build the full set of gluon antenna functions up to N$^3$LO. Although the knowledge 
of these multi-leg antenna functions might not
play a practical role at present for QCD calculation beyond leading
order, their knowledge could be of interest, for instance, in 
testing the infrared structure of recently introduced 
conjectures for gluon amplitudes at
all order in MSYM~\cite{Bern:2005iz}.

The special set of MHV rules relevant for building antenna functions,
can be easily recast in a recursive form, following the same approach already
introduced in Ref~\cite{Duhr:2006iq}.  This recursion is formulated in terms of currents and has a similar structure as the Berends-Giele recursive relations. Symbolically, the recursion can be written as
\unitlength=1mm
            \begin{fmffile}{SLDL}
               \begin{equation}
               \begin{split}
               J_1(n)&= \quad\parbox{10mm}{ \begin{fmfgraph}(7,7)
                \fmfleft{i1}
                \fmfright{o1,o2}
                \fmf{plain}{i1,v1}
                \fmf{plain}{v1,o1}
                \fmf{plain}{v1,o2}
\end{fmfgraph}}+\quad\parbox{10mm}{
                \begin{fmfgraph}(7,7)
                \fmfleft{i1}
                \fmfright{o1,o2}
                \fmf{plain}{i1,v1}
                \fmf{plain}{v1,o1}
                \fmf{dbl_plain}{v1,o2}
                \end{fmfgraph}},\\
                &\\
J_2(n)& = \quad\parbox{10mm}{ 
                \begin{fmfgraph}(7,7)
                \fmfleft{i1}
                \fmfright{o1,o2}
                \fmf{dbl_plain}{i1,v1}
                \fmf{plain}{v1,o1}
                \fmf{plain}{v1,o2}
                \end{fmfgraph}}
+\quad\parbox{1mm}{
                \begin{fmfgraph}(7,7)
                \fmfleft{i1}
                \fmfright{o1,o2}
                \fmf{dbl_plain}{i1,v1}
                \fmf{plain}{v1,o1}
                \fmf{dbl_plain}{v1,o2}
\end{fmfgraph}}\quad\quad\,\,,
\end{split}
\end{equation}
\end{fmffile}
where $J_1$ and $J_2$ denote the single and double-line currents (See
Appendix~\ref{appendix:B}). This recursion is formally equivalent to
the version of the Berends-Giele recursion where the four-point gluon
vertex is decomposed into three-point vertices involving a tensor
particle~\cite{Duhr:2006iq}, apart from the second term in the
equation for $J_2$, which arises only when vertices of multiplicity
five or higher are decomposed into three-point vertices. A tree-level
gluon amplitude can now be obtained in a straightforward
manner~\cite{Duhr:2006iq} by
\begin{fmffile}{ampj1}
\begin{equation}
A_{n+1} = J_1(n) = \parbox{5mm}{\begin{fmfgraph}(5,5)
\fmfleft{i1}
\fmfright{o1}
\fmf{plain}{i1,o1}
\fmfblob{10}{o1}
\end{fmfgraph}}
\quad.
\end{equation}
\end{fmffile}
The interesting and non-trivial
outcome of this formulation is that antenna and splitting amplitudes (and also
eikonal factors) can be built recursively in the same way as the amplitudes, 
\ie, without applying
any limiting procedure. 
For example, an antenna function for $(n-2)$ unresolved particles corresponds to the $n$-point double-line current\footnote{As it was discussed in Section~\ref{sec:antRR}, it is in fact a sum over all possible double-line currents.}
\begin{fmffile}{antj2}
\begin{equation}
\ant(1,2,\ldots,n) = V_4\,J_2(n) = V_4\,\parbox{5mm}{\begin{fmfgraph}(5,5)
\fmfleft{i1}
\fmfright{o1}
\fmf{dbl_plain}{i1,o1}
\fmfblob{10}{o1}
\end{fmfgraph}}
\quad,
\end{equation}
\end{fmffile}
whereas $n$-point splitting amplitudes are the sum of the $n$-point single and double-line currents,
\begin{fmffile}{splitj1j2}
\begin{equation}
\splitti(1,\ldots,n) = J_1(n) + V_c\,J_2(n) = 
\parbox{7mm}{\begin{fmfgraph}(5,5)
\fmfleft{i1}
\fmfright{o1}
\fmf{plain}{i1,o1}
\fmfblob{10}{o1}
\end{fmfgraph}}
+V_c\,
\parbox{5mm}{\begin{fmfgraph}(5,5)
\fmfleft{i1}
\fmfright{o1}
\fmf{dbl_plain}{i1,o1}
\fmfblob{10}{o1}
\end{fmfgraph}}
\quad.
\end{equation}
\end{fmffile}
The recursive relations obeyed by the
``currents'' needed for the full amplitudes and for the splitting
functions are formally exactly the same, the only differences being in
the definition of vertices and in the initial conditions.  In other
words, one finds again that the collinear splitting amplitudes enjoy
all the known properties of the full amplitudes: not only gauge
invariance, dual ward identities, Kleiss-Kuif relations,\ldots, but
also a fully recursive formulation.

In conclusion, we have shown how to efficiently obtain
compact multi-gluon antenna functions. Generalization of our approach to quark antenna functions as well as
to the computation of impact factors and Lipatov vertices relevant
in the high-energy limit~\cite{DelDuca:1999ha} is in progress.

\section*{Acknowledgments}

We would like to thank S.~Badger, N.~Glover, S.~Hoeche, D.~Kosower, D.~Maitre and R.~Roiban for useful discussions, as well as A.~Gehrmann, T.~Gehrmann and V.~Del Duca for valuable comments on the manuscript. FM thanks the Aspen Center for physics for the warm
hospitality at the later stage of this work.
CD is a Research Fellow of the \emph{Fonds National de la Recherche Scientifique}, Belgium.

%
%

\appendix

\section{Reconstruction functions}
\label{app:recfun}
In this appendix we present the explicit expressions for the reconstruction functions that have been used. Apart from a difference in the overall sign conventions they are the same as in Ref.~\cite{Kosower:2002su}. They read
\begin{eqnarray}
\hka & = & \frac{1}{2(K^2-t_{1\ldots nb})}\Bigg[(1+\rho)K^2+2R\cdot (k_a-k_b-K)+\frac{1}{s_{ab}}G\left(\begin{array}{cc}
k_a, & k_b \\
R, & P_{1,n}
\end{array}\right)\Bigg]\,k_a + R\nonumber\\
\label{eq:reconstruct1}
 & & +\frac{1}{2(K^2-t_{a1\ldots n})}\Bigg[(1-\rho)K^2+2R\cdot
 (k_a-k_b-K)+\frac{1}{s_{ab}}G\left(\begin{array}{cc}
k_a, & k_b \\ R, & P_{1,n}
\end{array}\right)\Bigg]\,k_b,\\
\hkb & = & \frac{1}{2(K^2-t_{1\ldots nb})}\Bigg[(1-\rho)K^2+2\tilde R\cdot (k_a-k_b-K)+\frac{1}{s_{ab}}G\left(\begin{array}{cc}
k_a, & k_b \\
\tilde R, & P_{1,n}
\end{array}\right)\Bigg]\,k_a+\tilde R\nonumber\\
\label{eq:reconstruct2}
 & & +\frac{1}{2(K^2-t_{a1\ldots n})}\Bigg[(1+\rho)K^2+2\tilde R\cdot
 (k_a-k_b-K)+\frac{1}{s_{ab}}G\left(\begin{array}{cc}
k_a, & k_b \\ \tilde R, & P_{1,n}
\end{array}\right)\Bigg]\,k_b,
\end{eqnarray}
where $t_{1\ldots nb} = (k_1+\ldots+k_n+k_b)^2$, $t_{a1\ldots n} = (k_a+k_1+\ldots+k_n)^2$ and $K=k_a+k_1+\ldots+k_n+k_b$. Furthermore 
\begin{equation}
\label{eq:reconstruct3}
R=\sum_{j=1}^nk_jr_j, \qquad \tilde R =P_{1,n}-R = \sum_{j=1}^nk_j(1-r_j),
\end{equation}
and
\begin{equation}
\label{eq:reconstruct4}
\rho = \left[1+\frac{2G\left(\begin{array}{ccc}
a, & R, & b\\
a, & \tilde R, & b
\end{array}\right)}{K^2s_{ab}^2 }
+\frac{\Delta(a,R,K,b)}{(K^2)^2s_{ab}^2}\right]^{\frac{1}{2}}.
\end{equation}
In these formulas, $G$ and $\Delta$ denote Gram determinants,
\begin{equation}
G\left(\begin{array}{ccc}
p_1, & \ldots,& p_n\\
q_1, & \ldots,& q_n
\end{array}\right)
\equiv \det\big(2\,p_i\cdot q_j\big), \qquad \Delta(p_1,\ldots,p_n)\equiv G\left(\begin{array}{ccc}
p_1, & \ldots,& p_n\\
p_1, & \ldots,& p_n
\end{array}\right).
\end{equation}
A suitable choice for the coefficients $r_j$ is~\cite{Kosower:2002su}
\begin{equation}
\label{eq:coeff}
r_j=\frac{k_j\cdot(P_{j+1,n}+k_b)}{k_j\cdot K}=\frac{t_{j\ldots nb}-t_{(j+1)\ldots nb}}{2k_j\cdot K}.
\end{equation}


\section{Gluon insertion rule}
\label{app:appendixa}
\begin{figure}[!t]
\begin{center}
            \begin{fmffile}{gluoninsertion}
            \parbox{50mm}{
            $(a)$\begin{fmfgraph}(30,30)
        \fmfcurved
        \fmfleft{i1}
        \fmfright{o0,o1,o2,o3,o4}
        \fmf{plain}{i1,v1}
        \fmf{dashes}{v1,o2}
        \fmf{plain}{v1,o1}
        \fmf{plain}{v1,o3}
        \fmfblob{10mm}{i1}
        \end{fmfgraph}}
         \parbox{50mm}{
            $(b)$\begin{fmfgraph}(30,30)
        \fmfcurved
        \fmfleft{i1}
        \fmfright{o1,o2,o3,o4}
        \fmf{plain}{i1,v1}
        \fmf{dashes}{v1,v2}
        \fmf{phantom}{v1,a1,o1}
        \fmf{phantom}{v1,a2,o4}
        \fmf{plain}{v1,a1}
        \fmf{plain}{v1,a2}
        \fmf{dashes}{v2,o2}
        \fmf{dashes}{v2,o3}
        \fmfblob{10mm}{i1}
        \end{fmfgraph}}
        \parbox{50mm}{
            $(c)$\begin{fmfgraph}(30,30)
        \fmfcurved
        \fmfleft{i1}
        \fmfright{o1,o2,o3,o4,o5}
        \fmf{plain}{i1,v1}
        \fmf{dashes}{v1,v2,o3}
        \fmf{phantom}{v1,a1,o1}
        \fmf{phantom}{v1,a2,o5}
        \fmf{plain}{v1,a1}
        \fmf{plain}{v1,a2}
        \fmf{dashes}{v2,o2}
        \fmf{dashes}{v2,o4}
        \fmfblob{10mm}{i1}
        \end{fmfgraph}}
                                     \end{fmffile}\caption{\label{fig:3V}The radiation of a soft gluon from a three-point vertex. Dashed lines indicate soft gluons. The blob represents any subdiagram contributing to this amplitude.}
\end{center}
\end{figure}
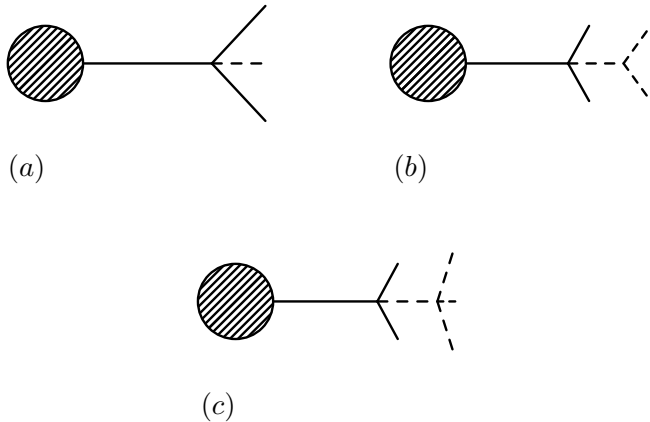

In this appendix we present the proof of the gluon insertion rule presented in Section~\ref{sec:antennaCSW}. 
The idea behind the gluon insertion rule is to start from the Feynman diagrams of the hard amplitude and to see where a soft gluon can be radiated from in such a way that the diagram becomes divergent. There are three places in a Feynman diagram where an additional soft gluon can be emitted from:
\begin{itemize}
\item[-]
a three-point vertex,
\item[-]
an internal line,
\item[-]
an external line.
\end{itemize}
Let us examine each case separately, and let us start with the situation where a soft gluon is emitted from a three-point vertex (See Fig.~\ref{fig:3V}). It is easy to see from Fig.~\ref{fig:3V}a that if a single soft gluon is radiated from a three-point vertex, there is no propagator going on-shell, and thus there is no divergence at all in this situation. But if more than one single gluon is emitted, then there may be a propagator going on-shell, and thus we can get a divergent diagram for more than one single soft gluon emitted from the three-point vertex (See Fig.~\ref{fig:3V}b and~\ref{fig:3V}c).\\
Let us first consider the situation where two soft gluons are emitted from the three-point vertex shown in Fig.~\ref{fig:3V}b. In this situation the propagator goes on-shell, and  the rescaling rule, Eq.~(\ref{eq:brij}), tells us that the propagator behaves as $t^{-2}$. However, as a three-point vertex is proportional to the momenta of the particles, it is easy to see that the second three-point vertex behaves in the soft limit as $t$, and thus the diagram shown in Fig.~\ref{fig:3V}b behaves as $t^{-2}\cdot t=t^{-1}$, and so it is not divergent enough to contribute to the soft factor.
Let us turn to the situation where three soft gluons are emitted from a three-point vertex (See Fig.~\ref{fig:3V}c). We get a divergent propagator which behaves as $t^{-2}$. The four-point vertex however is not divergent at all, and so the diagram shown in Fig.~\ref{fig:3V}c behaves as $t^{-2}$, and so it is not divergent enough to contribute to the soft limit.
Finally, we come to the conclusion that diagrams where a soft particle is radiated from a three-point vertex do not contribute in the soft limit.\\
Let's turn to the second case, where the soft particle is emitted from an internal line. Internal lines in a Feynman diagram correspond to off-shell propagators, \ie, if $p$ is the momentum carried by the internal line, then $p^2\neq 0$. If a soft particle with momentum $k$ is radiated from this internal line, then we get a propagator of the form $1/(p+k)^2=1/(p^2+2p\cdot k)$, which stays finite in the soft limit where $k\rightarrow 0$. So there will be no contribution from diagrams where the soft gluon is emitted from an internal line.\\
In the situation where a soft gluon with momentum $k$ is emitted from an external particle with momentum $p$, we get a propagator $1/(p+k)^2=1/(2p\cdot k) $. In the soft limit, $k\rightarrow 0$, this propagator behaves as $1/t$, and thus has the right divergence to contribute to the soft factor.\\
Finally, we see that only those Feynman diagrams contribute in the soft limit where  the soft gluons are radiated from the external legs of the hard amplitude, which finishes the proof.


\section{Recursive formulation of the CSW formalism}
\label{appendix:B}
\newcommand{\dst}{\displaystyle}
\newcommand{\sst}{\scriptstyle}
\newcommand{\abr}[1]{\langle #1\rangle}
\newcommand{\spp}[1]{\left| #1\right.\rangle}
\newcommand{\mq}[1]{\text{``}#1\text{''}}
\newcommand{\ldu}{0.3}

\newcommand{\mycaption}[2]{\refstepcounter{table} 
  \centerline{\small{\bf Table \thetable : }\parbox[t]{#1}{#2}}} 
\newcommand{\mytable}[5]{ 
  \begin{table}[#1!] 
    \centerline{\parbox{#2}{\begin{center}#4\end{center}}} 
    \centerline{\mycaption{#3}{#5}} 
  \end{table}} 

\newcolumntype{x}[1]{>{\raggedleft}p{#1}} 
\newcolumntype{y}[1]{@{}p{#1}}

\fmfset{dot_len}{1mm}
\newcommand{\cswv}{\parbox{2cm}{\begin{center}
\begin{fmfgraph*}(15,12)
  \fmftop{i,j}
  \fmfbottom{ij}
  \fmf{plain}{i,v1,j}
  \fmf{plain,tension=1.5}{v1,ij}
  \fmffreeze
  \fmf{phantom}{v2,v1}
  \fmf{phantom,tension=5}{ij,v2}
  \fmfv{d.s=0,l.d=\ldu,l.a=90,l=$\sst I$}{i}
  \fmfv{d.s=0,l.d=\ldu,l.a=90,l=$\sst J$}{j}
  \fmfv{d.shape=cross,d.size=3,l.a=0,l=$\sst K$}{ij}
\end{fmfgraph*}\end{center}}}
\newcommand{\cswvp}{\parbox{2cm}{\begin{center}
\begin{fmfgraph*}(15,12)
  \fmftop{i,j}
  \fmfbottom{ij}
  \fmf{plain}{i,v1,j}
  \fmf{double,tension=1.5}{v1,ij}
  \fmffreeze
  \fmf{phantom}{v2,v1}
  \fmf{phantom,tension=5}{ij,v2}
  \fmfv{l.d=\ldu,l.a=90,l=$\sst I$}{i}
  \fmfv{l.d=\ldu,l.a=90,l=$\sst J$}{j}
  \fmfv{d.shape=cross,d.size=3,l.a=0,l=$\sst K$}{ij}
\end{fmfgraph*}\end{center}}}
\newcommand{\cswpvp}{\parbox{2cm}{\begin{center}
\begin{fmfgraph*}(15,12)
  \fmftop{i,j}
  \fmfbottom{ij}
  \fmf{double}{i,v1}
  \fmf{plain}{v1,j}
  \fmf{double,tension=1.5}{v1,ij}
  \fmffreeze
  \fmf{phantom}{v2,v1}
  \fmf{phantom,tension=5}{ij,v2}
  \fmfv{l.d=\ldu,l.a=90,l=$\sst I$}{i}
  \fmfv{l.d=\ldu,l.a=90,l=$\sst J$}{j}
  \fmfv{d.shape=cross,d.size=3,l.a=0,l=$\sst K$}{ij}
\end{fmfgraph*}\end{center}}}
\newcommand{\cswpv}{\parbox{2cm}{\begin{center}
\begin{fmfgraph*}(15,12)
  \fmftop{i,j}
  \fmfbottom{ij}
  \fmf{double}{v1,i}
  \fmf{plain}{j,v1}
  \fmf{plain,tension=1.5}{ij,v1}
  \fmffreeze
  \fmf{phantom}{v2,v1}
  \fmf{phantom,tension=5}{ij,v2}
  \fmfv{l.d=\ldu,l.a=90,l=$\sst I$}{i}
  \fmfv{l.d=\ldu,l.a=90,l=$\sst J$}{j}
  \fmfv{d.shape=cross,d.size=3,l.a=0,l=$\sst K$}{ij}
\end{fmfgraph*}\end{center}}}
\newcommand{\lcswpvp}{\parbox{2cm}{\begin{center}
\begin{fmfgraph*}(15,12)
  \fmftop{i,j}
  \fmfbottom{ij}
  \fmf{double}{j,v1}
  \fmf{plain}{v1,i}
  \fmf{double,tension=1.5}{v1,ij}
  \fmffreeze
  \fmf{phantom}{v2,v1}
  \fmf{phantom,tension=5}{ij,v2}
  \fmfv{l.d=\ldu,l.a=90,l=$\sst I$}{i}
  \fmfv{l.d=\ldu,l.a=90,l=$\sst J$}{j}
  \fmfv{d.shape=cross,d.size=3,l.a=0,l=$\sst K$}{ij}
\end{fmfgraph*}\end{center}}}
\newcommand{\lcswpv}{\parbox{2cm}{\begin{center}
\begin{fmfgraph*}(15,12)
  \fmftop{i,j}
  \fmfbottom{ij}
  \fmf{double}{v1,j}
  \fmf{plain}{i,v1}
  \fmf{plain,tension=1.5}{ij,v1}
  \fmffreeze
  \fmf{phantom}{v2,v1}
  \fmf{phantom,tension=5}{ij,v2}
  \fmfv{l.d=\ldu,l.a=90,l=$\sst I$}{i}
  \fmfv{l.d=\ldu,l.a=90,l=$\sst J$}{j}
  \fmfv{d.shape=cross,d.size=3,l.a=0,l=$\sst K$}{ij}
\end{fmfgraph*}\end{center}}}

\paragraph{Decomposition of the MHV-amplitudes.}
\label{sec:MHVdecomp}
In this section we present a recursive method to calculate QCD tree-level amplitudes using the CSW construction. This recursive algorithm is equivalent to the recursion presented in Ref.~\cite{Duhr:2006iq}\footnote{A numerical study of the complexity of this algorithm can be found in Ref.~\cite{Duhr:2006iq}}. We will start by analysing this construction for MHV-amplitudes. In a second step we generalize this decomposition of MHV-amplitudes to CSW vertices, the fundamental building blocks in the CSW construction.

Let us start with the $n$-point gluon MHV-amplitude of the form 
\begin{equation}
A_n(1^+,2^+,\ldots,(n-2)^+,(n-1)^-,n^-)=\frac{\br{(n-1)\,n}^4}{\br{12}\ldots\br{n1}}.
\end{equation}
This amplitude can be written in terms of eikonal factors
\begin{eqnarray}
\label{eq:dipolefac}
 & & A_n(1^+,2^+,\ldots,(n-2)^+,(n-1)^-,n^-)  =  \\
  & & \qquad \qquad\qquad\qquad\qquad\nonumber \frac{\br{(n-1)\,n}^4}{\br{1\,(n-2)}\br{(n-2)\,(n-1)}\br{(n-1)\,n}\br{n1}}\,\prod_{k=3}^{n-2}D_{1,k}^{k-1},
\end{eqnarray}
where the eikonal factors are defined by $D_{ij}^k=\frac{\br{ij}}{\br{ik}\br{kj}}$.\\
Let us consider now the current $J_h(1,\ldots,n-1)$ defined as the sum of all CSW diagrams which have $n$ on-shell external legs and one off-shell leg with helicity $h$ \cite{Duhr:2006iq} \footnote{In the context of the CSW rules, it makes sense to talk 
  about the helicity of an off-shell particle. Note that, as the off-shell continuation 
  of the spinors involves an arbitrary reference spinor $\eta^{\dot a}$, 
  these currents are not gauge invariant objects. 
  However, the $\eta^{\dot a}$ dependence drops out in the end 
  \cite{Cachazo:2004kj}.}. We will refer to such a current as a single-line current.
A specific $(n+1)$-point amplitude can then be obtained by putting the off-shell leg on-shell,
\begin{equation}
\label{eq:putOS}
A_n(1,\ldots,n,(n+1)^h)=\lim_{P_{1,n}^2\rightarrow 0}P_{1,n}^2\,J_{-h}(1,\ldots,n).
\end{equation}
In particular, for an MHV-type current we can write
\begin{eqnarray}
 & & J_{+}(1^+,2^+,\ldots,(n-2)^+,(n-1)^-)  = \, \frac{1}{P_{1,n-1}^2}\,\frac{\br{(n-1),P_{1,n-1}}^4}{\br{1,2}\ldots\br{(n-1),P_{1,n-1}}\br{P_{1,n-1},1}}\nonumber\\
 & & \qquad \qquad = \, \frac{1}{P_{1,n-1}^2}\,\frac{\br{(n-1),P_{1,n-1}}^4}{\br{1,(n-2)}\br{(n-2),(n-1)}\br{(n-1),P_{1,n-1}}\br{P_{1,n-1},1}}\,\prod_{k=3}^{n-2}D_{1k}^{k-1}\nonumber\\
 \label{eq:MHVdecomp}
 & &  \qquad\qquad =\, \frac{1}{P_{1,n-1}^2}\,V_4(1,n-2,n-1,P_{1,n-1})\,\prod_{k=3}^{n-2}D_{1k}^{k-1},
\end{eqnarray}
where we define
\begin{equation}
V_4(a,b,c,d)=\frac{\br{cd}^4}{\br{ab}\br{bc}\br{cd}\br{da}}.
\end{equation}
In Fig.~\ref{fig:MHVdecomp}, we introduce a graphical representation
for Eq.~(\ref{eq:MHVdecomp}) in terms of an internal double-line to
which the external gluons may couple.  We also introduce double-line
currents $J_{uv}^m(1,\ldots,n)$ defined as the double-line diagram
with an external (off-shell) double-line to which the gluons $1$ to
$n$ are attached. The indices $u$ and $v$ refer to the indices of the
first particle one encounters if one follows each line of the
double-line into the current, and $m$ refers to the number of
negative-helicity gluons attached to the double-line \footnote{Up to
now these indices are redundant. The double-line currents contributing
to an MHV single-line current of the form
$J_+(1^+,\ldots,(n-2)^+,(n-1)^-)$ are all of the form
$J_{uv}^m(1^+,\ldots,k^+))$ with $u=1$, $v=k$ and $m=0$. The meaning
of these indices will become clear when we generalize double-line
currents to the CSW construction.}. As a double-line is part of an
MHV-amplitude with exactly two negative-helicity particles,
$J_{uv}^m(1,\ldots,n)=0$ if $m\ge3$.
\FIGURE[!t]{
\begin{fmffile}{MHVDecomp}
\parbox{22mm}{\begin{fmfgraph}(40,12)
\fmfleft{i1}
\fmfright{o1}
\fmf{phantom}{i1,o1}
\end{fmfgraph}}
\parbox{50mm}{\begin{fmfgraph*}(40,20)
\fmfstraight
\fmfleft{i1,i2,i3}
\fmfright{o1,o2,o3}
\fmfbottom{b1,b2,b3,b4,b5,b6,b7,b8}
\fmf{phantom}{i2,v1,v2,v3,v4,v5,v6,o2}
\fmffreeze
\fmf{dbl_plain}{v1,v2,v3,v4,v5,v6}
\fmf{plain}{v1,i1}
\fmf{plain}{v1,i3}
\fmf{plain}{v6,o1}
\fmf{plain}{v6,o3}
\fmffreeze
\fmf{plain}{v2,b3}
\fmf{phantom}{v4,a1,b3}
\fmfdot{a1}
\fmf{plain}{v4,b5}
\fmf{plain}{v5,b6}
\fmfv{l.d=0.2mm,l=$\scriptstyle 3$}{b6}
\fmfv{l.d=0.2mm,l=$\scriptstyle 4$}{b5}
\fmfv{l.d=0.1mm,l=$\scriptstyle n-2$}{b3}
\fmfv{l.d=0.5mm,l=$\scriptstyle n-1$}{i1}
\fmfv{l.d=1mm,l=$\scriptstyle n$,d.s=cross,d.a=45,d.size=2mm}{i3}
\fmfv{l.d=0.5mm,l=$\scriptstyle 1$}{o3}
\fmfv{l.d=0.5mm,l=$\scriptstyle 2$}{o1}
\end{fmfgraph*}}
\parbox{22mm}{\begin{fmfgraph}(40,12)
\fmfleft{i1}
\fmfright{o1}
\fmf{phantom}{i1,o1}
\end{fmfgraph}}
\vspace{3mm}
\end{fmffile}
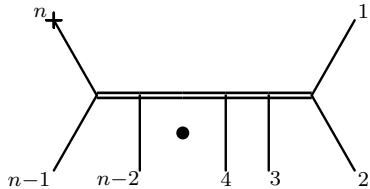
\caption{\label{fig:MHVdecomp}Decomposition of an MHV-amplitude where the MHV vertex has been stretched out into a double-line. The cross indicates the off-shell line.}
}

\begin{figure}[!t]
\begin{fmffile}{A6RR}
\begin{align*}
\begin{fmfgraph*}(25,20)
\fmfleft{i1}
\fmfright{o1,o2}
\fmf{dbl_plain}{i1,v1}
\fmfv{d.s=cross,d.a=0,d.size=2.5mm}{i1}
\fmf{plain}{v1,o1}
\fmf{plain}{v1,o2}
\fmflabel{$\scriptstyle 2^+$}{o1}
\fmflabel{$\scriptstyle 1^+$}{o2}
\end{fmfgraph*}
& \qquad J^0_{12}(1^+,2^+)=1\\
\phantom{a}&\phantom{b}\\
\begin{fmfgraph*}(30,20)
\fmfleft{i1}
\fmfbottom{b1,b2,b3,b4}
\fmfright{o1,o2,o3}
\fmf{phantom}{i1,v2,v3,o2}
\fmf{dbl_plain}{i1,v2,v3}
\fmfv{d.s=cross,d.a=0,d.size=2.5mm}{i1}
\fmf{plain}{v3,o1}
\fmf{plain}{v3,o3}
\fmffreeze
\fmf{plain}{v2,b2}
\fmflabel{$\scriptstyle 2^+$}{o1}
\fmflabel{$\scriptstyle 1^+$}{o3}
\fmflabel{$\scriptstyle 3^+$}{b2}
\end{fmfgraph*}
& \qquad J^0_{13}(1^+,2^+,3^+)=D^2_{13}\\
\phantom{a}&\phantom{b}\\
\begin{fmfgraph*}(35,20)
\fmfleft{i1}
\fmfbottom{b1,b2,b3,b4,b5}
\fmfright{o1,o2,o3}
\fmf{phantom}{i1,v2,v3,v4,o2}
\fmf{dbl_plain}{i1,v2,v3,v4}
\fmf{plain}{v4,o1}
\fmf{plain}{v4,o3}
\fmffreeze
\fmf{plain}{v2,b2}
\fmf{plain}{v3,b3}
\fmfv{d.s=cross,d.a=0,d.size=2.5mm}{i1}
\fmflabel{$\scriptstyle 2^+$}{o1}
\fmflabel{$\scriptstyle 1^+$}{o3}
\fmflabel{$\scriptstyle 4^+$}{b2}
\fmflabel{$\scriptstyle 3^+$}{b3}
\end{fmfgraph*}
& \qquad J^0_{14}(1^+,2^+,3^+,4^+)=D^3_{14}D^2_{13}\\
\phantom{a}&\phantom{b}\\
\begin{fmfgraph*}(40,20)
\fmfleft{i1,i2,i3}
\fmfbottom{b1,b2,b3,b4,b5,b6}
\fmfright{o1,o2,o3}
\fmf{phantom}{i2,v2,v3,v4,v5,o2}
\fmf{dbl_plain}{v2,v3,v4,v5}
\fmf{plain}{v5,o1}
\fmf{plain}{v5,o3}
\fmf{plain}{v2,i1}
\fmf{plain}{v2,i3}
\fmffreeze
\fmf{plain}{v3,b3}
\fmf{plain}{v4,b4}
\fmflabel{$\scriptstyle 2^+$}{o1}
\fmflabel{$\scriptstyle 1^+$}{o3}
\fmflabel{$\scriptstyle 5^-$}{i1}
\fmflabel{$\scriptstyle 6^-$}{i3}
\fmflabel{$\scriptstyle 4^+$}{b3}
\fmflabel{$\scriptstyle 3^+$}{b4}
\fmfv{d.s=cross,d.a=20,d.size=2.5mm}{i3}
\end{fmfgraph*}
& \qquad J^+(1^+,2^+,3^+,4^+,5^-)=V_4(1,4,5,6)D^3_{14}D^2_{13}
\end{align*}
\end{fmffile}
\vspace{3mm}
\caption{\label{fig:example}Construction of $A_6(1^+,2^+,3^+,4^+,5^-,6^-)$ using the recursion for the double-line current. A cross indicates an off-shell leg.}
\end{figure}
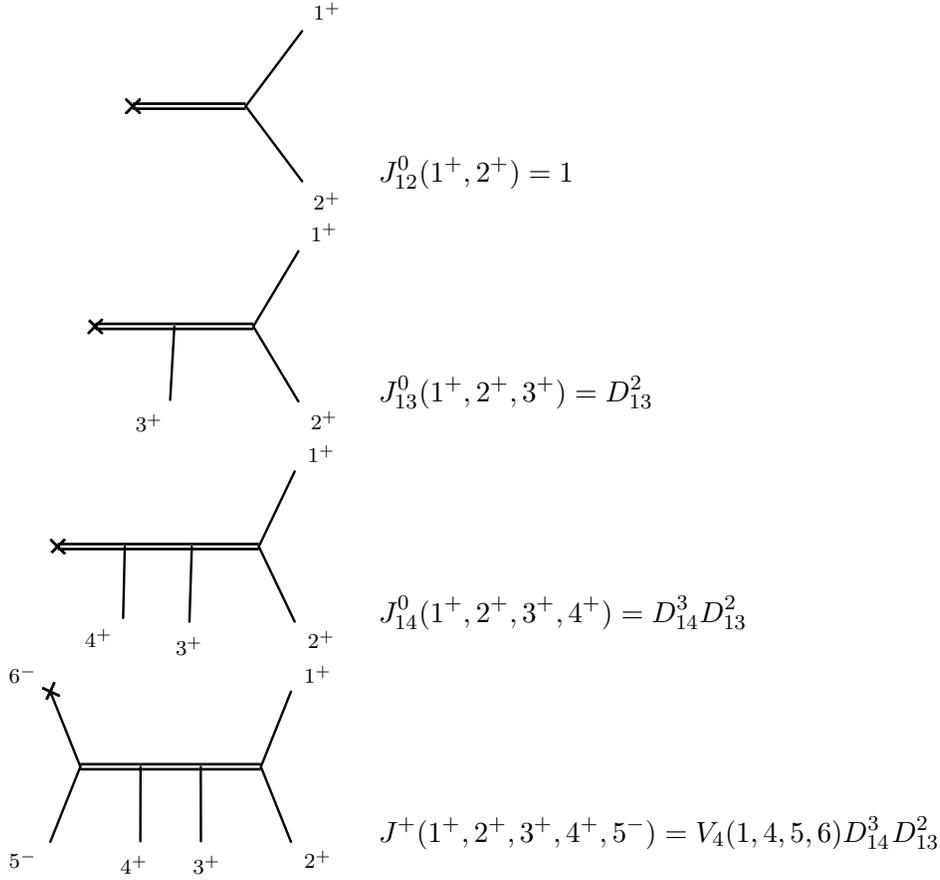

The double-line currents contributing to the MHV single-line current $J_+(1^+,\ldots,(n-2)^+,(n-1)^-)$ can now be easily constructed recursively by adding successively eikonals to the double-line current,
\begin{equation}
\label{eq:DLrec1}
J_{1k}^0(1^+,\ldots,k^+)=D_{1k}^{k-1}\,J_{1(k-1)}^0(1^+,\ldots,(k-1)^+),
\end{equation}
where we define that all one-point double-line currents are zero, and all two-point double-line currents are given by $J_{uv}^0(1^+,2^+)=\delta_{1u}\delta_{2v}$. The MHV single-line current is then given by
\begin{equation}
\label{eq:DLrec2}
J_+(1^+,\ldots,(n-2)^+,(n-1)^-)=V_4(1,n-2,n-1,P_{1,n})\,J_{1(n-2)}^0(1^+,\ldots,(n-2)^+),
\end{equation}
and the MHV-amplitude can then be obtained by putting the off-shell leg on-shell, Eq.~(\ref{eq:putOS}).
This procedure is illustrated in Fig.~\ref{fig:example} for the six-point MHV-amplitude .

Up to now we have only considered a very special class of MHV-amplitudes, namely those where the two negative-helicity gluons are adjacent in the MHV-amplitude. We now generalize this procedure to an arbitrary MHV-amplitude $A_n(1^+,\ldots,i^-,\ldots,j^-,\ldots,n^+)$. However, the factorisation~(\ref{eq:dipolefac}) stays no longer true in this case, because an MHV-amplitude only factorises when a soft positive-helicity gluon is radiated. A closer look at the recursive procedure we have introduced reveals that we only used this method to construct the denominator of the MHV-amplitude, whereas the numerator is completely encoded in $V_4(1,n-2,n-1,n)$. Thus we can generalize this procedure using the following rules:
\begin{enumerate}
\item Each time we add a negative-helicity particle $j$ to a double-line current containing already $m$ negative-helicity particles, add a $\delta_{jm_i}$ to the double-line current, where $i=1$ if $m$ passes from 0 to 1, and $i=2$ if $m$ passes from 1 to 2.
\item At the last step in the recursion sum over all possible values for $m_1$ and $m_2$,
\begin{equation}
J_h(1,\ldots,n-2,n-1)=\sum_{m_1,m_2}\,V_4(1,n-2,n-1,P_{1,n};m_1,m_2)\,J_{1(n-2)}^m(1,\ldots,n-2),
\end{equation}
where we defined
\begin{equation}
V_4(a,b,c,d;m_1,m_2)=\frac{\br{m_1,m_2}^4}{\br{ab}\br{bc}\br{cd}\br{da}}.
\end{equation}
\end{enumerate}

\paragraph{Generalisation to the CSW construction.}
\label{sec:CSWRR}
In this section we extend the  
recursive relations to the CSW formalism presented in Section~\ref{sec:twistreview}.   
We start by making the observation that each line in a CSW diagram, as well internal as external, contributing to an $n$-point amplitude can be uniquely labeled by a multiindex $I=(i_1,i_2)$, with $1\le i_1\le i_2\le n$. Indeed, each external line can be characterized by its momentum $p_i$, and so we associate to the external particle $i$ the multiindex $I=(i,i)$, and each internal line is characterized by an off-shell momentum $P_{i,j}$, and so we associate to this internal line the multiindex $I=(i,j)$. In general, we have
\begin{enumerate}
\item To each line with momentum $P_{i_1,i_2}$ we can associate in a unique way a multiindex $I=(i_1,i_2)$ with $i_1\le i_2$ defined by $P_I\equiv P_{i_1,i_2}$.
\item 
A line in a CSW diagram with multiindex $I=(i_1,i_2)$ is an external line if and only if $i_1=i_2$.
\end{enumerate}
For later convenience, we introduce the following definitions:
\begin{itemize}
\item[-]
An on-shell particle with momentum $p_i=P_{i,i}$ is labelled by the multiindex 
\begin{equation}
\label{eq:onshellmultind}\bar{\imath}\equiv(i,i).
\end{equation}
\item[-]
Kronecker-delta:
\begin{equation}
\delta^I_{J}=\delta^{i_1}_{j_1}\,\delta^{i_2}_{j_2}.
\end{equation}
\item[-]
Ordering relation:
\begin{equation}
I<J \Leftrightarrow i_1\le i_2 < j_1\le j_2.
\end{equation}
\end{itemize}

We want to write down a recursion relation for the CSW formalism. Looking back to ordinary Feynman rules and the Berends-Giele recursion, it is easy to write out a recursion for the single-line current defined in the previous section~\cite{Duhr:2006iq},
\begin{eqnarray}
J_h(1,\ldots,n)=\frac{1}{P_{1,n}^2} & \Bigg[ & \sum_{i=1}^{n-1}A_3\left(-P_{1,n}^{-h},\,P_{1,k}^{h_1},\,P_{k+1,n}^{h_2}\right)J_{h_1}(1,\ldots,i)\,J_{h_2}(i+1,\ldots,n)\nonumber\\
\label{eq:CSWrec}  & + & \sum_{i=1}^{n-2}\sum_{j=i+1}^{n-1}A_4\left(-P_{1,n}^{-h},\,P_{1,i}^{h_1},\,P_{i+1,j}^{h_2},\,P_{j+1,n}^{h_3}\right)J_{h_1}(1,\ldots,i)\\
 & & \qquad \qquad J_{h_2}(i+1,\ldots,j) J_{h_3}(j+1,\ldots,n)+\ldots\Bigg]\nonumber.
\end{eqnarray}
The dots indicate terms with higher order CSW vertices. A sum over the 
helicities $(h,h_1,h_2,\ldots)$ with $-h+h_1+h_2+\ldots=n-4$ is implicitly
understood. According to the CSW rules, the vertices $A_n$ correspond to 
off-shell continued $n$-point MHV-amplitudes, and so these vertices can themselves be constructed recursively using the recursive procedure introduced in the previous section. Merging the recursion relations~(\ref{eq:DLrec1} - \ref{eq:DLrec2}) and~(\ref{eq:CSWrec}) and replacing indices by multiindices, we obtain a recursion for the single and double-line currents,
\begin{eqnarray}
J_{UV}^m(1,\ldots,n) & = & \delta^{u_2}_{v_1-1}\,\delta_{(1,u_2)}^U\,\delta_{(v_1,n)}^V\,\epsilon_m^{h_1\,h_2}(P_U,P_V)\,J_{h_1}(1,\ldots,u_2)\,J_{h_2}(v_1,\ldots,n)\nonumber\\
\label{eq:rec1}
 & + & (1-\delta^{u_2}_{v_1-1})\,\sum_{\substack{W \\ w_2=v_1-1}}\,\delta_{(v_1,n)}^V\,\epsilon_{mm'}^h(P_V)\,D_{UW}^{V}\,J_{UW}^{m'}(1,\ldots,v_1-1)\\
 & & \qquad\qquad J_h(v_1,\ldots,n),\nonumber\\
J_h(1,\ldots,n) & = & \frac{1}{P_{1,n}^2}\sum_{k=1}^{n-1}\sum_{M_1,M_2}\Bigg[\epsilon^{h\,h_1\,h_2}(P_{1,n},P_{1,k},P_{k+1,n})\,V_3(P_{1,k},P_{k+1,n};M_1,M_2)\nonumber \\
\label{eq:rec2}
 & & \qquad\qquad J_{h_1}(1,\ldots,k)\,J_{h_2}(k+1,\ldots,n)\\
 & + & \sum_{\substack{U<V \\v_2=k}} \epsilon_{|2-m|}^{h\,h_1}(P_{1,n},P_{k+1,n})\,V_4(P_U,P_V,P_{k+1,n},P_{1,n};P_{M_1},P_{M_2})\nonumber\\
 & & \qquad\qquad J_{UV}^m(1,\ldots,k)\,J_{h_1}(k+1,\ldots,n)\Bigg],\nonumber
\end{eqnarray}
where a sum over all repeated helicity indices is understood. Capital letters indicate multiindices and the vertices are summarized in Table~\ref{tab:CSWvert}. A graphical representation of the recursion can be found in Fig.~\ref{fig:dipolerecCSW}. $V_3$ and $V_4$ correspond to the three and four-point CSW vertices, 
\begin{eqnarray}
V_3(A,B,C;M_1,M_2) & = & \frac{\br{M_1,M_2}^4}{\br{AB}\br{BC}\br{CA}},\\
V_4(A,B,C,D;M_1,M_2) & = & \frac{\br{M_1,M_2}^4}{\br{AB}\br{BC}\br{CD}\br{DA}},
\end{eqnarray}
and the epsilon functions appearing in the recursion keep track of the helicities, 
\begin{eqnarray}
\epsilon_{m\,m'}^{h}(I) & = & \left\{\begin{array}{ll}
\delta^{M_m}_I & ,\ifi h=-1, m=m'+1, m\le 2\\
0 & \elsi
\end{array}\right.\\
\epsilon_{m}^{h_I\,h_J}(I,J) & = & \left\{\begin{array}{ll}
\delta^{M_1}_I\,\delta_J^{M_2} & ,\ifi h_I=h_J=-1, m=2,\\
\delta^{M_1}_I& ,\ifi h_I=-h_J=-1, m=1,\\
\delta^{M_1}_J& ,\ifi h_I=-h_J=1, m=1,\\
1& ,\ifi h_I=h_J=1, m=0,\\
0 &, \elsi,
\end{array}\right.\\
\epsilon^{h_I\,h_J\,h_K}(I,J,K) & = & \left\{\begin{array}{ll}
\delta_X^{M_1}\,\delta_Y^{M_2} & \ifi h_X=h_Y=-1, h_1+h_2+h_3=-1,\\
0 &, \elsi. 
\end{array}\right.
\end{eqnarray}
\FIGURE[!t]{
\begin{fmffile}{recursion}
\qquad\qquad\parbox{22mm}{\begin{fmfgraph*}(15,7)
\fmfcurved
\fmfleft{i1}
\fmfright{o1}
\fmf{plain}{i1,o1}
\fmfblob{10mm}{o1}
\fmflabel{$(a)$}{i1}
\fmfv{d.s=cross,d.a=0,d.size=2mm}{i1}
\end{fmfgraph*}}
        =\,\,
\parbox{18mm}{\begin{fmfgraph}(15,10)
\fmfcurved
\fmfleft{i1}
\fmfright{o1,o2}
\fmf{plain}{i1,v1}
\fmf{plain}{v1,o1}
\fmf{plain}{v1,o2}
\fmfblob{5mm}{o1}
\fmfblob{5mm}{o2}
\fmfv{d.s=cross,d.a=0,d.size=2.5mm}{i1}
\end{fmfgraph}}
+\,          
\parbox{18mm}{\begin{fmfgraph}(15,10)
\fmfcurved
\fmfleft{i1}
\fmfright{o1,o2}
\fmf{plain}{i1,v1}
\fmf{dbl_plain}{v1,o2}
\fmf{plain}{v1,o1}
\fmfblob{5mm}{o1}
\fmfblob{5mm}{o2}
\fmfv{d.s=cross,d.a=0,d.size=2.5mm}{i1}
\end{fmfgraph}}
\vspace{10mm}
\\
\parbox{22mm}{\begin{fmfgraph*}(15,7)
\fmfcurved
\fmfleft{i1}
\fmfright{o1}
\fmf{dbl_plain}{i1,o1}
\fmfblob{10mm}{o1}
\fmflabel{$(b)$}{i1}
\fmfv{d.s=cross,d.a=0,d.size=2.5mm}{i1}
\end{fmfgraph*}}
        =\,\,
\parbox{18mm}{\begin{fmfgraph}(15,10)
\fmfcurved
\fmfleft{i1}
\fmfright{o1,o2}
\fmf{dbl_plain}{i1,v1}
\fmf{plain}{v1,o1}
\fmf{plain}{v1,o2}
\fmfblob{5mm}{o1}
\fmfblob{5mm}{o2}
\fmfv{d.s=cross,d.a=0,d.size=2.5mm}{i1}
\end{fmfgraph}}
+\,          
\parbox{18mm}{\begin{fmfgraph}(15,10)
\fmfcurved
\fmfleft{i1}
\fmfright{o1,o2}
\fmf{dbl_plain}{i1,v1}
\fmf{dbl_plain}{v1,o2}
\fmf{plain}{v1,o1}
\fmfblob{5mm}{o1}
\fmfblob{5mm}{o2}
\fmfv{d.s=cross,d.a=0,d.size=2.5mm}{i1}
\end{fmfgraph}}
\end{fmffile}
\vspace{3mm}
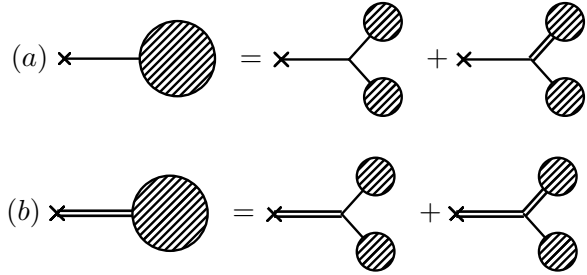
\caption{\label{fig:dipolerecCSW}Recursion for the single-line currents (a) and the double-line currents (b). A cross indicates an off-shell line.}
}
\begin{fmffile}{statistics_fg}
\mytable{h}{12cm}{11cm}{
\begin{tabular}{|c|c|}
  \hline
  \vphantom{\Large P}Vertex & Vertex factor  \\ 
  \hline
  \cswv & $\dst \epsilon^{h_K\,h_I\,h_J}(I,J,K)\frac{\abr{M_1M_2}^4}{\abr{IJ}\abr{JK}\abr{KI}}$ \\
  \cswvp & $\dst \epsilon_m^{h_I\,h_J}(U,V)\,\delta_{U}^I\delta_{V}^J$\\
 \cswpvp & $\dst \epsilon_{mm'}^h(I)\,D_{UV}^{I}\,\delta_{U}^{U'}\,\delta_{V}^I$\\
  \cswpv & $\dst \epsilon_{|2-m|}^{h_J\,h_K}(J,K)\,\frac{\abr{M_1M_2}^4}{\abr{UV}\abr{VJ}\abr{JK}\abr{JK}}$ \\
  \hline
\end{tabular}}{Vertices appearing in the recursion for the single and double-line currents.\label{tab:CSWvert}}
\end{fmffile}


\section{Recursive relations for soft factors}
\label{app:appendixc}
In this appendix we discuss how to build recusive relations for soft factors out of the recursion for antenna functions, Eq.~(\ref{eq:antform}), in a similar way as we did for the spilitting functions in Section~\ref{splitRR}. We start by giving some general considerations about the off-shell continuation that appears in the CSW formalism in the soft limit.
In general, a propagator involving more than two particles can be written as
\begin{equation}
\label{eq:pij}
P_{i,j}^2=\sum_{\substack{k,l=i\\ k<l}}^js_{kl}.
\end{equation}
In particular, if $i$ is a hard particle, say $i=a$, and all other momenta are soft, we can write
\begin{equation}
P_{a,j}^2=\sum_{l=s_1}^js_{al}+\sum_{k=s_1}^{j-1}\sum_{l=s_2}^{j}s_{kl}.
\end{equation}
Using the power counting~(\ref{eq:brij}), it is easy to see that the second term goes to zero much faster than the first one, so we have the following behavior in the soft limit
\begin{equation}
P_{a,j}^2\sim \sum_{l=s_1}^js_{al}.
\end{equation}
Similar arguments hold true for  spinor products involving off-shell momenta  in the CSW formalism,
\begin{equation}
\label{eq:CSW}
\br{k,P_{i,j}}=\langle k|P_{i,j}|\eta]=\sum_{l=i}^j\br{kl}\sq{l\eta}.
\end{equation}
Following exactly the same lines as for the propagators, we can derive the following rule for off-shell continued spinor products in the soft limit,
\begin{ruler}[Off-shell continuation in the soft limit]
In the soft limit, each spinor product of the form $\br{k,P_{a,j}}$ has to be interpreted as 
\begin{equation*}
\label{eq:softCSW}
\br{k\,P_{a,j}}\rightarrow \br{ka}\sq{a\eta},
\end{equation*}
except for $k=a$, because in this case we have trivially $\br{a,P_{a,j}}=\br{a,P_{1,j}}$, and equivalently for b.
\end{ruler}
In particular, this implies the following rules for off-shell continued momenta
\begin{eqnarray}
\br{a, P_{j,b}} & \rightarrow & \br{ab}\sq{b\eta},\nonumber\\
\br{P_{a,j},b} & \rightarrow & \br{ab}\sq{a\eta},\nonumber\\
\label{eq:CSWsoft2}
\br{P_{i,j}, P_{k,b}} & \rightarrow & \br{P_{i,j},b}\sq{b\eta},\\
\br{P_{a,i}, P_{j,k}} & \rightarrow & \br{a,P_{j,k}}\sq{a\eta},\nonumber\\
\br{P_{a,j}, P_{k,b}} & \rightarrow & \br{ab}\sq{a,\eta}\sq{b\eta}.\nonumber
\end{eqnarray}

\begin{figure}[!t]
\begin{fmffile}{SoftRR}
\begin{center}
\begin{tabular}{c}
\parbox{100mm}{
\begin{fmfgraph*}(100,10)
\fmfstraight
\fmfleft{t1}
\fmfright{t6}
\fmftop{w1,v1,v2,v3,w2,v4,v5,v6,w3,v7,v8,v9,w4,v10,v11,v12,w5,v13,v14,v15,w6}
\fmfbottom{b1,u1,u2,u3,b2,u4,u5,u6,b3,u7,u8,u9,b4,u10,u11,u12,b5,u13,u14,u15,b6}
\fmf{phantom}{t1,a1,a2,a3,t2,a4,a5,a6,t3,a7,a8,a9,t4,a10,a11,a12,t5,a13,a14,a15,t6}
\fmffreeze
\fmf{phantom,label=$\scriptstyle +\quad\quad\quad\quad -  +\quad\quad\quad\quad - + \quad\quad\quad\quad - + \quad\quad\quad\quad -\phantom{ \quad\quad+ \quad\quad -}$}{w1,w6}
\fmf{plain}{t1,t6}
\fmf{phantom}{w1,v1,v2,v3,w2,v4,v5,v6,w3,v7,v8,v9,w4,v10,v11,v12,w5,v13,v14,v15,w6}
\fmf{phantom}{b1,u1,u2,u3,b2,u4,u5,u6,b3,u7,u8,u9,b4,u10,u11,u12,b5,u13,u14,u15,b6}
\fmffreeze
\fmf{plain}{t2,u3}
\fmf{plain}{t2,u4}
\fmf{plain}{t3,u6}
\fmf{plain}{t3,u7}
\fmf{plain}{t4,u9}
\fmf{plain}{t4,u10}
\fmf{plain}{t5,u12}
\fmf{plain}{t5,u13}
\fmf{phantom}{t2,y2,x2,b2}
\fmf{phantom}{t3,y3,x3,b3}
\fmf{phantom}{t4,y4,x4,b4}
\fmf{phantom}{t5,y5,x5,b5}
\fmfdot{x2}
\fmfdot{x3}
\fmfdot{x4}
\fmfdot{x5}
\fmfdot{t1}
\fmflabel{$\scriptstyle a^+$}{t6}
\fmflabel{$(a)$}{t1}
\end{fmfgraph*}
}
\\
\parbox{100mm}{
\begin{fmfgraph*}(100,10)
\fmfstraight
\fmfleft{t1}
\fmfright{t6}
\fmftop{w1,v1,v2,v3,w2,v4,v5,v6,w3,v7,v8,v9,w4,v10,v11,v12,w5,v13,v14,v15,w6}
\fmfbottom{b1,u1,u2,u3,b2,u4,u5,u6,b3,u7,u8,u9,b4,u10,u11,u12,b5,u13,u14,u15,b6}
\fmf{phantom}{t1,a1,a2,a3,t2,a4,a5,a6,t3,a7,a8,a9,t4,a10,a11,a12,t5,a13,a14,a15,t6}
\fmffreeze
\fmf{phantom,label=$\scriptstyle +\quad\quad\quad\quad -  -\quad\quad\quad\quad + + \quad\quad\quad\quad - + \quad\quad\quad\quad -\phantom{ \quad\quad+ \quad\quad -}$}{w1,w6}
\fmf{plain}{t1,t6}
\fmf{phantom}{w1,v1,v2,v3,w2,v4,v5,v6,w3,v7,v8,v9,w4,v10,v11,v12,w5,v13,v14,v15,w6}
\fmf{phantom}{b1,u1,u2,u3,b2,u4,u5,u6,b3,u7,u8,u9,b4,u10,u11,u12,b5,u13,u14,u15,b6}
\fmffreeze
\fmf{plain}{t2,u3}
\fmf{plain}{t2,u4}
\fmf{plain}{t3,u6}
\fmf{plain}{t3,u7}
\fmf{plain}{t4,u9}
\fmf{plain}{t4,u10}
\fmf{plain}{t5,u12}
\fmf{plain}{t5,u13}
\fmf{phantom}{t2,y2,x2,b2}
\fmf{phantom}{t3,y3,x3,b3}
\fmf{phantom}{t4,y4,x4,b4}
\fmf{phantom}{t5,y5,x5,b5}
\fmfdot{x2}
\fmfdot{x3}
\fmfdot{x4}
\fmfdot{x5}
\fmfdot{t1}
\fmflabel{$\scriptstyle a^+$}{t6}
\fmflabel{$(b)$}{t1}
\end{fmfgraph*}
}
\end{tabular}
\end{center}
\end{fmffile}
\caption{\label{fig:antsoft}Diagrams contributing to $J_+(a^+,1,\ldots,k)$ in the soft limit.}
\end{figure}
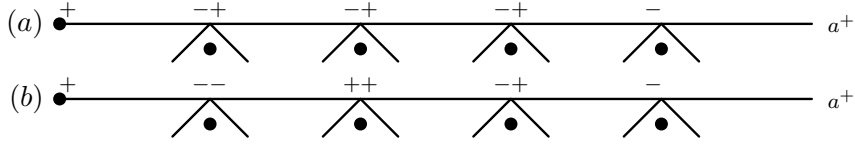

Let us turn now to the recursive relations. We know that in the limit where all the particles are soft, we have
\begin{equation}
\ant(\ha^{h_{\ha}},\hb^{h_{\hb}} \leftarrow a,1,\ldots,n,b)\longrightarrow \soft(a,1,\ldots,n,b).
\end{equation}
As the soft factor is independent of the helicities of the reference particles $a$ and $b$, some antenna functions are not divergent in this limit, but only those are divergent where $h_{\ha}=-h_a$ and $h_{\hb}=-h_b$. Taking the soft limit of Eq.~(\ref{eq:antform}), we can derive a formula for soft factors in terms of single and double-line currents. We will show this procedure explicitly for $\ant(\ha^{-},\hb^{-} \leftarrow a^+,1,\ldots,n,b^+)$.
 Applying the rules given in Section~\ref{sec:IRreview}, it is easy to see that in the soft limit
\begin{equation}
V_4(U,V,\ha,\hb;\ha,\hb)\rightarrow V_s(a,b,\eta)=\left\{\begin{array}{ll}
1 & ,\ifi U=(a,a)\hbox{ and }V=(b,b),\\
1/\sq{a\eta}^2 & ,\ifi U\neq (a,a)\hbox{ and }V=(b,b),\\
1/\sq{b\eta}^2 & ,\ifi U= (a,a)\hbox{ and }V\neq (b,b),\\
1/(\sq{a\eta}^2\sq{b\eta}^2) & ,\ifi U\neq (a,a)\hbox{ and }V\neq (n,n).
\end{array}\right.
\end{equation}
Finally, the contribution from an antenna function to a soft factor is
\begin{equation}
\label{eq:antsoft}
\sum_{\substack{U<V\\ v_2=b}}\,V_s(a,b,\eta)\,J_{UV}^{0}(a^+,1,\ldots,n,b^+).
\end{equation}
However, Eq.~(\ref{eq:antsoft}) generates not only diagrams that contribute to the soft limit, but also subleading diagrams. To see this, consider the diagrams contained in $J_+(a^+,1,\ldots,k)$ shown in Fig.~\ref{fig:antsoft}. 
Let us start with the situation where all external lines in Fig.~\ref{fig:antsoft} are on-shell (except the off-shell line $(a,k)$ indicated with a dot). Using the power counting rules~(\ref{eq:brij}), it follows that the denominators of both diagrams shown in Fig.~\ref{fig:antsoft} behave as $1/t^{2k+4k_-}$, where $k_-$ is the number of negative helicities in the set $\{1,\ldots,k\}$. We then come to the following conclusion:
\begin{itemize}
\item[-]
For diagrams of the form shown in Fig.~{\ref{fig:antsoft}.a}, the numerator behaves as $t^{4k_-}$, and so these diagrams behave as $t^{4k_-}/t^{2k+4k_-}=1/t^{2k}$, which is exactly the divergence we want.
\item[-]
For diagrams of the form shown in Fig.~{\ref{fig:antsoft}.b}, the numerator behaves as $t^{4k_-+8}$, and so these diagrams behave as $t^{4k_-+8}/t^{2k+4k_-}=1/t^{2k-8}$, and so these diagrams are not divergent enough to contribute to the soft factor.
\end{itemize}

Up to now we have only considered diagrams in Fig.~\ref{fig:antsoft} where all external lines are on-shell. Let us turn to the situation where some of these lines may themselves be off-shell, \ie, they correspond to single-line currents $J_h(i_1,\ldots,i_2)$. First, a simple power counting argument shows that if the particles $i_1,\ldots,i_2$ are soft,
\begin{equation}
J_+(i_1,\ldots,i_2)\sim 1/t^{2(i_2-i_1+1)-4},\qquad J_-(i_1,\ldots,i_2)\sim 1/t^{2(i_2-i_1+1)}.
\end{equation}
Using this result, it is easy to show that
\begin{itemize}
\item[-] Replacing a positive-helicity particle $i$ by a single-line current $J_+(i_1,\ldots,i_2)$ amounts to the replacement 
\begin{equation}
\begin{split}
\frac{1}{\br{Xi}\br{iY}} & \rightarrow  \frac{1}{\br{XI}\br{IY}}\, J_+(i_1,\ldots,i_2). \\
\sim 1/t^2\quad & \phantom{\rightarrow\quad} \sim 1/t^4\,1/t^{2(i_2-i_1+1)-4}
\end{split}
\end{equation}
So the divergence in $1/t^2$ of the soft particle $i$ gets replaced by the divergence in $1/t^{2(i_2-i_1+1)}$ of the soft particles $i_1,\ldots,i_2$.
\item[-]
Replacing a negative-helicity particle $i$ by a single-line current $J_-(i_1,\ldots,i_2)$ amounts to the replacement 
\begin{equation}
\begin{split}
\frac{\br{Zi}^4}{\br{Xi}\br{iY}} & \rightarrow  \frac{\br{ZI}^4}{\br{XI}\br{IY}}\, J_-(i_1,\ldots,i_2). \\
\sim t^2 \quad& \phantom{\rightarrow\quad} \sim t^4/t^{2(i_2-i_1+1)}
\end{split}
\end{equation}
Furthermore, there must be somewhere in the diagram a $1/t^4$ factor knocking the $t^2$ on the left-hand side down to a divergence in $1/t^2$ for the soft particle $i$. On the right-hand side, this same term now kills the $t^4$ in the numerator, giving the divergence in $1/t^{2(j_2-j_1+1)}$ for the soft particles $i_1,\ldots,i_2$.
\end{itemize}
 So we come to the conclusion that only the diagrams in $J_+(a^+,1\ldots,k)$ of the form shown in {Fig.~\ref{fig:antsoft}.a} contribute in the soft limit. These diagrams are exactly those which do not contain any  subcurrent of the form $J_-(a^+,1,\ldots,l)$. A similar result holds of course for $b$, \ie, only those diagrams contribute to the soft limit which do not contain a subcurrent of the form $J_-(l,\ldots,n,b^+)$. Finally we can write
\begin{equation}
\soft(a,1,\ldots,n,b)=\sum_{\substack{U<V\\ v_2=b}}\,V_s(a,b,\eta)\,\bar J_{UV}^{0}(a^+,1,\ldots,n,b^+),
\end{equation}
where $\bar J_{UV}^{0}(a^+,1,\ldots,n,b^+)$ denotes a double-line current which does not contain any subcurrent of the form $J_-(a^+,1,\ldots,l)$ or $J_-(l,\ldots,n,b^+)$.

Of course we could have used a different antenna function to extract the soft factors. In general, we get the result
\begin{equation}
\soft(a,1,\ldots,n,b)=\sum_{\substack{U<V\\ v_2=b}}\,V_s(a,b,\eta)\,\epsilon^{h_{a}\,h_{b}}\,\bar J_{UV}^{(2-h_{a}-h_{b})/2}(a^{h_{a}},1,\ldots,n,b^{h_{b}}),
\end{equation}
where $\bar J_{UV}^{0}(a^{h_{a}},1,\ldots,n,b^{h_{b}})$ is the double-line current which does not contain any subcurrents of the form $J_{-h_{a}}(a^{h_{a}},1,\ldots,l)$ or $J_{-h_{b}}(l,\ldots,n,b^{h_{b}})$, and 
\begin{equation}
\begin{split}
\epsilon^{++}=1, \,\,\,\quad\quad &\quad \epsilon^{+-}=\frac{\br{aM}^4}{\br{ab}^4},\\
\epsilon^{+-}=\frac{\br{Mb}^4}{\br{ab}^4}, &\quad \epsilon^{--}=\frac{\br{M_1M_2}^4}{\br{ab}^4}.
\end{split}
\end{equation}
We calculated numerically the soft factors for one, two or three soft particles, and we checked numerically that these quantities are independent of the arbitrary spinor $\eta$ as well as of the helicities of the spectator particles $a$ and $b$.

\section{Explicit results}
\label{appendix:results}
In this appendix we present our results for all independent NLO, NNLO and NNNLO antenna functions. These results are obtained from the general formulas in Section~\ref{sec:results}. We give explicit results for  antenna functions of the form $\ant(\ha^{-},\hb^{-} \leftarrow a,1,\ldots,n,b)$ and $\ant(\ha^{-},\hb^{+} \leftarrow a,1,\ldots,n,b)$. The remaining antenna functions are related to the previous ones by parity and reflection identity~(\ref{eq:antreflec}). At NNNLO we have to include also antenna functions of the form $\ant(\ha^{+},\hb^{+} \leftarrow a,1,\ldots,n,b)$.

\paragraph{NLO antenna functions}
All NLO antenna functions are just MHV-type antenna functions.
\begin{align*}
\ant(\ha^{-},\hb^{-} \leftarrow a^+,1^+,b^+)&=\frac{\br{\ha\hb}^3}{\br{a1}\br{1b}\br{b\ha}\br{\hb a}}.\\
\ant(\ha^{-},\hb^{-} \leftarrow a^-,1^+,b^+)&=-\frac{\sq{1b}^3}{\sq{a1}\sq{b\ha}\sq{\ha\hb}\sq{\hb a}}.\nonumber\\
\ant(\ha^{-},\hb^{-} \leftarrow a^+,1^-,b^+)&=-\frac{\sq{ab}^4}{\sq{a1}\sq{1b}\sq{b\ha}\sq{\ha\hb}\sq{\hb a}}.\nonumber\\
\ant(\ha^{-},\hb^{-} \leftarrow a^+,1^+,b^-)&=-\frac{\sq{a1}^3}{\sq{1b}\sq{b\ha}\sq{\ha\hb}\sq{\hb a}}.\nonumber\\
\ant(\ha^{-},\hb^{-} \leftarrow a^-,1^-,b^+)&=0.\nonumber\\
\ant(\ha^{-},\hb^{-} \leftarrow a^-,1^+,b^-)&=0.\nonumber\\
\ant(\ha^{-},\hb^{-} \leftarrow a^+,1^-,b^-)&=0.\nonumber\\
\ant(\ha^{-},\hb^{-} \leftarrow a^-,1^-,b^-)&=0.\nonumber
\end{align*}
\begin{align*}
\ant(\ha^{-},\hb^{+} \leftarrow a^+,1^+,b^+)&=0.\\
\ant(\ha^{-},\hb^{+} \leftarrow a^-,1^+,b^+)&=0.\nonumber\\
\ant(\ha^{-},\hb^{+} \leftarrow a^+,1^-,b^+)&=\frac{\br{\ha1}^4}{\br{a1}\br{1b}\br{b\ha}\br{\ha\hb}\br{\hb a}}.\nonumber\\
\ant(\ha^{-},\hb^{+} \leftarrow a^+,1^+,b^-)&=\frac{\br{b\ha}^3}{\br{a1}\br{1b}\br{\ha\hb}\br{\hb a}}.\nonumber\\
\ant(\ha^{-},\hb^{+} \leftarrow a^-,1^-,b^+)&=0.\nonumber\\
\ant(\ha^{-},\hb^{+} \leftarrow a^-,1^+,b^-)&=-\frac{\sq{\ha1}^4}{\sq{a1}\sq{1b}\sq{b\ha}\sq{\ha\hb}\sq{\hb a}}.\nonumber\\
\ant(\ha^{-},\hb^{+} \leftarrow a^+,1^-,b^-)&=-\frac{\sq{b\ha}^3}{\sq{a1}\sq{1b}\sq{\ha\hb}\sq{\hb a}}.\nonumber\\
\ant(\ha^{-},\hb^{+} \leftarrow a^-,1^-,b^-)&=0.\nonumber
\end{align*}


\paragraph{NNLO antenna functions}
All NNLO antenna functions are just MHV and NMHV-type antenna functions.
\begin{align*}
\ant(\ha^{-},\hb^{-} \leftarrow a^+,1^+,2^+,b^+)&=\frac{\br{\ha\hb}^3}{\br{a1}\br{12}\br{2b}\br{b\ha}\br{\hb a}}.\\
\ant(\ha^{-},\hb^{-} \leftarrow a^-,1^-,2^+,b^+)&=\frac{\sq{2b}^3}{\sq{a1}\sq{12}\sq{b\ha}\sq{\ha\hb}\sq{\hb a}}.\nonumber\\
\ant(\ha^{-},\hb^{-} \leftarrow a^-,1^+,2^-,b^+)&=\frac{\sq{1b}^4}{\sq{a1}\sq{12}\sq{2b}\sq{b\ha}\sq{\ha\hb}\sq{\hb a}}.\nonumber\\
\ant(\ha^{-},\hb^{-} \leftarrow a^-,1^+,2^+,b^-)&=\frac{\sq{12}^3}{\sq{a1}\sq{2b}\sq{b\ha}\sq{\ha\hb}\sq{\hb a}}.\nonumber\\
\ant(\ha^{-},\hb^{-} \leftarrow a^+,1^-,2^-,b^+)&=\frac{\sq{ab}^4}{\sq{a1}\sq{12}\sq{2b}\sq{b\ha}\sq{\ha\hb}\sq{\hb a}}.\nonumber\\
\ant(\ha^{-},\hb^{-} \leftarrow a^+,1^-,2^+,b^-)&=\frac{\sq{a2}^4}{\sq{a1}\sq{12}\sq{2b}\sq{b\ha}\sq{\ha\hb}\sq{\hb a}}.\nonumber\\
\ant(\ha^{-},\hb^{-} \leftarrow a^+,1^+,2^-,b^-)&=\frac{\sq{a1}^3}{\sq{12}\sq{2b}\sq{b\ha}\sq{\ha\hb}\sq{\hb a}}.\nonumber\\
\ant(\ha^{-},\hb^{-} \leftarrow a^-,1^-,2^-,b^+)&=0.\nonumber\\
\ant(\ha^{-},\hb^{-} \leftarrow a^-,1^-,2^+,b^-)&=0.\nonumber\\
\ant(\ha^{-},\hb^{-} \leftarrow a^-,1^+,2^-,b^-)&=0.\nonumber\\
\ant(\ha^{-},\hb^{-} \leftarrow a^+,1^-,2^-,b^-)&=0.\nonumber\\
\ant(\ha^{-},\hb^{-} \leftarrow a^-,1^-,2^-,b^-)&=0.\nonumber
\end{align*}
\begin{align*}
\ant(\ha^{-},\hb^{+} \leftarrow a^+,1^+,2^+,b^+)&=0.\\
\ant(\ha^{-},\hb^{+} \leftarrow a^-,1^+,2^+,b^+)&=0.\nonumber\\
\ant(\ha^{-},\hb^{+} \leftarrow a^+,1^-,2^+,b^+)&=\frac{\br{\ha1}^4}{\br{a1}\br{12}\br{2b}\br{b\ha}\br{\ha\hb}\br{\hb a}}.\nonumber\\
\ant(\ha^{-},\hb^{+} \leftarrow a^+,1^+,2^-,b^+)&=\frac{\br{\ha2}^4}{\br{a1}\br{12}\br{2b}\br{b\ha}\br{\ha\hb}\br{\hb a}}.\nonumber\\
\ant(\ha^{-},\hb^{+} \leftarrow a^+,1^+,2^+,b^-)&=\frac{\br{b\ha}^3}{\br{a1}\br{12}\br{2b}\br{\ha\hb}\br{\hb a}}.\nonumber\\
\ant(\ha^{-},\hb^{+} \leftarrow a^-,1^-,2^-,b^+)&=0.\nonumber\\
\ant(\ha^{-},\hb^{+} \leftarrow a^-,1^-,2^+,b^-)&=\frac{\sq{2\hb}^4}{\sq{a1}\sq{12}\sq{2b}\sq{b\ha}\sq{\ha\hb}\sq{\hb a}}.\nonumber\\
\ant(\ha^{-},\hb^{+} \leftarrow a^-,1^+,2^-,b^-)&=\frac{\sq{1\hb}^4}{\sq{a1}\sq{12}\sq{2b}\sq{b\ha}\sq{\ha\hb}\sq{\hb a}}.\nonumber\\
\ant(\ha^{-},\hb^{+} \leftarrow a^+,1^-,2^-,b^-)&=\frac{\sq{\hb a}^4}{\sq{a1}\sq{12}\sq{2b}\sq{b\ha}\sq{\ha\hb}}.\nonumber\\
\ant(\ha^{-},\hb^{+} \leftarrow a^-,1^-,2^-,b^-)&=0.\nonumber
\end{align*}
\begin{align*}
\ant(\ha^{-},&\hb^{-} \leftarrow a^-,1^+,2^+,b^+)= \frac{\langle \hat{a} \hat{b} \rangle^3}{\langle \hat{a} b \rangle \langle
   2|a+1|\eta]}\\
  & \left(-\frac{\langle
   a|1+2|\eta]^3}{s_{a12} \langle 1 2 \rangle \langle
   a 1 \rangle \langle b|a+1+2|\eta] \langle
   \hat{b}|a+1+2|\eta]}-\frac{[1 \eta]^3}{\langle 2 b
   \rangle \langle \hat{b}|a+1|\eta] [a 1] [a
   \eta]}\right)\nonumber\\
\ant(\ha^{-},&\hb^{-} \leftarrow a^+,1^-,2^+,b^+)= -\frac{\langle \hat{a} \hat{b} \rangle^3 \langle
   1|2+b|\eta]^3}{s_{b12} \langle 1 2 \rangle \langle
   2 b \rangle \langle a \hat{b} \rangle \langle
   b|1+2|\eta] \langle a|1+2+b|\eta] \langle
   \hat{a}|1+2+b|\eta]}\nonumber\\
   &-\frac{\langle \hat{a} \hat{b} \rangle^3
   \langle 1|a+2|\eta]^4}{s_{a12} \langle 1 2 \rangle
   \langle a 1 \rangle \langle \hat{a} b \rangle \langle
   2|a+1|\eta] \langle a|1+2|\eta] \langle
   b|a+1+2|\eta] \langle
   \hat{b}|a+1+2|\eta]}\nonumber\\
   &+\frac{\langle \hat{a} \hat{b} \rangle^3 [2
   \eta]^3}{\langle a \hat{b} \rangle \langle \hat{a} b \rangle
   \langle a|1+2|\eta] \langle b|1+2|\eta]
   [1 2] [1 \eta]}\nonumber\\
   &-\frac{\langle \hat{a} \hat{b} \rangle^3
   [a \eta]^3}{\langle 2 b \rangle \langle \hat{a} b
   \rangle \langle 2|a+1|\eta] \langle
   \hat{b}|a+1|\eta] [1 \eta] [a 1]}\nonumber\\
\ant(\ha^{-},&\hb^{-} \leftarrow a^+,1^+,2^-,b^+)=\ant(\hb^{-},\ha^{-} \leftarrow b^+,2^-,1^+,a^+).\nonumber\\
\ant(\ha^{-},&\hb^{-} \leftarrow a^+,1^+,2^+,b^-)=\ant(\hb^{-},\ha^{-} \leftarrow b^-,2^+,1^+,a^+).\nonumber\\
\ant(\ha^{-},&\hb^{+} \leftarrow a^-,1^-,2^+,b^+)=0.\\
\ant(\ha^{-},&\hb^{+} \leftarrow a^-,1^+,2^-,b^+)=-\frac{1}{\langle \hat{a} b \rangle
   \langle \hat{a} \hat{b} \rangle \langle 2|a+1|\eta]}\nonumber\\
   &\left(\frac{\langle a 2 \rangle^4 \langle
   \hat{a}|1+2|\eta]^4}{s_{a12} \langle 1 2 \rangle \langle
   a 1 \rangle \langle a|1+2|\eta] \langle
   b|a+1+2|\eta] \langle
   \hat{b}|a+1+2|\eta]}+\frac{\langle \hat{a} 2 \rangle^4 [1
   \eta]^3}{\langle 2 b \rangle \langle \hat{b}|a+1|\eta]
   [a 1] [a \eta]}\right)\nonumber
   \end{align*}
\begin{align*}
\ant(\ha^{-},&\hb^{+} \leftarrow a^-,1^+,2^+,b^-)= \frac{\langle \hat{a} b \rangle^3 }{\langle \hat{a} \hat{b} \rangle \langle
   2|a+1|\eta]}\nonumber\\
   &\left(-\frac{\langle
   a|1+2|\eta]^3}{s_{a12} \langle 1 2 \rangle \langle
   a 1 \rangle \langle b|a+1+2|\eta] \langle
   \hat{b}|a+1+2|\eta]}-\frac{[1 \eta]^3}{\langle 2 b
   \rangle \langle \hat{b}|a+1|\eta] [a 1] [a
   \eta]}\right)\nonumber\\
\ant(\ha^{-},&\hb^{+} \leftarrow a^+,1^-,2^-,b^+)= -\frac{\langle 1 2 \rangle^3 \langle
   \hat{a}|1+2+b|\eta]^3}{s_{b12} \langle 2 b \rangle
   \langle a \hat{b} \rangle \langle \hat{a} \hat{b} \rangle \langle
   1|2+b|\eta] \langle b|1+2|\eta] \langle
   a|1+2+b|\eta]}\nonumber\\
   &-\frac{\langle 1 2 \rangle^3 \langle
   \hat{a}|1+2|\eta]^4}{s_{a12} \langle a 1 \rangle \langle
   \hat{a} b \rangle \langle \hat{a} \hat{b} \rangle \langle
   2|a+1|\eta] \langle a|1+2|\eta] \langle
   b|a+1+2|\eta] \langle
   \hat{b}|a+1+2|\eta]}\nonumber\\
   &+\frac{\langle
   \hat{a}|1+2|\eta]^4}{\langle a \hat{b} \rangle \langle \hat{a} b
   \rangle \langle \hat{a} \hat{b} \rangle \langle a|1+2|\eta]
   \langle b|1+2|\eta] [1 2] [1 \eta] [2
   \eta]}\nonumber\\
   &-\frac{\langle \hat{a} 2 \rangle^4 [a
   \eta]^3}{\langle 2 b \rangle \langle \hat{a} b \rangle
   \langle \hat{a} \hat{b} \rangle \langle 2|a+1|\eta]
   \langle \hat{b}|a+1|\eta] [1 \eta] [a
   1]}\nonumber\\
   &-\frac{\langle \hat{a} 1 \rangle^4 [b
   \eta]^3}{\langle a 1 \rangle \langle a \hat{b} \rangle
   \langle \hat{a} \hat{b} \rangle \langle 1|2+b|\eta]
   \langle \hat{a}|2+b|\eta] [2 b] [2 \eta]}\nonumber
   \end{align*}
   \begin{align*}
\ant(\ha^{-},&\hb^{+} \leftarrow a^+,1^-,2^+,b^-)=-\frac{\langle 1 b \rangle^4 \langle
   \hat{a}|1+2+b|\eta]^3}{s_{b12} \langle 1 2 \rangle
   \langle 2 b \rangle \langle a \hat{b} \rangle \langle
   \hat{a} \hat{b} \rangle \langle 1|2+b|\eta] \langle
   b|1+2|\eta] \langle
   a|1+2+b|\eta]}\nonumber\\
   &-\frac{\langle \hat{a} b \rangle^3 \langle
   1|a+2|\eta]^4}{s_{a12} \langle 1 2 \rangle \langle
   a 1 \rangle \langle \hat{a} \hat{b} \rangle \langle
   2|a+1|\eta] \langle a|1+2|\eta] \langle
   b|a+1+2|\eta] \langle
   \hat{b}|a+1+2|\eta]}\nonumber\\
   &+\frac{\langle \hat{a} b \rangle^3 [2
   \eta]^3}{\langle a \hat{b} \rangle \langle \hat{a} \hat{b} \rangle
   \langle a|1+2|\eta] \langle b|1+2|\eta]
   [1 2] [1 \eta]}-\frac{\langle \hat{a} b \rangle^3
   [a \eta]^3}{\langle 2 b \rangle \langle \hat{a} \hat{b}
   \rangle \langle 2|a+1|\eta] \langle
   \hat{b}|a+1|\eta] [1 \eta] [a 1]}\nonumber\\
   &-\frac{\langle
   \hat{a} 1 \rangle^4 [2 \eta]^3}{\langle a 1 \rangle
   \langle a \hat{b} \rangle \langle \hat{a} \hat{b} \rangle \langle
   1|2+b|\eta] \langle \hat{a}|2+b|\eta] [2 b]
   [b \eta]}\nonumber\\
\ant(\ha^{-},&\hb^{+} \leftarrow a^+,1^+,2^-,b^-)= -\frac{\langle 2 b \rangle^3 \langle
   \hat{a}|1+2+b|\eta]^3}{s_{b12} \langle 1 2 \rangle
   \langle a \hat{b} \rangle \langle \hat{a} \hat{b} \rangle \langle
   1|2+b|\eta] \langle b|1+2|\eta] \langle
   a|1+2+b|\eta]}\nonumber\\
   &-\frac{\langle \hat{a} b \rangle^3 \langle
   2|a+1|\eta]^3}{s_{a12} \langle 1 2 \rangle \langle
   a 1 \rangle \langle \hat{a} \hat{b} \rangle \langle
   a|1+2|\eta] \langle b|a+1+2|\eta] \langle
   \hat{b}|a+1+2|\eta]}\nonumber\\
   &+\frac{\langle \hat{a} b \rangle^3 [1
   \eta]^3}{\langle a \hat{b} \rangle \langle \hat{a} \hat{b} \rangle
   \langle a|1+2|\eta] \langle b|1+2|\eta]
   [1 2] [2 \eta]}-\frac{\langle
   \hat{a}|2+b|\eta]^3}{\langle a 1 \rangle \langle a \hat{b}
   \rangle \langle \hat{a} \hat{b} \rangle \langle 1|2+b|\eta]
   [2 b] [2 \eta] [b \eta]}\nonumber
\end{align*}


\paragraph{NNNLO antenna functions}
All NNNLO antenna functions are just MHV and NMHV-type antenna functions.

\begin{align*}
\ant(\ha^{-},\hb^{-} \leftarrow a^-,1^-,2^-,3^-,b^-)&=0.\\
\ant(\ha^{-},\hb^{-} \leftarrow a^-,1^-,2^-,3^-,b^+)&=0.\nonumber\\
\ant(\ha^{-},\hb^{-} \leftarrow a^-,1^-,2^-,3^+,b^-)&=0.\nonumber\\
\ant(\ha^{-},\hb^{-} \leftarrow a^-,1^-,2^+,3^-,b^-)&=0.\nonumber\\
\ant(\ha^{-},\hb^{-} \leftarrow a^-,1^+,2^-,3^-,b^-)&=0.\nonumber\\
\ant(\ha^{-},\hb^{-} \leftarrow a^+,1^-,2^-,3^-,b^-)&=0.\nonumber
\end{align*}
\begin{align*}
\ant(\ha^{-},\hb^{-} \leftarrow a^+,1^+,2^+,3^+,b^+)&=\frac{\br{\ha\hb}^3}{\br{a1}\br{12}\br{23}\br{3b}\br{b\ha}\br{\hb a}}.\nonumber\\
\ant(\ha^{-},\hb^{-} \leftarrow a^-,1^-,2^-,3^+,b^+)&=\frac{-\sq{3b}^3}{\sq{a1}\sq{12}\sq{23}\sq{b\ha}\sq{\ha\hb}\sq{\hb a}}.\nonumber\\
\ant(\ha^{-},\hb^{-} \leftarrow a^-,1^-,2^+,3^-,b^+)&=\frac{-\sq{2b}^4}{\sq{a1}\sq{12}\sq{23}\sq{3b}\sq{b\ha}\sq{\ha\hb}\sq{\hb a}}.\nonumber\\
\ant(\ha^{-},\hb^{-} \leftarrow a^-,1^+,2^-,3^-,b^+)&=\frac{-\sq{1b}^4}{\sq{a1}\sq{12}\sq{23}\sq{3b}\sq{b\ha}\sq{\ha\hb}\sq{\hb a}}.\nonumber\\
\ant(\ha^{-},\hb^{-} \leftarrow a^+,1^-,2^-,3^-,b^+)&=\frac{-\sq{ab}^4}{\sq{a1}\sq{12}\sq{23}\sq{3b}\sq{b\ha}\sq{\ha\hb}\sq{\hb a}}.\nonumber\\
\ant(\ha^{-},\hb^{-} \leftarrow a^-,1^-,2^+,3^+,b^-)&=\frac{-\sq{23}^3}{\sq{a1}\sq{12}\sq{3b}\sq{b\ha}\sq{\ha\hb}\sq{\hb a}}.\nonumber\\
\ant(\ha^{-},\hb^{-} \leftarrow a^-,1^+,2^-,3^+,b^-)&=\frac{-\sq{13}^4}{\sq{a1}\sq{12}\sq{23}\sq{3b}\sq{b\ha}\sq{\ha\hb}\sq{\hb a}}.\nonumber
\end{align*}
\begin{align*}
\ant&(\ha^{-},\hb^{-} \leftarrow a^+,1^-,2^-,3^+,b^-)=\frac{-\sq{a3}^4}{\sq{a1}\sq{12}\sq{23}\sq{3b}\sq{b\ha}\sq{\ha\hb}\sq{\hb a}}.\nonumber\\
\ant&(\ha^{-},\hb^{-} \leftarrow a^-,1^+,2^+,3^-,b^-)=\frac{-\sq{12}^3}{\sq{a1}\sq{23}\sq{3b}\sq{b\ha}\sq{\ha\hb}\sq{\hb a}}.\nonumber\\
\ant&(\ha^{-},\hb^{-} \leftarrow a^+,1^-,2^+,3^-,b^-)=\frac{-\sq{a2}^4}{\sq{a1}\sq{12}\sq{23}\sq{3b}\sq{b\ha}\sq{\ha\hb}\sq{\hb a}}.\nonumber\\
\ant&(\ha^{-},\hb^{-} \leftarrow a^+,1^+,2^-,3^-,b^-)=\frac{-\sq{a1}^3}{\sq{12}\sq{23}\sq{3b}\sq{b\ha}\sq{\ha\hb}\sq{\hb a}}.\nonumber
\end{align*}
   \begin{align*}
   \ant&(\ha^{-},\hb^{-} \leftarrow a^+,1^+,2^-,3^+,b^+)=\nonumber\\
   &-\frac{\langle \hat{a} \hat{b} \rangle^3 \langle
   2|3+b|\eta]^3}{s_{b23} \langle 2 3 \rangle \langle
   3 b \rangle \langle a 1 \rangle \langle a \hat{b} \rangle
   \langle b|2+3|\eta] \langle 1|2+3+b|\eta]
   \langle \hat{a}|2+3+b|\eta]}\nonumber\\
   &+\frac{\langle \hat{a} \hat{b}
   \rangle^3 \langle 2|1+3|\eta]^4}{s_{123} \langle 1
   2 \rangle \langle 2 3 \rangle \langle a \hat{b} \rangle
   \langle \hat{a} b \rangle \langle 1|2+3|\eta]
   \langle 3|1+2|\eta] \langle a|1+2+3|\eta]
   \langle b|1+2+3|\eta]}\nonumber\\
   &-\frac{\langle \hat{a} \hat{b}
   \rangle^3 \langle 2|a+1|\eta]^3}{s_{a12} \langle 1
   2 \rangle \langle 3 b \rangle \langle a 1 \rangle
   \langle \hat{a} b \rangle \langle a|1+2|\eta]
   \langle 3|a+1+2|\eta] \langle
   \hat{b}|a+1+2|\eta]}\nonumber\\
   &-\frac{\langle \hat{a} \hat{b} \rangle^3
   \langle 2|1+3+b|\eta]^4}{s_{b123} \langle 1 2
   \rangle \langle 2 3 \rangle \langle 3 b \rangle
   \langle a \hat{b} \rangle \langle 1|2+3+b|\eta]
   \langle b|1+2+3|\eta] \langle a|1+2+3+b|\eta]
   \langle \hat{a}|1+2+3+b|\eta]}\nonumber\\
   &-\frac{\langle \hat{a} \hat{b}
   \rangle^3 \langle 2|a+1+3|\eta]^4}{s_{a123} \langle
   1 2 \rangle \langle 2 3 \rangle \langle a 1 \rangle
   \langle \hat{a} b \rangle \langle 3|a+1+2|\eta]
   \langle a|1+2+3|\eta] \langle b|a+1+2+3|\eta]
   \langle \hat{b}|a+1+2+3|\eta]}\nonumber\\
   &+\frac{\langle \hat{a} \hat{b}
   \rangle^3 [1 \eta]^3}{\langle 3 b \rangle \langle
   a \hat{b} \rangle \langle \hat{a} b \rangle \langle
   3|1+2|\eta] \langle a|1+2|\eta] [1 2]
   [2 \eta]}\nonumber\\
   &+\frac{\langle \hat{a} \hat{b} \rangle^3 [3
   \eta]^3}{\langle a 1 \rangle \langle a \hat{b} \rangle
   \langle \hat{a} b \rangle \langle 1|2+3|\eta]
   \langle b|2+3|\eta] [2 3] [2 \eta]}\nonumber
   \end{align*}
   \begin{align*}
\ant&(\ha^{-},\hb^{-} \leftarrow a^+,1^-,2^+,3^+,b^+)=\nonumber\\
&\frac{\langle \hat{a} \hat{b} \rangle^3 \langle
   1|2+3|\eta]^3}{s_{123} \langle 1 2 \rangle \langle
   2 3 \rangle \langle a \hat{b} \rangle \langle \hat{a} b \rangle
   \langle 3|1+2|\eta] \langle a|1+2+3|\eta]
   \langle b|1+2+3|\eta]}\nonumber\\
   &-\frac{\langle \hat{a} \hat{b}
   \rangle^3 \langle 1|a+2|\eta]^4}{s_{a12} \langle 1
   2 \rangle \langle 3 b \rangle \langle a 1 \rangle
   \langle \hat{a} b \rangle \langle 2|a+1|\eta]
   \langle a|1+2|\eta] \langle 3|a+1+2|\eta]
   \langle \hat{b}|a+1+2|\eta]}\nonumber\\
   &-\frac{\langle \hat{a} \hat{b}
   \rangle^3 \langle 1|2+3+b|\eta]^3}{s_{b123} \langle
   1 2 \rangle \langle 2 3 \rangle \langle 3 b \rangle
   \langle a \hat{b} \rangle \langle b|1+2+3|\eta]
   \langle a|1+2+3+b|\eta] \langle
   \hat{a}|1+2+3+b|\eta]}\nonumber\\
   &-\frac{\langle \hat{a} \hat{b} \rangle^3
   \langle 1|a+2+3|\eta]^4}{s_{a123} \langle 1 2
   \rangle \langle 2 3 \rangle \langle a 1 \rangle
   \langle \hat{a} b \rangle \langle 3|a+1+2|\eta]
   \langle a|1+2+3|\eta] \langle b|a+1+2+3|\eta]
   \langle \hat{b}|a+1+2+3|\eta]}\nonumber\\
   &+\frac{\langle \hat{a} \hat{b}
   \rangle^3 [2 \eta]^3}{\langle 3 b \rangle \langle
   a \hat{b} \rangle \langle \hat{a} b \rangle \langle
   3|1+2|\eta] \langle a|1+2|\eta] [1 2]
   [1 \eta]}-\frac{\langle \hat{a} \hat{b} \rangle^3 [a
   \eta]^3}{\langle 2 3 \rangle \langle 3 b \rangle
   \langle \hat{a} b \rangle \langle 2|a+1|\eta]
   \langle \hat{b}|a+1|\eta] [1 \eta] [a 1]}\nonumber
   \end{align*}
   \begin{align*}
\ant&(\ha^{-},\hb^{-} \leftarrow a^-,1^+,2^+,3^+,b^+)=\nonumber\\
 &-\frac{\langle \hat{a} \hat{b} \rangle^3 \langle
   a|1+2|\eta]^3}{s_{a12} \langle 1 2 \rangle \langle
   3 b \rangle \langle a 1 \rangle \langle \hat{a} b \rangle
   \langle 2|a+1|\eta] \langle 3|a+1+2|\eta]
   \langle \hat{b}|a+1+2|\eta]}\nonumber\\
   &-\frac{\langle \hat{a} \hat{b}
   \rangle^3 \langle a|1+2+3|\eta]^3}{s_{a123} \langle
   1 2 \rangle \langle 2 3 \rangle \langle a 1 \rangle
   \langle \hat{a} b \rangle \langle 3|a+1+2|\eta]
   \langle b|a+1+2+3|\eta] \langle
   \hat{b}|a+1+2+3|\eta]}\nonumber\\
   &-\frac{\langle \hat{a} \hat{b} \rangle^3 [1
   \eta]^3}{\langle 2 3 \rangle \langle 3 b \rangle
   \langle \hat{a} b \rangle \langle 2|a+1|\eta]
   \langle \hat{b}|a+1|\eta] [a 1] [a \eta]}
\nonumber
\end{align*}
\begin{align*}
\ant(\ha^{-},\hb^{-} \leftarrow a^+,1^+,2^+,3^-,b^+)&=-\ant(\hb^{-},\ha^{-} \leftarrow b^-,3^+,2^+,1^+,a^+)\nonumber\\
\ant(\ha^{-},\hb^{-} \leftarrow a^+,1^+,2^+,3^+,b^-)&=-\ant(\hb^{-},\ha^{-} \leftarrow b^+,3^-,2^+,1^+,a^+)\nonumber\\
\ant(\ha^{-},\hb^{+} \leftarrow a^+,1^+,2^+,3^+,b^+)&=0.\\
\ant(\ha^{-},\hb^{+} \leftarrow a^-,1^+,2^+,3^+,b^+)&=0.\nonumber\\
\ant(\ha^{-},\hb^{+} \leftarrow a^+,1^-,2^+,3^+,b^+)&=\frac{\br{\ha 1}^4}{\br{a1}\br{12}\br{23}\br{3b}\br{b\ha}\br{\ha\hb}\br{\hb a}}.\nonumber\\
\ant(\ha^{-},\hb^{+} \leftarrow a^+,1^+,2^-,3^+,b^+)&=\frac{\br{\ha 2}^4}{\br{a1}\br{12}\br{23}\br{3b}\br{b\ha}\br{\ha\hb}\br{\hb a}}.\nonumber\nonumber\\
\ant(\ha^{-},\hb^{+} \leftarrow a^+,1^+,2^+,3^-,b^+)&=\frac{\br{\ha 3}^4}{\br{a1}\br{12}\br{23}\br{3b}\br{b\ha}\br{\ha\hb}\br{\hb a}}.\nonumber\nonumber\\
\ant(\ha^{-},\hb^{+} \leftarrow a^+,1^+,2^+,3^+,b^-)&=\frac{\br{\ha b}^4}{\br{a1}\br{12}\br{23}\br{3b}\br{b\ha}\br{\ha\hb}\br{\hb a}}.\nonumber\nonumber
\end{align*}
\begin{align*}
\ant(\ha^{-},\hb^{+} \leftarrow a^-,1^-,2^-,3^-,b^+)&=0.\nonumber\\
\ant(\ha^{-},\hb^{+} \leftarrow a^-,1^-,2^-,3^+,b^-)&=\frac{-\sq{3\hb}^4}{\sq{a1}\sq{12}\sq{23}\sq{3b}\sq{b\ha}\sq{\ha\hb}\sq{\hb a}}.\nonumber\\
\ant(\ha^{-},\hb^{+} \leftarrow a^-,1^-,2^+,3^-,b^-)&=\frac{-\sq{2\hb}^4}{\sq{a1}\sq{12}\sq{23}\sq{3b}\sq{b\ha}\sq{\ha\hb}\sq{\hb a}}.\nonumber\\
\ant(\ha^{-},\hb^{+} \leftarrow a^-,1^+,2^-,3^-,b^-)&=\frac{-\sq{1\hb}^4}{\sq{a1}\sq{12}\sq{23}\sq{3b}\sq{b\ha}\sq{\ha\hb}\sq{\hb a}}.\nonumber\\
\ant(\ha^{-},\hb^{+} \leftarrow a^+,1^-,2^-,3^-,b^-)&=\frac{-\sq{3\hb}^4}{\sq{a1}\sq{12}\sq{23}\sq{3b}\sq{b\ha}\sq{\ha\hb}\sq{\hb a}}.\nonumber\\
\ant(\ha^{-},\hb^{+} \leftarrow a^-,1^-,2^-,3^-,b^-)&=0.\nonumber\\
\ant(\ha^{-},\hb^{+} \leftarrow a^-,1^-,2^+,3^+,b^+)&=0.\nonumber
\end{align*}
\begin{align*}
\ant&(\ha^{-},\hb^{+} \leftarrow a^-,1^+,2^-,3^+,b^+)=\nonumber\\
&-\frac{\langle a 2 \rangle^4 \langle
   \hat{a}|1+2|\eta]^4}{s_{a12} \langle 1 2 \rangle \langle
   3 b \rangle \langle a 1 \rangle \langle \hat{a} b \rangle
   \langle \hat{a} \hat{b} \rangle \langle 2|a+1|\eta]
   \langle a|1+2|\eta] \langle 3|a+1+2|\eta]
   \langle \hat{b}|a+1+2|\eta]}\nonumber\\
   &-\frac{\langle a 2 \rangle^4
   \langle \hat{a}|1+2+3|\eta]^4}{s_{a123} \langle 1 2
   \rangle \langle 2 3 \rangle \langle a 1 \rangle
   \langle \hat{a} b \rangle \langle \hat{a} \hat{b} \rangle \langle
   3|a+1+2|\eta] \langle a|1+2+3|\eta] \langle
   b|a+1+2+3|\eta] \langle
   \hat{b}|a+1+2+3|\eta]}\nonumber\\
   &-\frac{\langle \hat{a} 2 \rangle^4 [1
   \eta]^3}{\langle 2 3 \rangle \langle 3 b \rangle
   \langle \hat{a} b \rangle \langle \hat{a} \hat{b} \rangle \langle
   2|a+1|\eta] \langle \hat{b}|a+1|\eta] [a 1]
   [a \eta]}\nonumber
   \end{align*}
\begin{align*}
\ant&(\ha^{-},\hb^{+} \leftarrow a^-,1^+,2^+,3^-,b^+)=\nonumber\\
&-\frac{\langle \hat{a} 3 \rangle^4 \langle
   a|1+2|\eta]^3}{s_{a12} \langle 1 2 \rangle \langle
   3 b \rangle \langle a 1 \rangle \langle \hat{a} b \rangle
   \langle \hat{a} \hat{b} \rangle \langle 2|a+1|\eta]
   \langle 3|a+1+2|\eta] \langle
   \hat{b}|a+1+2|\eta]}\nonumber\\
   &-\frac{\langle a 3 \rangle^4 \langle
   \hat{a}|1+2+3|\eta]^4}{s_{a123} \langle 1 2 \rangle
   \langle 2 3 \rangle \langle a 1 \rangle \langle \hat{a}
   b \rangle \langle \hat{a} \hat{b} \rangle \langle
   3|a+1+2|\eta] \langle a|1+2+3|\eta] \langle
   b|a+1+2+3|\eta] \langle
   \hat{b}|a+1+2+3|\eta]}\nonumber\\
   &-\frac{\langle \hat{a} 3 \rangle^4 [1
   \eta]^3}{\langle 2 3 \rangle \langle 3 b \rangle
   \langle \hat{a} b \rangle \langle \hat{a} \hat{b} \rangle \langle
   2|a+1|\eta] \langle \hat{b}|a+1|\eta] [a 1]
   [a \eta]}\nonumber
   \end{align*}
   \begin{align*}
\ant&(\ha^{-},\hb^{+} \leftarrow a^-,1^+,2^+,3^+,b^-)=\nonumber\\
&-\frac{\langle \hat{a} b \rangle^3 \langle
   a|1+2|\eta]^3}{s_{a12} \langle 1 2 \rangle \langle
   3 b \rangle \langle a 1 \rangle \langle \hat{a} \hat{b} \rangle
   \langle 2|a+1|\eta] \langle 3|a+1+2|\eta]
   \langle \hat{b}|a+1+2|\eta]}\nonumber\\
   &-\frac{\langle \hat{a} b
   \rangle^3 \langle a|1+2+3|\eta]^3}{s_{a123} \langle
   1 2 \rangle \langle 2 3 \rangle \langle a 1 \rangle
   \langle \hat{a} \hat{b} \rangle \langle 3|a+1+2|\eta]
   \langle b|a+1+2+3|\eta] \langle
   \hat{b}|a+1+2+3|\eta]}\nonumber\\
   &-\frac{\langle \hat{a} b \rangle^3 [1
   \eta]^3}{\langle 2 3 \rangle \langle 3 b \rangle
   \langle \hat{a} \hat{b} \rangle \langle 2|a+1|\eta]
   \langle \hat{b}|a+1|\eta] [a 1] [a \eta]}\nonumber
   \end{align*}
   \begin{align*}
\ant&(\ha^{-},\hb^{+} \leftarrow a^+,1^-,2^+,3^+,b^-)=\nonumber\\
&-\frac{\langle \hat{a} 1 \rangle^4 \langle
   b|2+3|\eta]^3}{s_{b23} \langle 2 3 \rangle \langle
   3 b \rangle \langle a 1 \rangle \langle a \hat{b} \rangle
   \langle \hat{a} \hat{b} \rangle \langle 2|3+b|\eta]
   \langle 1|2+3+b|\eta] \langle
   \hat{a}|2+3+b|\eta]}\nonumber\\
   &+\frac{\langle \hat{a} b \rangle^3
   \langle 1|2+3|\eta]^3}{s_{123} \langle 1 2 \rangle
   \langle 2 3 \rangle \langle a \hat{b} \rangle \langle
   \hat{a} \hat{b} \rangle \langle 3|1+2|\eta] \langle
   a|1+2+3|\eta] \langle
   b|1+2+3|\eta]}\nonumber\\
   &-\frac{\langle \hat{a} b \rangle^3 \langle
   1|a+2|\eta]^4}{s_{a12} \langle 1 2 \rangle \langle
   3 b \rangle \langle a 1 \rangle \langle \hat{a} \hat{b} \rangle
   \langle 2|a+1|\eta] \langle a|1+2|\eta]
   \langle 3|a+1+2|\eta] \langle
   \hat{b}|a+1+2|\eta]}\nonumber\\
   &-\frac{\langle 1 b \rangle^4 \langle
   \hat{a}|1+2+3+b|\eta]^3}{s_{b123} \langle 1 2 \rangle
   \langle 2 3 \rangle \langle 3 b \rangle \langle a
   \hat{b} \rangle \langle \hat{a} \hat{b} \rangle \langle
   1|2+3+b|\eta] \langle b|1+2+3|\eta] \langle
   a|1+2+3+b|\eta]}\nonumber\\
   &-\frac{\langle \hat{a} b \rangle^3
   \langle 1|a+2+3|\eta]^4}{s_{a123} \langle 1 2
   \rangle \langle 2 3 \rangle \langle a 1 \rangle
   \langle \hat{a} \hat{b} \rangle \langle 3|a+1+2|\eta]
   \langle a|1+2+3|\eta] \langle b|a+1+2+3|\eta]
   \langle \hat{b}|a+1+2+3|\eta]}\nonumber\\
   &+\frac{\langle \hat{a} b
   \rangle^3 [2 \eta]^3}{\langle 3 b \rangle \langle
   a \hat{b} \rangle \langle \hat{a} \hat{b} \rangle \langle
   3|1+2|\eta] \langle a|1+2|\eta] [1 2]
   [1 \eta]}\nonumber\\
   &-\frac{\langle \hat{a} b \rangle^3 [a
   \eta]^3}{\langle 2 3 \rangle \langle 3 b \rangle
   \langle \hat{a} \hat{b} \rangle \langle 2|a+1|\eta]
   \langle \hat{b}|a+1|\eta] [1 \eta] [a
   1]}-\frac{\langle \hat{a} 1 \rangle^4 [3
   \eta]^3}{\langle 1 2 \rangle \langle a 1 \rangle
   \langle a \hat{b} \rangle \langle \hat{a} \hat{b} \rangle \langle
   2|3+b|\eta] \langle \hat{a}|3+b|\eta] [3 b]
   [b \eta]}\nonumber
\end{align*}
   \begin{align*}
\ant&(\ha^{-},\hb^{+} \leftarrow a^+,1^-,2^+,3^-,b^+)=\nonumber\\
&-\frac{\langle \hat{a} 1 \rangle^4 \langle
   3|2+b|\eta]^4}{s_{b23} \langle 2 3 \rangle \langle
   3 b \rangle \langle a 1 \rangle \langle a \hat{b} \rangle
   \langle \hat{a} \hat{b} \rangle \langle 2|3+b|\eta]
   \langle b|2+3|\eta] \langle 1|2+3+b|\eta]
   \langle \hat{a}|2+3+b|\eta]}\nonumber\\
   &+\frac{\langle 1 3 \rangle^4
   \langle \hat{a}|1+2+3|\eta]^4}{s_{123} \langle 1 2
   \rangle \langle 2 3 \rangle \langle a \hat{b} \rangle
   \langle \hat{a} b \rangle \langle \hat{a} \hat{b} \rangle \langle
   1|2+3|\eta] \langle 3|1+2|\eta] \langle
   a|1+2+3|\eta] \langle
   b|1+2+3|\eta]}\nonumber\\
   &-\frac{\langle \hat{a} 3 \rangle^4 \langle
   1|a+2|\eta]^4}{s_{a12} \langle 1 2 \rangle \langle
   3 b \rangle \langle a 1 \rangle \langle \hat{a} b \rangle
   \langle \hat{a} \hat{b} \rangle \langle 2|a+1|\eta]
   \langle a|1+2|\eta] \langle 3|a+1+2|\eta]
   \langle \hat{b}|a+1+2|\eta]}\nonumber\\
   &-\frac{\langle 1 3 \rangle^4
   \langle \hat{a}|1+2+3+b|\eta]^3}{s_{b123} \langle 1 2
   \rangle \langle 2 3 \rangle \langle 3 b \rangle
   \langle a \hat{b} \rangle \langle \hat{a} \hat{b} \rangle \langle
   1|2+3+b|\eta] \langle b|1+2+3|\eta] \langle
   a|1+2+3+b|\eta]}\nonumber\\
   &-\frac{\langle 1 3 \rangle^4
   \langle \hat{a}|1+2+3|\eta]^4}{s_{a123} \langle 1 2
   \rangle \langle 2 3 \rangle \langle a 1 \rangle
   \langle \hat{a} b \rangle \langle \hat{a} \hat{b} \rangle \langle
   3|a+1+2|\eta] \langle a|1+2+3|\eta] \langle
   b|a+1+2+3|\eta] \langle
   \hat{b}|a+1+2+3|\eta]}\nonumber\\
   &+\frac{\langle \hat{a} 3 \rangle^4 [2
   \eta]^3}{\langle 3 b \rangle \langle a \hat{b} \rangle
   \langle \hat{a} b \rangle \langle \hat{a} \hat{b} \rangle \langle
   3|1+2|\eta] \langle a|1+2|\eta] [1 2]
   [1 \eta]}\nonumber\\
   &+\frac{\langle \hat{a} 1 \rangle^4 [2
   \eta]^3}{\langle a 1 \rangle \langle a \hat{b} \rangle
   \langle \hat{a} b \rangle \langle \hat{a} \hat{b} \rangle \langle
   1|2+3|\eta] \langle b|2+3|\eta] [2 3]
   [3 \eta]}\nonumber\\
   &-\frac{\langle \hat{a} 3 \rangle^4 [a
   \eta]^3}{\langle 2 3 \rangle \langle 3 b \rangle
   \langle \hat{a} b \rangle \langle \hat{a} \hat{b} \rangle \langle
   2|a+1|\eta] \langle \hat{b}|a+1|\eta] [1 \eta]
   [a 1]}\nonumber\\
   &-\frac{\langle \hat{a} 1 \rangle^4 [b
   \eta]^3}{\langle 1 2 \rangle \langle a 1 \rangle
   \langle a \hat{b} \rangle \langle \hat{a} \hat{b} \rangle \langle
   2|3+b|\eta] \langle \hat{a}|3+b|\eta] [3 b]
   [3 \eta]}\nonumber
\end{align*}
   \begin{align*}
\ant&(\ha^{-},\hb^{+} \leftarrow a^+,1^-,2^-,3^+,b^+)=\nonumber\\
&-\frac{\langle \hat{a} 1 \rangle^4 \langle
   2|3+b|\eta]^3}{s_{b23} \langle 2 3 \rangle \langle
   3 b \rangle \langle a 1 \rangle \langle a \hat{b} \rangle
   \langle \hat{a} \hat{b} \rangle \langle b|2+3|\eta]
   \langle 1|2+3+b|\eta] \langle
   \hat{a}|2+3+b|\eta]}\nonumber\\
   &+\frac{\langle 1 2 \rangle^3 \langle
   \hat{a}|1+2+3|\eta]^4}{s_{123} \langle 2 3 \rangle
   \langle a \hat{b} \rangle \langle \hat{a} b \rangle \langle
   \hat{a} \hat{b} \rangle \langle 1|2+3|\eta] \langle
   3|1+2|\eta] \langle a|1+2+3|\eta] \langle
   b|1+2+3|\eta]}\nonumber\\
   &-\frac{\langle 1 2 \rangle^3 \langle
   \hat{a}|1+2|\eta]^4}{s_{a12} \langle 3 b \rangle \langle
   a 1 \rangle \langle \hat{a} b \rangle \langle \hat{a} \hat{b} \rangle
   \langle 2|a+1|\eta] \langle a|1+2|\eta]
   \langle 3|a+1+2|\eta] \langle
   \hat{b}|a+1+2|\eta]}\nonumber\\
   &-\frac{\langle 1 2 \rangle^3 \langle
   \hat{a}|1+2+3+b|\eta]^3}{s_{b123} \langle 2 3 \rangle
   \langle 3 b \rangle \langle a \hat{b} \rangle \langle
   \hat{a} \hat{b} \rangle \langle 1|2+3+b|\eta] \langle
   b|1+2+3|\eta] \langle
   a|1+2+3+b|\eta]}\nonumber\\
   &-\frac{\langle 1 2 \rangle^3
   \langle \hat{a}|1+2+3|\eta]^4}{s_{a123} \langle 2 3
   \rangle \langle a 1 \rangle \langle \hat{a} b \rangle
   \langle \hat{a} \hat{b} \rangle \langle 3|a+1+2|\eta]
   \langle a|1+2+3|\eta] \langle b|a+1+2+3|\eta]
   \langle \hat{b}|a+1+2+3|\eta]}\nonumber\\
   &+\frac{\langle
   \hat{a}|1+2|\eta]^4}{\langle 3 b \rangle \langle a \hat{b}
   \rangle \langle \hat{a} b \rangle \langle \hat{a} \hat{b} \rangle
   \langle 3|1+2|\eta] \langle a|1+2|\eta]
   [1 2] [1 \eta] [2 \eta]}\nonumber\\
   &+\frac{\langle \hat{a}
   1 \rangle^4 [3 \eta]^3}{\langle a 1 \rangle
   \langle a \hat{b} \rangle \langle \hat{a} b \rangle \langle
   \hat{a} \hat{b} \rangle \langle 1|2+3|\eta] \langle
   b|2+3|\eta] [2 3] [2 \eta]}\nonumber\\
   &-\frac{\langle
   \hat{a} 2 \rangle^4 [a \eta]^3}{\langle 2 3 \rangle
   \langle 3 b \rangle \langle \hat{a} b \rangle \langle
   \hat{a} \hat{b} \rangle \langle 2|a+1|\eta] \langle
   \hat{b}|a+1|\eta] [1 \eta] [a 1]}\nonumber
   \end{align*}
   \begin{align*}
   \ant&(\ha^{-},\hb^{+} \leftarrow a^+,1^+,2^-,3^-,b^+)=\nonumber\\
   &-\frac{\langle 2 3 \rangle^3 \langle
   \hat{a}|2+3+b|\eta]^3}{s_{b23} \langle 3 b \rangle
   \langle a 1 \rangle \langle a \hat{b} \rangle \langle
   \hat{a} \hat{b} \rangle \langle 2|3+b|\eta] \langle
   b|2+3|\eta] \langle
   1|2+3+b|\eta]}\nonumber\\
   &+\frac{\langle 2 3 \rangle^3 \langle
   \hat{a}|1+2+3|\eta]^4}{s_{123} \langle 1 2 \rangle
   \langle a \hat{b} \rangle \langle \hat{a} b \rangle \langle
   \hat{a} \hat{b} \rangle \langle 1|2+3|\eta] \langle
   3|1+2|\eta] \langle a|1+2+3|\eta] \langle
   b|1+2+3|\eta]}\nonumber\\
   &-\frac{\langle \hat{a} 3 \rangle^4 \langle
   2|a+1|\eta]^3}{s_{a12} \langle 1 2 \rangle \langle
   3 b \rangle \langle a 1 \rangle \langle \hat{a} b \rangle
   \langle \hat{a} \hat{b} \rangle \langle a|1+2|\eta]
   \langle 3|a+1+2|\eta] \langle
   \hat{b}|a+1+2|\eta]}\nonumber\\
   &-\frac{\langle 2 3 \rangle^3 \langle
   \hat{a}|1+2+3+b|\eta]^3}{s_{b123} \langle 1 2 \rangle
   \langle 3 b \rangle \langle a \hat{b} \rangle \langle
   \hat{a} \hat{b} \rangle \langle 1|2+3+b|\eta] \langle
   b|1+2+3|\eta] \langle
   a|1+2+3+b|\eta]}\nonumber\\
   &-\frac{\langle 2 3 \rangle^3
   \langle \hat{a}|1+2+3|\eta]^4}{s_{a123} \langle 1 2
   \rangle \langle a 1 \rangle \langle \hat{a} b \rangle
   \langle \hat{a} \hat{b} \rangle \langle 3|a+1+2|\eta]
   \langle a|1+2+3|\eta] \langle b|a+1+2+3|\eta]
   \langle \hat{b}|a+1+2+3|\eta]}\nonumber\\
   &+\frac{\langle \hat{a} 3
   \rangle^4 [1 \eta]^3}{\langle 3 b \rangle \langle
   a \hat{b} \rangle \langle \hat{a} b \rangle \langle \hat{a} \hat{b} \rangle
   \langle 3|1+2|\eta] \langle a|1+2|\eta]
   [1 2] [2 \eta]}\nonumber\\
   &+\frac{\langle
   \hat{a}|2+3|\eta]^4}{\langle a 1 \rangle \langle a \hat{b}
   \rangle \langle \hat{a} b \rangle \langle \hat{a} \hat{b} \rangle
   \langle 1|2+3|\eta] \langle b|2+3|\eta]
   [2 3] [2 \eta] [3 \eta]}\nonumber\\
   &-\frac{\langle \hat{a}
   2 \rangle^4 [b \eta]^3}{\langle 1 2 \rangle
   \langle a 1 \rangle \langle a \hat{b} \rangle \langle
   \hat{a} \hat{b} \rangle \langle 2|3+b|\eta] \langle
   \hat{a}|3+b|\eta] [3 b] [3 \eta]}\nonumber
   \end{align*}
   \begin{align*}
   \ant&(\ha^{-},\hb^{+} \leftarrow a^+,1^+,2^-,3^+,b^-)=\nonumber\\
   &-\frac{\langle 2 b \rangle^4 \langle
   \hat{a}|2+3+b|\eta]^3}{s_{b23} \langle 2 3 \rangle
   \langle 3 b \rangle \langle a 1 \rangle \langle a
   \hat{b} \rangle \langle \hat{a} \hat{b} \rangle \langle
   2|3+b|\eta] \langle b|2+3|\eta] \langle
   1|2+3+b|\eta]}\nonumber\\
   &+\frac{\langle \hat{a} b \rangle^3 \langle
   2|1+3|\eta]^4}{s_{123} \langle 1 2 \rangle \langle
   2 3 \rangle \langle a \hat{b} \rangle \langle \hat{a} \hat{b} \rangle
   \langle 1|2+3|\eta] \langle 3|1+2|\eta]
   \langle a|1+2+3|\eta] \langle
   b|1+2+3|\eta]}\nonumber\\
   &-\frac{\langle \hat{a} b \rangle^3 \langle
   2|a+1|\eta]^3}{s_{a12} \langle 1 2 \rangle \langle
   3 b \rangle \langle a 1 \rangle \langle \hat{a} \hat{b} \rangle
   \langle a|1+2|\eta] \langle 3|a+1+2|\eta]
   \langle \hat{b}|a+1+2|\eta]}\nonumber\\
   &-\frac{\langle 2 b \rangle^4
   \langle \hat{a}|1+2+3+b|\eta]^3}{s_{b123} \langle 1 2
   \rangle \langle 2 3 \rangle \langle 3 b \rangle
   \langle a \hat{b} \rangle \langle \hat{a} \hat{b} \rangle \langle
   1|2+3+b|\eta] \langle b|1+2+3|\eta] \langle
   a|1+2+3+b|\eta]}\nonumber\\
   &-\frac{\langle \hat{a} b \rangle^3
   \langle 2|a+1+3|\eta]^4}{s_{a123} \langle 1 2
   \rangle \langle 2 3 \rangle \langle a 1 \rangle
   \langle \hat{a} \hat{b} \rangle \langle 3|a+1+2|\eta]
   \langle a|1+2+3|\eta] \langle b|a+1+2+3|\eta]
   \langle \hat{b}|a+1+2+3|\eta]}\nonumber\\
   &+\frac{\langle \hat{a} b
   \rangle^3 [1 \eta]^3}{\langle 3 b \rangle \langle
   a \hat{b} \rangle \langle \hat{a} \hat{b} \rangle \langle
   3|1+2|\eta] \langle a|1+2|\eta] [1 2]
   [2 \eta]}\nonumber\\
   &+\frac{\langle \hat{a} b \rangle^3 [3
   \eta]^3}{\langle a 1 \rangle \langle a \hat{b} \rangle
   \langle \hat{a} \hat{b} \rangle \langle 1|2+3|\eta]
   \langle b|2+3|\eta] [2 3] [2
   \eta]}\nonumber\\
   &-\frac{\langle \hat{a} 2 \rangle^4 [3
   \eta]^3}{\langle 1 2 \rangle \langle a 1 \rangle
   \langle a \hat{b} \rangle \langle \hat{a} \hat{b} \rangle \langle
   2|3+b|\eta] \langle \hat{a}|3+b|\eta] [3 b]
   [b \eta]}\nonumber
   \end{align*}
\begin{align*}
   \ant&(\ha^{-},\hb^{+} \leftarrow a^+,1^+,2^+,3^-,b^-)=\nonumber\\
&-\frac{\langle 3 b \rangle^3 \langle
   \hat{a}|2+3+b|\eta]^3}{s_{b23} \langle 2 3 \rangle
   \langle a 1 \rangle \langle a \hat{b} \rangle \langle
   \hat{a} \hat{b} \rangle \langle 2|3+b|\eta] \langle
   b|2+3|\eta] \langle
   1|2+3+b|\eta]}\nonumber\\
   &+\frac{\langle \hat{a} b \rangle^3 \langle
   3|1+2|\eta]^3}{s_{123} \langle 1 2 \rangle \langle
   2 3 \rangle \langle a \hat{b} \rangle \langle \hat{a} \hat{b} \rangle
   \langle 1|2+3|\eta] \langle a|1+2+3|\eta]
   \langle b|1+2+3|\eta]}\nonumber\\
   &-\frac{\langle 3 b \rangle^3
   \langle \hat{a}|1+2+3+b|\eta]^3}{s_{b123} \langle 1 2
   \rangle \langle 2 3 \rangle \langle a \hat{b} \rangle
   \langle \hat{a} \hat{b} \rangle \langle 1|2+3+b|\eta]
   \langle b|1+2+3|\eta] \langle
   a|1+2+3+b|\eta]}\nonumber\\
   &-\frac{\langle \hat{a} b \rangle^3
   \langle 3|a+1+2|\eta]^3}{s_{a123} \langle 1 2
   \rangle \langle 2 3 \rangle \langle a 1 \rangle
   \langle \hat{a} \hat{b} \rangle \langle a|1+2+3|\eta]
   \langle b|a+1+2+3|\eta] \langle
   \hat{b}|a+1+2+3|\eta]}\nonumber\\
   &+\frac{\langle \hat{a} b \rangle^3 [2
   \eta]^3}{\langle a 1 \rangle \langle a \hat{b} \rangle
   \langle \hat{a} \hat{b} \rangle \langle 1|2+3|\eta]
   \langle b|2+3|\eta] [2 3] [3
   \eta]}\nonumber\\
   &-\frac{\langle \hat{a}|3+b|\eta]^3}{\langle 1 2
   \rangle \langle a 1 \rangle \langle a \hat{b} \rangle
   \langle \hat{a} \hat{b} \rangle \langle 2|3+b|\eta] [3
   b] [3 \eta] [b \eta]}\nonumber
      \end{align*}
\begin{align*}
\ant(\ha^{+},\hb^{+} \leftarrow a^+,1^+,2^+,3^+,b^+)&=0.\\
\ant(\ha^{+},\hb^{+} \leftarrow a^-,1^+,2^+,3^+,b^+)&=0.\nonumber\\
\ant(\ha^{+},\hb^{+} \leftarrow a^+,1^-,2^+,3^+,b^+)&=0.\nonumber\\
\ant(\ha^{+},\hb^{+} \leftarrow a^+,1^+,2^-,3^+,b^+)&=0.\nonumber\\
\ant(\ha^{+},\hb^{+} \leftarrow a^+,1^+,2^+,3^-,b^+)&=0.\nonumber\\
\ant(\ha^{+},\hb^{+} \leftarrow a^+,1^+,2^+,3^+,b^-)&=0.\nonumber
   \end{align*}
\begin{align*}
\ant(\ha^{+},\hb^{+} \leftarrow a^-,1^-,2^+,3^+,b^+)&=\frac{\br{a1}^3}{\br{12}\br{23}\br{3b}\br{b\ha}\br{\ha\hb}\br{\hb a}}.\nonumber\\
\ant(\ha^{+},\hb^{+} \leftarrow a^-,1^+,2^-,3^+,b^+)&=\frac{\br{a2}^4}{\br{a1}\br{12}\br{23}\br{3b}\br{b\ha}\br{\ha\hb}\br{\hb a}}.\nonumber\\
\ant(\ha^{+},\hb^{+} \leftarrow a^-,1^+,2^+,3^-,b^+)&=\frac{\br{a3}^4}{\br{a1}\br{12}\br{23}\br{3b}\br{b\ha}\br{\ha\hb}\br{\hb a}}.\nonumber\\
\ant(\ha^{+},\hb^{+} \leftarrow a^-,1^+,2^+,3^+,b^-)&=\frac{\br{ab}^4}{\br{a1}\br{12}\br{23}\br{3b}\br{b\ha}\br{\ha\hb}\br{\hb a}}.\nonumber\\
\ant(\ha^{+},\hb^{+} \leftarrow a^+,1^-,2^-,3^+,b^+)&=\frac{\br{12}^3}{\br{a1}\br{23}\br{3b}\br{b\ha}\br{\ha\hb}\br{\hb a}}.\nonumber\\
\ant(\ha^{+},\hb^{+} \leftarrow a^+,1^-,2^+,3^-,b^+)&=\frac{\br{13}^4}{\br{a1}\br{12}\br{23}\br{3b}\br{b\ha}\br{\ha\hb}\br{\hb a}}.\nonumber\\
\ant(\ha^{+},\hb^{+} \leftarrow a^+,1^-,2^+,3^+,b^-)&=\frac{\br{1b}^4}{\br{a1}\br{12}\br{23}\br{3b}\br{b\ha}\br{\ha\hb}\br{\hb a}}.\nonumber\\
\ant(\ha^{+},\hb^{+} \leftarrow a^+,1^+,2^-,3^-,b^+)&=\frac{\br{23}^3}{\br{a1}\br{12}\br{3b}\br{b\ha}\br{\ha\hb}\br{\hb a}}.\nonumber\\
\ant(\ha^{+},\hb^{+} \leftarrow a^+,1^+,2^-,3^+,b^-)&=\frac{\br{2b}^4}{\br{a1}\br{12}\br{23}\br{3b}\br{b\ha}\br{\ha\hb}\br{\hb a}}.\nonumber
\end{align*}
\begin{align*}
\ant(\ha^{+},\hb^{+} \leftarrow a^+,1^+,2^+,3^-,b^-)&=\frac{\br{3b}^4}{\br{a1}\br{12}\br{23}\br{3b}\br{b\ha}\br{\ha\hb}\br{\hb a}}.\nonumber\\
\ant(\ha^{+},\hb^{+} \leftarrow a^-,1^-,2^-,3^-,b^-)&=\frac{-\sq{\ha\hb}^3}{\sq{a1}\sq{12}\sq{23}\sq{3b}\sq{b\ha}\sq{\hb a}}.\nonumber
\end{align*}
\begin{align*}
\ant&(\ha^{+},\hb^{+} \leftarrow a^-,1^-,2^-,3^+,b^+)=\nonumber\\
&-\frac{\langle a 1 \rangle^3 \langle
   2|3+b|\eta]^3}{s_{b23} \langle 2 3 \rangle \langle
   3 b \rangle \langle a \hat{b} \rangle \langle \hat{a} \hat{b} \rangle
   \langle b|2+3|\eta] \langle 1|2+3+b|\eta]
   \langle \hat{a}|2+3+b|\eta]}\nonumber\\
   &+\frac{\langle 1 2 \rangle^3
   \langle a|1+2+3|\eta]^3}{s_{123} \langle 2 3
   \rangle \langle a \hat{b} \rangle \langle \hat{a} b \rangle
   \langle \hat{a} \hat{b} \rangle \langle 1|2+3|\eta]
   \langle 3|1+2|\eta] \langle
   b|1+2+3|\eta]}\nonumber\\
   &-\frac{\langle 1 2 \rangle^3 \langle
   a|1+2+3+b|\eta]^3}{s_{b123} \langle 2 3 \rangle
   \langle 3 b \rangle \langle a \hat{b} \rangle \langle
   \hat{a} \hat{b} \rangle \langle 1|2+3+b|\eta] \langle
   b|1+2+3|\eta] \langle
   \hat{a}|1+2+3+b|\eta]}\nonumber\\
   &+\frac{\langle
   a|1+2|\eta]^3}{\langle 3 b \rangle \langle a \hat{b}
   \rangle \langle \hat{a} b \rangle \langle \hat{a} \hat{b} \rangle
   \langle 3|1+2|\eta] [1 2] [1 \eta] [2
   \eta]}\nonumber\\
   &+\frac{\langle a 1 \rangle^3 [3
   \eta]^3}{\langle a \hat{b} \rangle \langle \hat{a} b \rangle
   \langle \hat{a} \hat{b} \rangle \langle 1|2+3|\eta]
   \langle b|2+3|\eta] [2 3] [2
   \eta]}\nonumber\\
   &-\frac{\langle 2|a+1|\eta]^3}{\langle 2 3
   \rangle \langle 3 b \rangle \langle \hat{a} b \rangle
   \langle \hat{a} \hat{b} \rangle \langle \hat{b}|a+1|\eta] [1
   \eta] [a 1] [a \eta]}\nonumber
\end{align*}
\begin{align*}
\ant&(\ha^{+},\hb^{+} \leftarrow a^-,1^-,2^+,3^-,b^+)=\nonumber\\
&-\frac{\langle a 1 \rangle^3 \langle
   3|2+b|\eta]^4}{s_{b23} \langle 2 3 \rangle \langle
   3 b \rangle \langle a \hat{b} \rangle \langle \hat{a} \hat{b} \rangle
   \langle 2|3+b|\eta] \langle b|2+3|\eta]
   \langle 1|2+3+b|\eta] \langle
   \hat{a}|2+3+b|\eta]}\nonumber\\
   &+\frac{\langle 1 3 \rangle^4 \langle
   a|1+2+3|\eta]^3}{s_{123} \langle 1 2 \rangle
   \langle 2 3 \rangle \langle a \hat{b} \rangle \langle
   \hat{a} b \rangle \langle \hat{a} \hat{b} \rangle \langle
   1|2+3|\eta] \langle 3|1+2|\eta] \langle
   b|1+2+3|\eta]}\nonumber\\
   &-\frac{\langle a 1 \rangle^3 \langle
   3|a+1+2|\eta]^3}{s_{a12} \langle 1 2 \rangle
   \langle 3 b \rangle \langle \hat{a} b \rangle \langle
   \hat{a} \hat{b} \rangle \langle 2|a+1|\eta] \langle
   a|1+2|\eta] \langle
   \hat{b}|a+1+2|\eta]}\nonumber\\
   &-\frac{\langle 1 3 \rangle^4 \langle
   a|1+2+3+b|\eta]^3}{s_{b123} \langle 1 2 \rangle
   \langle 2 3 \rangle \langle 3 b \rangle \langle a
   \hat{b} \rangle \langle \hat{a} \hat{b} \rangle \langle
   1|2+3+b|\eta] \langle b|1+2+3|\eta] \langle
   \hat{a}|1+2+3+b|\eta]}\nonumber\\
   &+\frac{\langle a 3 \rangle^4 [2
   \eta]^3}{\langle 3 b \rangle \langle a \hat{b} \rangle
   \langle \hat{a} b \rangle \langle \hat{a} \hat{b} \rangle \langle
   3|1+2|\eta] \langle a|1+2|\eta] [1 2]
   [1 \eta]}\nonumber\\
   &+\frac{\langle a 1 \rangle^3 [2
   \eta]^3}{\langle a \hat{b} \rangle \langle \hat{a} b \rangle
   \langle \hat{a} \hat{b} \rangle \langle 1|2+3|\eta]
   \langle b|2+3|\eta] [2 3] [3
   \eta]}\nonumber\\
   &-\frac{\langle 3|a+1|\eta]^4}{\langle 2 3
   \rangle \langle 3 b \rangle \langle \hat{a} b \rangle
   \langle \hat{a} \hat{b} \rangle \langle 2|a+1|\eta]
   \langle \hat{b}|a+1|\eta] [1 \eta] [a 1]
   [a \eta]}\nonumber\\
   &-\frac{\langle a 1 \rangle^3 [b
   \eta]^3}{\langle 1 2 \rangle \langle a \hat{b} \rangle
   \langle \hat{a} \hat{b} \rangle \langle 2|3+b|\eta]
   \langle \hat{a}|3+b|\eta] [3 b] [3 \eta]}\nonumber
\end{align*}
\begin{align*}
\ant&(\ha^{+},\hb^{+} \leftarrow a^-,1^-,2^+,3^+,b^-)=\nonumber\\
&-\frac{\langle a 1 \rangle^3 \langle
   b|2+3|\eta]^3}{s_{b23} \langle 2 3 \rangle \langle
   3 b \rangle \langle a \hat{b} \rangle \langle \hat{a} \hat{b} \rangle
   \langle 2|3+b|\eta] \langle 1|2+3+b|\eta]
   \langle \hat{a}|2+3+b|\eta]}\nonumber\\
   &+\frac{\langle a b \rangle^4
   \langle 1|2+3|\eta]^3}{s_{123} \langle 1 2 \rangle
   \langle 2 3 \rangle \langle a \hat{b} \rangle \langle
   \hat{a} b \rangle \langle \hat{a} \hat{b} \rangle \langle
   3|1+2|\eta] \langle a|1+2+3|\eta] \langle
   b|1+2+3|\eta]}\nonumber\\
   &-\frac{\langle a 1 \rangle^3 \langle
   b|a+1+2|\eta]^4}{s_{a12} \langle 1 2 \rangle
   \langle 3 b \rangle \langle \hat{a} b \rangle \langle
   \hat{a} \hat{b} \rangle \langle 2|a+1|\eta] \langle
   a|1+2|\eta] \langle 3|a+1+2|\eta] \langle
   \hat{b}|a+1+2|\eta]}\nonumber\\
   &-\frac{\langle 1 b \rangle^4 \langle
   a|1+2+3+b|\eta]^3}{s_{b123} \langle 1 2 \rangle
   \langle 2 3 \rangle \langle 3 b \rangle \langle a
   \hat{b} \rangle \langle \hat{a} \hat{b} \rangle \langle
   1|2+3+b|\eta] \langle b|1+2+3|\eta] \langle
   \hat{a}|1+2+3+b|\eta]}\nonumber\\
   &-\frac{\langle a 1 \rangle^3
   \langle b|a+1+2+3|\eta]^3}{s_{a123} \langle 1 2
   \rangle \langle 2 3 \rangle \langle \hat{a} b \rangle
   \langle \hat{a} \hat{b} \rangle \langle 3|a+1+2|\eta]
   \langle a|1+2+3|\eta] \langle
   \hat{b}|a+1+2+3|\eta]}\nonumber\\
   &+\frac{\langle a b \rangle^4 [2
   \eta]^3}{\langle 3 b \rangle \langle a \hat{b} \rangle
   \langle \hat{a} b \rangle \langle \hat{a} \hat{b} \rangle \langle
   3|1+2|\eta] \langle a|1+2|\eta] [1 2]
   [1 \eta]}\nonumber\\
   &-\frac{\langle
   b|a+1|\eta]^4}{\langle 2 3 \rangle \langle 3 b
   \rangle \langle \hat{a} b \rangle \langle \hat{a} \hat{b} \rangle
   \langle 2|a+1|\eta] \langle \hat{b}|a+1|\eta]
   [1 \eta] [a 1] [a \eta]}\nonumber\\
   &-\frac{\langle a
   1 \rangle^3 [3 \eta]^3}{\langle 1 2 \rangle
   \langle a \hat{b} \rangle \langle \hat{a} \hat{b} \rangle \langle
   2|3+b|\eta] \langle \hat{a}|3+b|\eta] [3 b]
   [b \eta]}\nonumber
\end{align*}
\begin{align*}
\ant&(\ha^{+},\hb^{+} \leftarrow a^-,1^+,2^-,3^-,b^+)=\nonumber\\
&-\frac{\langle 2 3 \rangle^3 \langle
   a|2+3+b|\eta]^4}{s_{b23} \langle 3 b \rangle
   \langle a 1 \rangle \langle a \hat{b} \rangle \langle
   \hat{a} \hat{b} \rangle \langle 2|3+b|\eta] \langle
   b|2+3|\eta] \langle 1|2+3+b|\eta] \langle
   \hat{a}|2+3+b|\eta]}\nonumber\\
   &+\frac{\langle 2 3 \rangle^3 \langle
   a|1+2+3|\eta]^3}{s_{123} \langle 1 2 \rangle
   \langle a \hat{b} \rangle \langle \hat{a} b \rangle \langle
   \hat{a} \hat{b} \rangle \langle 1|2+3|\eta] \langle
   3|1+2|\eta] \langle
   b|1+2+3|\eta]}\nonumber\\
   &-\frac{\langle a 2 \rangle^4 \langle
   3|a+1+2|\eta]^3}{s_{a12} \langle 1 2 \rangle
   \langle 3 b \rangle \langle a 1 \rangle \langle \hat{a}
   b \rangle \langle \hat{a} \hat{b} \rangle \langle 2|a+1|\eta]
   \langle a|1+2|\eta] \langle
   \hat{b}|a+1+2|\eta]}\nonumber\\
   &-\frac{\langle 2 3 \rangle^3 \langle
   a|1+2+3+b|\eta]^3}{s_{b123} \langle 1 2 \rangle
   \langle 3 b \rangle \langle a \hat{b} \rangle \langle
   \hat{a} \hat{b} \rangle \langle 1|2+3+b|\eta] \langle
   b|1+2+3|\eta] \langle
   \hat{a}|1+2+3+b|\eta]}\nonumber\\
   &+\frac{\langle a 3 \rangle^4 [1
   \eta]^3}{\langle 3 b \rangle \langle a \hat{b} \rangle
   \langle \hat{a} b \rangle \langle \hat{a} \hat{b} \rangle \langle
   3|1+2|\eta] \langle a|1+2|\eta] [1 2]
   [2 \eta]}\nonumber\\
   &+\frac{\langle
   a|2+3|\eta]^4}{\langle a 1 \rangle \langle a \hat{b}
   \rangle \langle \hat{a} b \rangle \langle \hat{a} \hat{b} \rangle
   \langle 1|2+3|\eta] \langle b|2+3|\eta]
   [2 3] [2 \eta] [3 \eta]}\nonumber\\
   &-\frac{\langle 2
   3 \rangle^3 [1 \eta]^3}{\langle 3 b \rangle
   \langle \hat{a} b \rangle \langle \hat{a} \hat{b} \rangle \langle
   2|a+1|\eta] \langle \hat{b}|a+1|\eta] [a 1]
   [a \eta]}\nonumber\\
   &-\frac{\langle a 2 \rangle^4 [b
   \eta]^3}{\langle 1 2 \rangle \langle a 1 \rangle
   \langle a \hat{b} \rangle \langle \hat{a} \hat{b} \rangle \langle
   2|3+b|\eta] \langle \hat{a}|3+b|\eta] [3 b]
   [3 \eta]}\nonumber
   \end{align*}
   \begin{align*}
\ant&(\ha^{+},\hb^{+} \leftarrow a^-,1^+,2^-,3^+,b^-)=\nonumber\\
&-\frac{\langle 2 b \rangle^4 \langle
   a|2+3+b|\eta]^4}{s_{b23} \langle 2 3 \rangle
   \langle 3 b \rangle \langle a 1 \rangle \langle a
   \hat{b} \rangle \langle \hat{a} \hat{b} \rangle \langle
   2|3+b|\eta] \langle b|2+3|\eta] \langle
   1|2+3+b|\eta] \langle
   \hat{a}|2+3+b|\eta]}\nonumber\\
   &+\frac{\langle a b \rangle^4 \langle
   2|1+3|\eta]^4}{s_{123} \langle 1 2 \rangle \langle
   2 3 \rangle \langle a \hat{b} \rangle \langle \hat{a} b \rangle
   \langle \hat{a} \hat{b} \rangle \langle 1|2+3|\eta]
   \langle 3|1+2|\eta] \langle a|1+2+3|\eta]
   \langle b|1+2+3|\eta]}\nonumber\\
   &-\frac{\langle a 2 \rangle^4
   \langle b|a+1+2|\eta]^4}{s_{a12} \langle 1 2
   \rangle \langle 3 b \rangle \langle a 1 \rangle
   \langle \hat{a} b \rangle \langle \hat{a} \hat{b} \rangle \langle
   2|a+1|\eta] \langle a|1+2|\eta] \langle
   3|a+1+2|\eta] \langle
   \hat{b}|a+1+2|\eta]}\nonumber\\
   &-\frac{\langle 2 b \rangle^4 \langle
   a|1+2+3+b|\eta]^3}{s_{b123} \langle 1 2 \rangle
   \langle 2 3 \rangle \langle 3 b \rangle \langle a
   \hat{b} \rangle \langle \hat{a} \hat{b} \rangle \langle
   1|2+3+b|\eta] \langle b|1+2+3|\eta] \langle
   \hat{a}|1+2+3+b|\eta]}\nonumber\\
   &-\frac{\langle a 2 \rangle^4
   \langle b|a+1+2+3|\eta]^3}{s_{a123} \langle 1 2
   \rangle \langle 2 3 \rangle \langle a 1 \rangle
   \langle \hat{a} b \rangle \langle \hat{a} \hat{b} \rangle \langle
   3|a+1+2|\eta] \langle a|1+2+3|\eta] \langle
   \hat{b}|a+1+2+3|\eta]}\nonumber\\
   &+\frac{\langle a b \rangle^4 [1
   \eta]^3}{\langle 3 b \rangle \langle a \hat{b} \rangle
   \langle \hat{a} b \rangle \langle \hat{a} \hat{b} \rangle \langle
   3|1+2|\eta] \langle a|1+2|\eta] [1 2]
   [2 \eta]}\nonumber\\
   &+\frac{\langle a b \rangle^4 [3
   \eta]^3}{\langle a 1 \rangle \langle a \hat{b} \rangle
   \langle \hat{a} b \rangle \langle \hat{a} \hat{b} \rangle \langle
   1|2+3|\eta] \langle b|2+3|\eta] [2 3]
   [2 \eta]}\nonumber\\
   &-\frac{\langle 2 b \rangle^4 [1
   \eta]^3}{\langle 2 3 \rangle \langle 3 b \rangle
   \langle \hat{a} b \rangle \langle \hat{a} \hat{b} \rangle \langle
   2|a+1|\eta] \langle \hat{b}|a+1|\eta] [a 1]
   [a \eta]}\nonumber\\
   &-\frac{\langle a 2 \rangle^4 [3
   \eta]^3}{\langle 1 2 \rangle \langle a 1 \rangle
   \langle a \hat{b} \rangle \langle \hat{a} \hat{b} \rangle \langle
   2|3+b|\eta] \langle \hat{a}|3+b|\eta] [3 b]
   [b \eta]}\nonumber
   \end{align*}
    \begin{align*}
\ant&(\ha^{+},\hb^{+} \leftarrow a^+,1^-,2^-,3^-,b^+)=\nonumber\\
&-\frac{\langle 2 3 \rangle^3 \langle
   1|2+3+b|\eta]^3}{s_{b23} \langle 3 b \rangle
   \langle a 1 \rangle \langle a \hat{b} \rangle \langle
   \hat{a} \hat{b} \rangle \langle 2|3+b|\eta] \langle
   b|2+3|\eta] \langle
   \hat{a}|2+3+b|\eta]}\nonumber\\
   &-\frac{\langle 1 2 \rangle^3 \langle
   3|a+1+2|\eta]^3}{s_{a12} \langle 3 b \rangle
   \langle a 1 \rangle \langle \hat{a} b \rangle \langle
   \hat{a} \hat{b} \rangle \langle 2|a+1|\eta] \langle
   a|1+2|\eta] \langle
   \hat{b}|a+1+2|\eta]}\nonumber\\
   &+\frac{\langle
   3|1+2|\eta]^3}{\langle 3 b \rangle \langle a \hat{b}
   \rangle \langle \hat{a} b \rangle \langle \hat{a} \hat{b} \rangle
   \langle a|1+2|\eta] [1 2] [1 \eta] [2
   \eta]}\nonumber\\
   &+\frac{\langle 1|2+3|\eta]^3}{\langle a 1
   \rangle \langle a \hat{b} \rangle \langle \hat{a} b \rangle
   \langle \hat{a} \hat{b} \rangle \langle b|2+3|\eta] [2
   3] [2 \eta] [3 \eta]}\nonumber\\
   &-\frac{\langle 2 3
   \rangle^3 [a \eta]^3}{\langle 3 b \rangle \langle
   \hat{a} b \rangle \langle \hat{a} \hat{b} \rangle \langle
   2|a+1|\eta] \langle \hat{b}|a+1|\eta] [1 \eta]
   [a 1]}\nonumber\\
   &-\frac{\langle 1 2 \rangle^3 [b
   \eta]^3}{\langle a 1 \rangle \langle a \hat{b} \rangle
   \langle \hat{a} \hat{b} \rangle \langle 2|3+b|\eta]
   \langle \hat{a}|3+b|\eta] [3 b] [3 \eta]}\nonumber
   \end{align*}
 \begin{align*}
\ant(\ha^{+},\hb^{+} \leftarrow a^+,1^-,2^-,3^+,b^-)&=-\ant(\hb^{+},\ha^{+} \leftarrow b^-,3^+,2^-,1^-,a^+)\nonumber\\
\ant(\ha^{+},\hb^{+} \leftarrow a^+,1^-,2^+,3^-,b^-)&=-\ant(\hb^{+},\ha^{+} \leftarrow b^-,3^-,2^+,1^-,a^+)\nonumber\\
\ant(\ha^{+},\hb^{+} \leftarrow a^+,1^+,2^-,3^-,b^-)&=-\ant(\hb^{+},\ha^{+} \leftarrow b^-,3^-,2^-,1^+,a^+)\nonumber
\end{align*}

\bibliography{database}

\end{document}